\documentclass[%
reprint,
superscriptaddress,
amsmath,amssymb,
aps,
prb,
]{revtex4-1}

\usepackage{graphicx}
\usepackage{dcolumn}
\usepackage{bm}
\usepackage{color}

\begin{document}

\preprint{APS/123-QED}

\title{Ground-state atlas of a three-dimensional semimetal in the quantum limit}

\author{Zhiming Pan}
\affiliation{International Center for Quantum Materials, School of Physics, Peking University, Beijing 100871, China}
\affiliation{Collaborative Innovation Center of Quantum Matter, Beijing 100871, China}%
\author{Ryuichi Shindou}%
\email{rshindou@pku.edu.cn}
\affiliation{International Center for Quantum Materials,  School of Physics, Peking University, Beijing 100871, China}%
\affiliation{Collaborative Innovation Center of Quantum Matter, Beijing 100871, China}%

\date{\today}

\begin{abstract}
An interplay between electron correlation and reduced dimensionality due to 
the Landau quantization gives rise to exotic electronic phases in three-dimensional 
semimetals under high magnetic field. Using an unbiased theoretical method, 
we clarify for the first time comprehensive ground-state phase diagrams of a 
three-dimensional semimetal with a pair of electron and hole pockets in the quantum 
limit. For the electron interaction, we consider 
either screened Coulomb repulsive interaction or an attractive 
electron-electron interaction mediated by a screened electron-phonon coupling, 
where a screening length is generally given by a dimensionless constant times 
magnetic length $l$. By solving the parquet RG equation numerically and employing 
a mean-field argument, we construct comprehensive ground-state phase 
diagrams of the semimetal in the quantum limit 
for these two cases, as a function of the Fermi wave length and the screening length 
(both normalized by $l$). In the repulsive interaction case, the ground state is either 
excitonic insulator (EI) in strong screening regime or 
Ising-type spin density wave in weak screening regime. In the attractive interaction case, 
the ground state is either EI that breaks the translational symmetries (strong screening regime), 
topological EI, charge Wigner crystal (intermediate screening regime), plain charge density 
wave or possible non-Fermi liquid (weak screening regime). We show that the topological EI 
supports a single copy of massless Dirac fermion at its side surface, and thereby exhibit 
a $\sqrt{H_{\perp}}$-type surface Shubnikov-de Haas (SdH) oscillation in in-plane surface 
transports as a function of a canted magnetic field $H_{\perp}$. Armed with these 
theoretical knowledge, we discuss implications of recent transport experiments on 
graphite under the high field.

\end{abstract}


\maketitle

\section{introduction}
One of the fundamental challenge in condensed matter physics 
is a realization of three-dimensional unconventional electronic phases 
in the quantum limit~\cite{halperin87}. Recent experimental 
discoveries of Dirac, Weyl and nodal Dirac 
semimetal materials~\cite{weyl-dirac18} lead to growing research interests on novel 
quantum transports and quantum phase transitions under high magnetic field 
in these new compounds~\cite{tang19,zhang19,fujioka19,zhang17,ramshaw18} 
as well as celebrated semimetal compounds such 
as bismuth~\cite{zhu12,zhu17a} and graphite~\cite{fauque13,zhu19,arnold17,leboeuf17}. 
In fact, these semimetallic compounds under the high 
field often exhibit low-temperature metal-insulator transitions within wide ranges 
of the field~\cite{tang19,fujioka19,yaguchi09,fauque13,akiba15,arnold17,leboeuf17,zhu17b,taen18,zhu19}.

The quantizing effect of the strong magnetic field confines electrons into cyclotron 
motions in the Landau levels, while the electron's kinetic energy along the field direction remains 
unaffected. This leaves the system with pristine one-dimensional momentum-energy dispersions 
along the field, making the system extremely sensitive to various instabilities~\cite{gruner00}. 
Previous theories proposed a number of spontaneous symmetry broken (SSB) phases as well as 
non-Fermi liquid phase\cite{yakovenko93}. The SSB phases proposed include charge-density 
wave~\cite{halperin87,macdonald87,celli65,lee69,fukuyama78,yoshioka81,tesanovic87,zhang17}, 
three-dimensional quantum Hall~\cite{halperin87,balents96,bernevig07,zhang17}, 
charge-Wigner crystal~\cite{kleppmann75,fukuyama78,biagini01,tsai02a,tsai02b,alicea09}, 
spin-density wave\cite{takahashi94,takada98,yaguchi09,pan18}, 
excitonic insulator~\cite{fenton68,jerome67,pan18,song17,pan18}, 
valley-density wave, and three-dimensional topological excitonic insulator~\cite{pan18}. 
Recent theoretical efforts on Dirac and Weyl semimetal models can be  
found in Ref.~\onlinecite{song17,trescher17}. In spite of these efforts during last decades, 
indentities of the low-temperature insulating phases in the experiments are still veiled in 
mystery due to a lack of comprehensive microscopic theory based on an unbiased theoretical 
method.

An electronic state of the prototypical semimetal materials can be captured 
by a  pair of electron and hole band. Under the magnetic field 
($\parallel$ $z$), the electron/hole's motions in the $xy$ plane are confined 
into clockwise/anticlockwise cyclotron orbits around the field respectively. 
The counter-propagating cyclotron motion in the $xy$ plane inspires 
an `electron-hole' analogy of the two-dimensional quantum spin Hall 
physics~\cite{qi11,hasan10}. The one-dimensional dispersions along $z$ of 
the electron and hole bands go across the Fermi level at several Fermi points 
in the Brillouin zone. The experimental Hall conductivity measurements 
conclude that the relevant semimetal material within the 
relevant field regime~\cite{yaguchi09,fauque13,akiba15,arnold17,leboeuf17,zhu17b,taen18,zhu19} 
are in the charge neutrality region, where electron and hole densities compensate 
with each other completely~\cite{uji98,kopelevich09,kumar10,akiba15,pan18,zhu19}. 
Thereby, to uncover the identities of the low-temperature insulator phases in the 
experiments~\cite{yaguchi09,fauque13,akiba15,arnold17,leboeuf17,zhu17b,taen18,zhu19,liang19}, 
it is vital to understand a ground-state phase diagram of a microscopic  
Hamiltonian for the semimetal material in the quantum limit at their  
charge neutrality point. 

\begin{figure}[tp]
\begin{center}
	\includegraphics[width=1.0\linewidth]{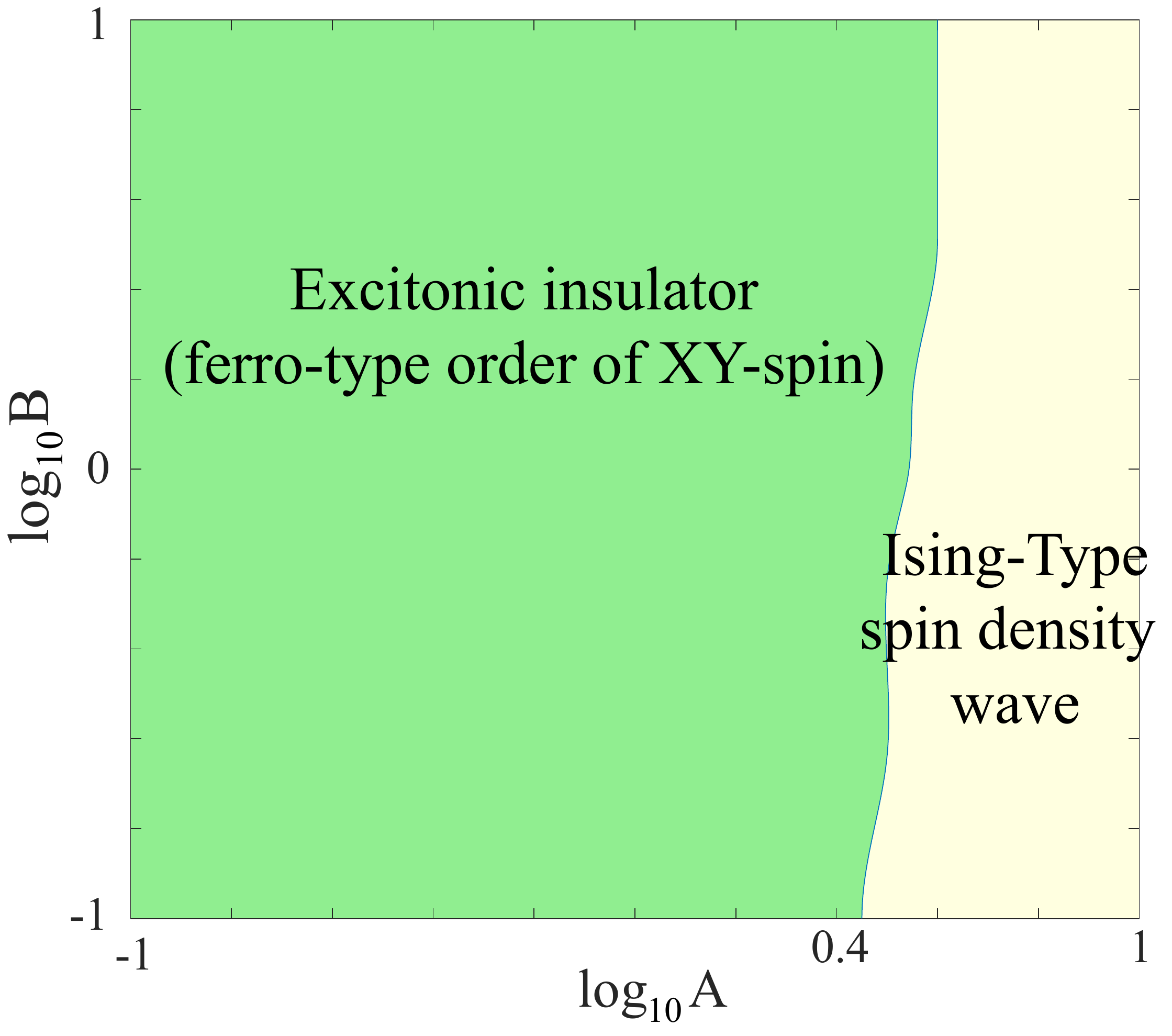}
	\caption{Electronic phase diagram of a semimetal model in the quantum limit 
at the charge neutrality point in the presence of the repulsive Coulomb interaction. 
The phase diagram is obtained from numerical solutions of the parquet RG equations. 
The vertical axis is $B \equiv 2\log_{10} (2k_{F}l)$ [$2k_F$ is a distance between the 
right and left Fermi points along the field, and $l$ is the magnetic length]. 
The horizontal axis is $\log_{10} A$, where $\sqrt{A}$ is the screening length divided by 
the magnetic length. We set $A=A^{\prime}$; see the main text. 
Within the RPA, $A$ and $A^{\prime}$ are evaluated as in Eqs.~(\ref{0-pi},\ref{2kf-pi}) 
respectively.}
	\label{fig:1}
\end{center}
\end{figure} 

\begin{figure}[bp]
\begin{center}	\includegraphics[width=1.0\linewidth]{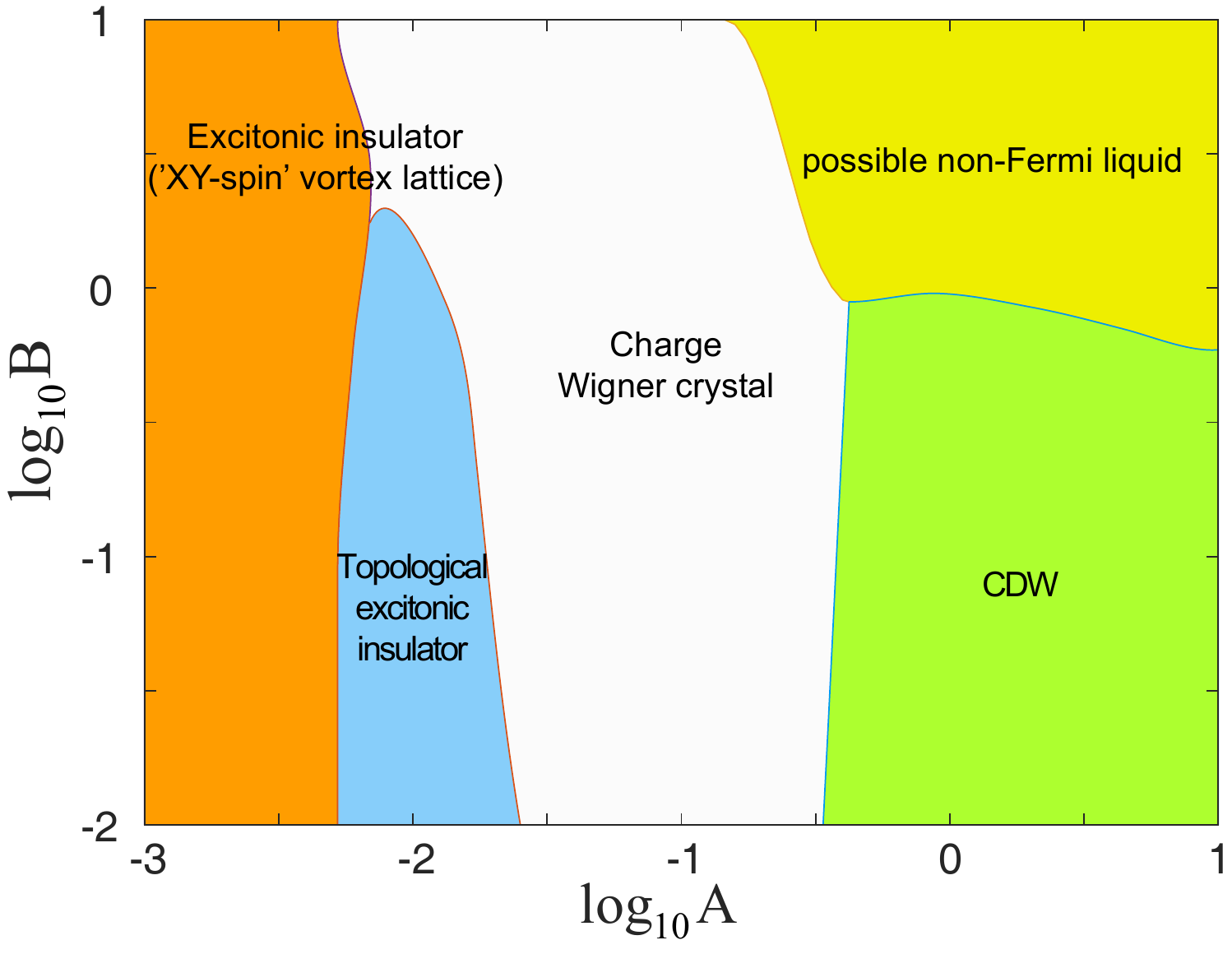}
	\caption{Electronic phase diagram of a semimetal model in the quantum limit 
at the charge neutrality point in the presence of the electron-phonon 
interaction ($A=A^{\prime}$). 
The phase diagram is obtained from numerical solutions of 
the parquet RG equations for the semimetal model with an effective attractive 
electron-electron interaction, Eq.~(\ref{effective-attractive}). The vertical axis is 
$B \equiv 2\log_{10} (2k_{F}l)$. The horizontal axis is $\log_{10} A$.}
	\label{fig:2}
\end{center}
\end{figure}

In this paper, using an unbiased theoretical method, we clarify the comprehensive 
ground-state phase diagrams of a prototype model 
for a three-dimensional semimetal under the magnetic field $H$ ($\parallel \!\ z$). 
The semimetal model has a pair of electron pocket with $\uparrow$ spin
and hole pocket with $\downarrow$ spin. We study two limiting cases 
at the charge neutrality point, (i) the model with screened Coulomb 
interaction and (ii) the model with an effective attractive interaction 
mediated by the screened electron-phonon interaction.  
We rederive parquet renormalization group (RG) equations, that was originally 
given by Brazovskii ~\cite{abrikosov70, brazovskii72,yakovenko93}, 
solve numerically the parquet RG equations, and complete the ground-state 
electronic phase diagrams for these two cases. 
The ground-state phase diagrams thus 
obtained exhibit rich phase diagram structures as a function of three 
important length scales in the system, ($\alpha$) magnetic length 
$l \equiv \sqrt{\hbar/eH}$ , ($\beta$) 
Fermi wave length (along the field) $2\pi/(2k_F)$, and ($\gamma$) 
screening length of the long-range Coulomb interaction, $\sqrt{A}l$ [see Eq.~(\ref{screen-0}) for 
the definition of a dimensionless parameter $A$, see Eq.~(\ref{0-pi}) for its RPA evaluation]. 
In the repulsive interaction case, we found that the ground state is either Ising-type spin density wave 
phase or excitonic insulator (EI) phase (Fig.~\ref{fig:1}). In the attractive interaction case, we 
found that the ground state is 
either charge Wigner crystal phase, 
charge density wave phase, possible non-Fermi liquid~\cite{yakovenko93} 
phase, EI phase that breaks the translational symmetry within the $xy$ plane or three-dimensional 
topological EI phase with a topological surface massless Dirac 
state~\cite{pan18} (Fig.~\ref{fig:2}). 
We show in the paper that a canted 
magnetic field $H_{\perp}$ splits the surface Dirac state into surface Landau levels (sLL). 
The result suggests that when in-plane transport in the topological EI phase is dominated 
by surface transport, the in-plane resistivity must show a $\sqrt{H_{\perp}}$-type 
surface Shubnikov-de Haas (SdH) oscillation under the canted magnetic field.

\subsection{highlight of the paper}
In the next section, we introduce a model Hamiltonian for the semimetal material 
in the quantum limit, where a pair of the electron pocket with $\uparrow$ spin 
and hole pocket with $\downarrow$ spin go across the Fermi level at $p_z=\pm k_{F}$ 
at the charge neutrality point [$p_z$ is momentum along the field direction].  
By using the RPA approximation, we discuss in Sec.~III how the long-range 
Coulomb interaction $V({\bm r})\equiv e^2/{\varepsilon r}$ 
is screened by a density fluctuation at $p_z=0$ and a fast 
mode of the density fluctuation at $p_z=2k_F$. The screened 
Coulomb interactions thus obtained take the following forms in the 
momentum space~\cite{tsvelik18}, 
\begin{align}
V(q_{\perp},p_z\simeq 0) &= \frac{4\pi e^2 l^2}{\varepsilon}\frac{1}{q^2_{\perp} l^2 
+ \frac{1}{A} e^{-\frac{1}{2}q^2_{\perp}l^2}}, \label{screen-0} \\
 V(q_{\perp},p_z\simeq 2k_F) &= \frac{4\pi e^2 l^2}{\varepsilon}\frac{1}{(q^2_{\perp} + 4k^2_F) l^2 
+ \frac{1}{A^{\prime}} e^{-\frac{1}{2}q^2_{\perp}l^2}}, \label{screen-2kf} 
\end{align}  
where $q_{\perp}$ is momentum within the $xy$ plane. Here two dimensionless 
parameters $1/A$ and $1/A^{\prime}$ are nothing but bare polarization 
functions associated with the density fluctuation at $p_z=0$ and 
the fast mode of the density fluctuation at $p_z=2k_F$ respectively. Eq.~(\ref{screen-0})  
especially indicates that the screening length of the long-range Coulomb interaction is 
given by the magnetic length times the dimensionless parameter $\sqrt{A}$; larger/smaller  
$A$ stands for the weak/strong screening respectively.   
 
The polarization function associated with the slow density fluctuation at $p_z=2k_F$ as 
well as the polarization function for an excitonic 
fluctuation have logarithmic singularities~\cite{abrikosov70,brazovskii72}. These singularities  
indicate several competing instabilities 
in the semimetal model at lower temperature. To identify the most dominant 
instability in the model within 
a controlled theoretical framework, we rederive in Sec.~IV the one-loop 
parquet renormalization group (RG) equations, where the 
slow $2k_F$ density fluctuation as well as other 
fluctuations with the logarithmic singularity are taken into account on the same footing~\cite{brazovskii72,zheleznyak97}. 
Using the screened Coulomb interaction [Eqs.~(\ref{screen-0},\ref{screen-2kf})] 
as an `initial' interaction form for the RG equations, 
we solve the parquet RG equations numerically and identify the 
most relevant fluctuations (instabilities) at lower temperature for different values of 
$A$, $A^{\prime}$ and $2k_F l$ [an overall factor of $V(q_{\perp},p_z)$, 
$(4\pi e^2 l^2)/\varepsilon$, can be absorbed into a RG scale change at the 
one-loop RG equation; it does not change the phase diagram]. 
By combining mean-field arguments with the numerical 
RG solutions, we construct in Sec.~V a comprehensive ground-state phase diagram in the 
presence of the repulsive Coulomb interaction (Fig.~\ref{fig:1}).  

As the complimentary aspect for the semimetal in the quantum limit, we also 
study in Sec.~VI an effect of electron-phonon interaction in the semimetal model under 
high magnetic field. Thereby, we employ a correspondence between an 
electron-phonon coupled system and a system with an electron-electron interaction, and 
adopt an effective attractive electron-electron interaction~\cite{fetter03,mahan00}. 
The effective interaction is mediated by the {\it screened} 
Coulomb interaction between electron and (acoustic) phonon, and thereby 
it takes the following form in the momentum space,
\begin{align}
V_{\rm eff}(q_{\perp},p_z) = - \frac{\rho_0}{Mc^2}\bigg(\frac{4\pi Z e^2 l^2}{\varepsilon} 
\frac{1}{(q^2_{\perp}+p^2_z) l^2 + \frac{1}{A} e^{-\frac{1}{2}q^2_{\perp} l^2}} \bigg)^2 
\label{effective-attractive}
\end{align} 
Here $Z$ and $c$ are an electron valence of positively charged nucleus  ion and a sound velocity 
of the acoustic phonon respectively, $\rho_0$ and $M$ are the density of the charged nucleus ions 
and a mass of the ion respectively. 
To clarify possible instabilities in the semimetal 
model in the presence of the electron-phonon interaction, 
we use this effective attractive interaction [Eq.~(\ref{effective-attractive})] as an initial 
interaction form for the parquet RG equations, and solve the RG equations numerically for different 
values of $A$, $A^{\prime}$ and $2k_F l$. The numerical solutions 
in combination with mean-field arguments gives out a comprehensive ground-state phase 
diagram in the presence of the effective attractive interaction (Fig.~\ref{fig:2}). 

The two ground-state phase diagrams thus obtained accommodate a rich variety of 
electronic phases as a function of $2k_F l$ and $A$. In the repulsive 
Coulomb interaction case (Sec.~V), the ground state (GS) for strong screening regime 
[$A \le 3$ for $2k_F l \simeq 1$] is an EI phase with a spatially even-parity 
excitonic pairing between electron and hole at the same momentum $p_z$. Since the 
pairing is between the electron with $\uparrow$ spin and the hole with $\downarrow$ spin 
and it is between the electron and hole at the {\it same} spatial location within 
the $xy$ plane, the excitonic pairing field leads to a long-ranged ferro-type order of a $XY$ component 
of the spin-1 moment. For the weak screening regime 
[$A \ge 3$ for $2k_F l \simeq 1$], the GS is a plain superposition of a density 
wave (DW) of the electron band with $\uparrow$ spin and a DW of the hole band 
with $\downarrow$ spin, that have the $2\pi/(2k_F)$ spatial pitch along the field.  
Due to the Coulomb interaction between the two DWs, a relative phase between 
the two DWs is locked to $\pi$.  Such superposition leads to an Ising-type spin density 
wave without any charge density modulation; the spatial pitch 
of the Ising-type antiferromagnetic order is $2\pi/(2k_F)$.  


In the attractive interaction case (Sec.~VI), the GS for a strong screening regime 
[$A\le 0.3$ for $2k_F l \simeq 1$] is either one of two distinct EI phases or 
the charge Wigner crystal. In one of the EI phases, the excitonic pairing is 
the spatially even-parity excitonic pairing between electron and hole at the same momentum 
$p_z$, but it is between the electron and hole at the {\it different} spatial location within 
the $xy$ plane [the field $\parallel \!\ z$]. As a result, the EI phase forms a two-dimensional 
texture of the $XY$ component of the spin-1 moment, breaking the translational symmetries 
within the $xy$ plane. The charge Wigner crystal phase breaks both the 
translational symmetries along the field and within the $xy$ plane by a 
three-dimensional texture of the charge density. The GS for the weak screening regime 
[$A\ge 0.3$ for $2k_F l \simeq 1$] is either a plain charge density wave 
phase with the $2\pi/(2k_F)$ spatial pitch along the field 
[$2k_F l < 1$ for $A \simeq 1$] or  a possible non-Fermi 
liquid phase [$2k_F l>1$ for $A \simeq 1$].  
 
The other EI phase found in the strong screening regime is a 
three-dimensional topological band insulator~\cite{pan18}; the EI phase supports a single copy 
of massless surface Dirac fermion state at its side surface [side surface 
is parallel to the field; $zx$ and $yz$ planes with the field along $z$]. 
The EI phase in the 
bulk is characterized by a spatially {\it odd}-parity pairing between electron 
and hole at the same momentum $p_z$ and at the same spatial location within 
the $xy$ plane. As a result, the EI phase does not break any translational symmetries.
Besides, it has no local $XY$ component of the spin-1 moment, since the odd parity 
leads to a cancellation between $p_z$ and $-p_z$. 
Meanwhile, the odd-parity excitonic pairing in the bulk reconstructs a surface chiral 
Fermi arc state of the electron band with $\uparrow$ spin and that of the hole band 
with $\downarrow$ spin into 
the massless surface Dirac state with a helical spin texture. 
According to the so-called `periodic table' of non-interacting topological insulator 
and topological crystalline insulator~\cite{schnyder08,kitaev09,shiozaki14,kruthoff17}, the 
EI phase can be classified as topological `magnetic crystalline' 
insulator~\cite{shiozaki14,sato19}, where the massless nature of the surface 
Dirac fermion is protected by a magnetic point group symmetry $C_{2,\perp}T$ 
[$C_{2,\perp}$ denotes a $\pi$ rotation that changes $z$ to $-z$, and $T$ is 
the time reversal].  To give a physical characterization to the topological 
EI phase, we show in Sec.~VII that a canted magnetic field $H_{\perp}$ 
splits the massless surface Dirac state into surface Landau levels (sLL), whose energy 
spacing is proportional to $\sqrt{H_{\perp}\Delta_0}$ [$\Delta_0$ is a 
strength of the excitonic pairing]. The result suggests that the longitudinal 
electric surface transport in the $xy$ direction can show a $\sqrt{H_{\perp}}$-type 
surface Shubnikov-de Haas (SdH) oscillation under the canted magnetic field. Based on these 
finding, we give a brief summary and discussion on the semimetal experiments in 
Sec. VIII.

\section{model Hamiltonian} 
An interacting electron model with a pair of electron pocket and hole pocket under 
high magnetic field $H$ ($\parallel z$) is considered;
\begin{align}
&\hat{H}_{T} = \int d^3 {\bm r} \hat{h}_{0}({\bm r}) + \frac{1}{2}
\int d^3 {\bm r} d^3 {\bm r}^{\prime} \hat{\rho}({\bm r}) \hat{\rho}({\bm r}^{\prime}) 
V({\bm r}-{\bm r}^{\prime}) \label{HT} \\
&\hat{h}_{0}({\bm r})  \equiv \sum_{\tau=+ (\uparrow), - (\downarrow)}\nonumber \\  
&\hspace{0.1cm} \bigg[ 
\Psi^{\dagger}_{e,\tau}({\bm r}) \Big\{ - E_g + \frac{(-\hbar^2 \nabla^2_z+{\bm \pi}^2)}{2m_e}
 - H_{\rm z} \tau \Big\} \Psi_{e,\tau}({\bm r}) \nonumber  \\
& \hspace{0.3cm} + \Psi^{\dagger}_{h,\tau}({\bm r}) 
\Big\{E_g -\frac{(-\hbar^2 \nabla^2_z+{\bm \pi}^2)}{2m_h}
 - H_{\rm z} \tau \Big\} \Psi_{h,\tau}({\bm r}) \bigg]  \label{h0} \\ 
&(\pi_x,\pi_y) = \left\{\begin{array}{c} 
(-i\hbar \partial_x, -i\hbar \partial_y + eHx)  \\
(-i\hbar \partial_x - eHy, -i\hbar \partial_y)  \\ 
\end{array}\right. \label{landau-gauge}
\end{align}
with ${\bm \pi} \equiv (\pi_x,\pi_y)$. $-E_g$ and $E_g$ are charge state energies of 
an electron-type and hole-type bands at the $\Gamma$ point. $m_e$ and $m_h$ are effective 
masses of the electron and hole band respectively, $m_e>0$ and $m_h>0$. $\tau=\pm$ refers 
to the spin $1/2$ degree of freedom ($+\equiv\!\ \uparrow$, $-\equiv \!\ \downarrow$). 
$H_z$ denotes the Zeeman field and we assume that the $g$ factor is isotropic in spin and 
same for electron and hole band.

An electron density $\hat{\rho}({\bm r})$ is a sum of the density of the electron 
band and that of the hole band, $\rho({\bm r}) \equiv 
\sum_{a=e,h} \sum_{\tau} \Psi^{\dagger}_{a,\tau}({\bm r})\Psi_{a,\tau}({\bm r})$. 
In this paper, we consider as the electron 
correlation $V({\bm r})$ either repulsive Coulomb interaction (Secs.~III,V) or an effective 
attractive electron-electron interaction mediated by the screened electron-phonon 
interaction (Sec.~VI).

\begin{figure}[t]
	\centering
	\includegraphics[width=0.9\linewidth]{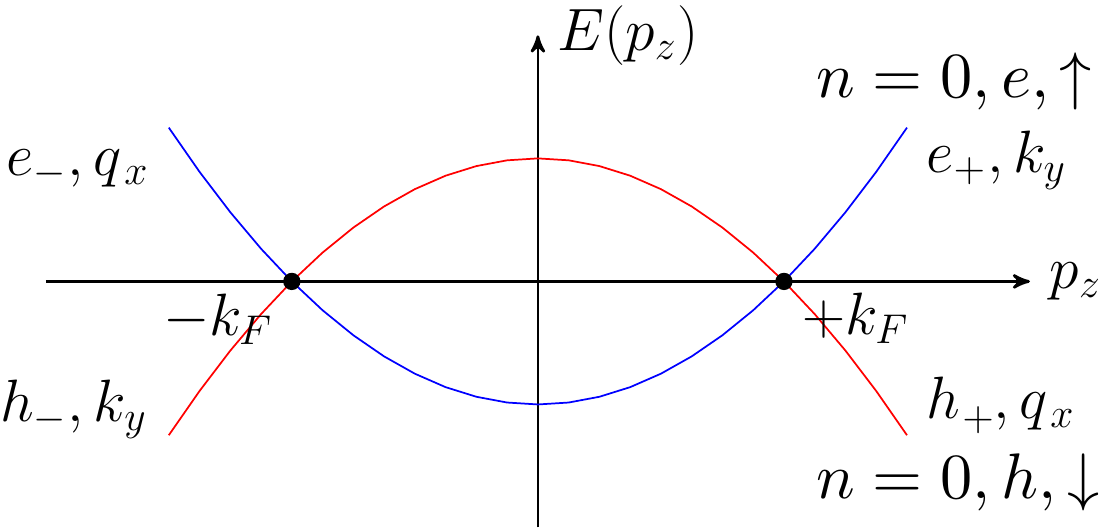}
	\caption{energy-momentum dispersion of the electron and hole pocket along the field direction.} 
	\label{fig:3}
\end{figure}

Due to the Landau quantization, the kinetic energy within 
a plane perpendicular to the field is quenched, where the electron band and hole band 
form a sequence of the Landau levels respectively;
\begin{align}
\left\{\begin{array}{c}
E^{e}_{n,\tau}(k_z) = - E_g + \frac{\hbar^2 p^2_z}{2m_e} -H_z \tau + 
\big(n+\frac{1}{2} \big) \hbar \omega_e \\
E^{h}_{n,\tau}(k_z) = E_g - \frac{\hbar^2 p^2_z}{2m_h}  - H_{z} \tau -  
\big(n+\frac{1}{2}\big) \hbar \omega_h \\
\end{array} \right.  \label{LL-bulk}
\end{align} 
with the cyclotron frequency $\hbar \omega_{e,h} \equiv eH/m_{e,h}$. 
we consider the charge neutrality region in the quantum limit, where only 
the lowest Landau levels ($n=0$) with $\uparrow$-spin electron and $\downarrow$-spin hole 
bands go across the Fermi level at the same Fermi points ($p_{z}=\pm k_F$) , 
while all the others Landau levels leave the Fermi level (Fig.~\ref{fig:3}). 
For simplicity, we assume that the electron mass and hole 
mass are same, $m_e=m_h \equiv m$, $\omega_e=\omega_h \equiv \omega$.

We linearize the kinetic energy along the field direction around the two Fermi points. 
This leads to the following low-energy Hamiltonian for $\hat{h}_0$,   
\begin{align}
&\int d^3{\bm r} \hat{h}_{0}({\bm r}) - \mu \hat{N} = 2\pi l \sum_{\sigma=\pm} 
\int^{+\Lambda}_{-\Lambda} \frac{dp}{2\pi} \!\ \sigma v_{F} p
\nonumber \\
&\hspace{0.2cm}   \int d(lQ)  \Big\{ e^{\dagger}_{\sigma}(Q,p) 
e_{\sigma}(Q,p)  -  h^{\dagger}_{\sigma}(Q,p) h_{\sigma}(Q,p)  \Big\} \label{kin}  
\end{align}
where the magnetic length $l \equiv \sqrt{\hbar/eH}$ and $v_F=\frac{\hbar^2 k_F}{m}$. 
$\sigma=\pm$ distinguishes two Fermi points, $k_z=\pm k_F$. $Q$ in Eq.~(\ref{kin}) 
denotes momentum within 
the $xy$ plane. For the `right-mover' fermion with positive velocity along the field, $e_{+}(Q,p)$ 
and $h_{-}(Q,p)$, we use a Landau gauge with an eigenstate localized along $x$-direction 
($x$-gauge); $Q$ is momentum along $y$-direction, $Q=k_y$. For the `left-mover' fermion 
with the negative velocity, $e_{-}(Q,p)$ and $h_{+}(Q,p)$, we use the Landau gauge 
with an eigenstate localized along $y$ ($y$-gauge);  $Q$ is the momentum along $x$, 
$Q=q_x$~\cite{yakovenko93}. To be more specific, 
electron-band and hole-band creation operators in Eq.~(\ref{h0}) are expanded as,  
\begin{align}
\Psi_{e,\uparrow}({\bm r}) &= \int^{+\Lambda}_{-\Lambda} 
\frac{dp}{2\pi} e^{i(k_F+p) z}\int d(l k_y) \psi_{k_y}(x,y) e_{+}(k_y,p) \nonumber \\
& \hspace{-0.7cm} + \int^{+\Lambda}_{-\Lambda} 
\frac{dp}{2\pi} e^{i(-k_F+p) z}\int d(l q_x) \phi_{q_x}(x,y) e_{-}(q_x,p), \label{Psi-e-up} \\ 
\Psi_{h,\downarrow}({\bm r}) &= \int^{+\Lambda}_{-\Lambda} 
\frac{dp}{2\pi} e^{i(k_F+p) z}\int d(l q_x) \phi_{q_x}(x,y) h_{+}(q_x,p) \nonumber \\
& \hspace{-0.8cm} + \int^{+\Lambda}_{-\Lambda} 
\frac{dp}{2\pi} e^{i(-k_F+p) z}\int d(l k_y) \psi_{k_y}(x,y) h_{-}(k_y,p), \label{Psi-h-down} 
\end{align}    
with the eigenstates in the LLL, 
\begin{align}
\psi_{k_y}(x,y) &\equiv \frac{1}{\sqrt{\sqrt{\pi} l}} e^{-\frac{1}{2l^2}(x-k_yl^2)^2} e^{-ik_y y}, \label{x-gauge} \\ 
\phi_{q_x}(x,y) &\equiv \frac{1}{\sqrt{\sqrt{\pi} l}} e^{-\frac{1}{2l^2}(y-q_xl^2)^2} e^{i(q_x - \frac{y}{l^2})x}. 
\label{y-gauge} 
\end{align}
For later convenience, note that the $x$-gauge eigenstates 
and $y$-gauge eigenstates are transformed to each other by a Fourier transformation;
\begin{align}
&\int^{\infty}_{-\infty}\frac{d(l k_y)}{\sqrt{2\pi}} \!\ 
e^{iq_x k_y l^2} \!\ \psi_{k_y}(x,y) = \phi_{q_x}(x,y), \label{ytox} \\
& \int^{\infty}_{-\infty}\frac{d(l q_x)}{\sqrt{2\pi}} \!\ 
e^{-iq_x k_y l^2} \!\ \phi_{q_x}(x,y) = \psi_{k_y}(x,y). \label{xtoy} 
\end{align} 

By substituting Eqs.~(\ref{Psi-e-up},\ref{Psi-h-down}) into the electron-electron interaction 
in Eq.~(\ref{HT}), we obtain, 
\begin{align}
&\frac{1}{2} 
\int d^3{\bm r} d^3 {\bm r}^{\prime} \hat{\rho}({\bm r}) \hat{\rho}({\bm r}^{\prime}) V({\bm r}-{\bm r}^{\prime}) 
 \equiv H_1 + H_2, \label{int}
\end{align} 
\begin{widetext}
\begin{align}
&
H_1 \equiv \frac{1}{2} \sum_{\mu,\nu=e_{+},e_{-},h_{+},h_{-}} \int \frac{dp}{2\pi} 
\int d(lQ_1) \int d(lQ^{\prime}_1) \int d(lQ^{\prime}_2) 
\int d(lQ_2)  \nonumber \\
& \hspace{2.7cm} 
\Gamma_{\mu\nu}(Q_1,Q^{\prime}_1,Q^{\prime}_{2},Q_{2};I_0) 
\int \frac{dp_2}{2\pi}
a^{\dagger}_{\mu}(Q^{\prime}_1,p_2+p) a_{\mu}(Q^{\prime}_2,p_2)  \!\ 
\int \frac{dp_1}{2\pi}
a^{\dagger}_{\nu}(Q_1,p_1-p) a_{\nu}(Q_2,p_1),  \label{int-h1} \\  
&  H_2 \equiv  \sum_{\mu,\nu=e,h} \int \frac{dp}{2\pi} 
\int d(lQ_1) \int d(lQ^{\prime}_1) \int d(lQ^{\prime}_2) 
\int d(lQ_2)  \nonumber \\
& \hspace{1.4cm} 
\Phi^{+-}_{\mu\nu}(Q_1,Q^{\prime}_1,Q^{\prime}_{2},Q_{2};I_{2k_{F}}) 
\int \frac{dp_2}{2\pi}
a^{\dagger}_{\mu_{+}}(Q^{\prime}_1,p_2+p) a_{\mu_{-}}(Q^{\prime}_2,p_2)  \!\ 
\int \frac{dp_1}{2\pi}
a^{\dagger}_{\nu_{-}}(Q_1,p_1-p) a_{\nu_{+}}(Q_2,p_1),   \label{int-h2}
\end{align}
\end{widetext}
with a notation of 
\begin{align} 
\left\{\begin{array}{c}
a_{e_{+}}(Q,p) \equiv e_{+}(k_y, p), \\
a_{e_{-}}(Q,p) \equiv e_{-}(q_x,p), \\
a_{h_{+}}(Q,p) \equiv h_{+}(q_x,p),  \\
a_{h_{-}}(Q,p) \equiv h_{-}(k_y,p). \\ 
\end{array}\right. \label{notation-a}
\end{align}
$H_1$ is a sum of all the interactions that carry the zero 
momentum along the field (Fig.~\ref{fig:4}(a)), while $H_2$ is a sum of all the interactions 
that carry the $2k_F$ momentum along the field (Fig.~\ref{fig:4}(c)). The respective interaction 
potentials are given by {\it functionals} of following two 
`bare' functions of the in-plane momentum $(q_x,k_y)$, 
\begin{align}
I_{0}(q_x,k_y) & \equiv V(q_x,k_y,p_z=0) e^{-\frac{1}{2} (q^2_x+k^2_y)l^2}, \label{I0-bare} \\
I_{2k_F}(q_x,k_y) & \equiv V(q_x,k_y,p_z=2k_F) e^{-\frac{1}{2} (q^2_x+k^2_y)l^2}, \label{I2kf-bare} 
\end{align}
with $V(q_x,k_y,p_z)\equiv \frac{4\pi e^2}{\varepsilon (q^2_x+k^2_y+p^2_z)}$. 
Specifically, $\Gamma_{\mu\nu}$ for $(\mu,\nu)=(e_{+},e_{+})$, $(h_{-},h_{-})$, $(e_{+},h_{-})$, $(h_{-},e_{+})$, 
$\Gamma_{\mu\nu}$ for $(\mu,\nu)=(e_{-},e_{-})$, $(h_{+},h_{+})$, $(e_{-},h_{+})$, $(h_{+},e_{-})$, 
$\Gamma_{\mu\nu}$ for $(\mu,\nu)=(e_{+},e_{-})$, $(h_{-},h_{+})$, $(e_{+},h_{+})$, $(h_{-},e_{-})$, 
and $\Gamma_{\mu\nu}$ for $(\mu,\nu)=(e_{-},e_{+})$, $(h_{+},h_{-})$, $(h_{+},e_{+})$, $(e_{-},h_{-})$  
are given by the same functionals of $I_{0}(q_x,k_y)$ respectively;  
\begin{align}
&\Gamma_{e_{+}e_{+}}(k_1,k^{\prime}_1,k^{\prime}_2,k_2;I_0) =  \cdots = \Gamma_{h_{-}e_{+}}
(k_1,k^{\prime}_1,k^{\prime}_2,k_2;I_0) \nonumber \\
&\hspace{0.6cm}  = \delta(k_1+k^{\prime}_1-k_2-k^{\prime}_{2})  \int dq_x \nonumber \\
&  \hspace{1.8cm}  I_0(q_x,-k_1+k_2) 
e^{-i\frac{1}{2}q_x(k_1+k_2-k^{\prime}_1-k^{\prime}_2)l^2}, \label{e+e+} \\
&\Gamma_{e_{-}e_{-}}(q_1,q^{\prime}_1,q^{\prime}_2,q_2;I_0) =  \cdots = 
\Gamma_{h_{+}e_{-}}(q_1,q^{\prime}_1,q^{\prime}_2,q_2;I_0) \nonumber \\
&\hspace{0.6cm} = \delta(q_1+q^{\prime}_1-q_2-q^{\prime}_{2}) \int dk_y  \nonumber \\
&  \hspace{1.8cm}  I_0(q_1-q_2,k_y) 
e^{i\frac{1}{2}k_y(q_1+q_2-q^{\prime}_1-q^{\prime}_2)l^2}, \label{e-e-} \\ 
&\Gamma_{e_{+}e_{-}}(q_1,k_1,k_2,q_2;I_0) =  \cdots = 
\Gamma_{h_{-}e_{-}}(q_1,k_1,k_2,q_2;I_0) \nonumber \\
&  \hspace{0.6cm} = e^{i{\bm k}_1 \wedge {\bm k}_2 l^2} I_0(q_1-q_2,-k_1+k_2),  \label{e+e-} \\
%
&\Gamma_{e_{-}e_{+}}(k_1,q_1,q_2,k_2;I_0) =  \cdots = 
\Gamma_{e_{-}h_{-}}(k_1,q_1,q_2,k_2;I_0) \nonumber \\
&  \hspace{0.6cm} = e^{i{\bm k}_1 \wedge {\bm k}_2 l^2} I_0(q_1-q_2,-k_1+k_2), \label{e-e+} 
\end{align}
with 
\begin{align}
&{\bm k}_1 \wedge {\bm k}_2 \equiv \left(\begin{array}{c}
k_1 \\
q_1 \\
\end{array}\right) \wedge  \left(\begin{array}{c}
k_2 \\
q_2 \\
\end{array}\right) = k_1 q_2 - q_1 k_2. \label{wedge}  
\end{align}   
$\Phi^{+-}_{\mu\nu}$ for $(\mu,\nu)=(e,e),(h,h)$, $\Phi^{+-}_{\mu\nu}$ for $(\mu,\nu)=(e,h)$, 
and $\Phi^{+-}_{\mu\nu}$ for $(\mu,\nu)=(h,e)$ are given by the 
following functionals of $I_{2k_F}(q_x,k_y)$, 
\begin{align}
&\Phi^{+-}_{ee}(q_1,k_1,q_2,k_2;I_{2k_F}) = \Phi^{+-}_{hh}(k_1,q_1,k_2,q_2;I_{2k_F})   \nonumber \\
&  = l^2 e^{i{\bm k}_1 \wedge {\bm k}_2 l^2} \int \frac{dq_x dk_y}{2\pi} I_{2k_F}(q_x,k_y) 
e^{i(k_1-k_2)q_x l^2 -i(q_1-q_2)k_y l^2}, \label{2kF-1} \\
&\Phi^{+-}_{eh}(k_2,k_1,q_1,q_2;I_{2k_F})  = l^2 e^{i(k_1 q_1+k_2 q_2) l^2} \int \frac{dq_x dk_y}{2\pi}  \nonumber \\
& \hspace{0.9cm}  
 I_{2k_F}(q_x,k_y) e^{iq_x k_y l^2 + iq_x(k_1-k_2)l^2 + ik_y(q_1-q_2)l^2}, \label{2kF-2} \\ 
& \Phi^{+-}_{he}(q_1,q_2,k_2,k_1;I_{2k_F})  = l^2 e^{-i(k_1 q_1+k_2 q_2) l^2} \int \frac{dq_x dk_y}{2\pi} 
\nonumber \\
& \hspace{0.9cm}   I_{2k_F}(q_x,k_y) 
e^{-iq_x k_y l^2 + iq_x(k_1-k_2)l^2 + ik_y(q_1-q_2)l^2}. \label{2kF-3} 
\end{align}   

For later convenience, note that these functionals can be regarded homomorphic 
mappings of the functions. Namely, a product 
between two functionals of functions $f$ and $g$ is a functional of $fg$,
\begin{align}
&\int d(lQ) d(lQ^{\prime}) \Gamma_{\mu\nu}(Q,Q_2,Q_1,Q^{\prime};f) 
\Gamma_{\nu\lambda}(Q^{\prime}_2,Q^{\prime},Q,Q^{\prime}_{1};g) \nonumber \\
& \hspace{0.7cm} 
= 2\pi \Gamma_{\mu\lambda}(Q^{\prime}_2,Q_2,Q_1,Q^{\prime}_1;f g), \label{homomorphic-0}  
\end{align}
for {\it any} $\mu, \nu, \lambda =e_{+},e_{-},h_{+},h_{-}$. Here a 
summation over $\nu$ is {\it not} taken in their left-hand sides. Similarly,  
\begin{align}
&\int d(lQ) d(lQ^{\prime}) \Phi^{+-}_{\mu\nu}(Q,Q_2,Q_1,Q^{\prime};f) 
\Phi^{+-}_{\nu\lambda}(Q^{\prime}_2,Q^{\prime},Q,Q^{\prime}_{1};g) \nonumber \\
&\hspace{0.7cm} 
= 2\pi \Phi^{+-}_{\mu\lambda}(Q^{\prime}_2,Q_2,Q_1,Q^{\prime}_1;f g), \label{homomorphic-1} 
\end{align}
for any $\mu, \nu, \lambda = e,h$. These homomorphic natures of the functionals are 
useful in the next section. 

In the following, we consider Eqs.~(\ref{kin},\ref{int},\ref{int-h1},\ref{int-h2}) as the prototype 
model Hamiltonian for a semimetal in the quantum limit at the charge neutrality point.   

\section{screened Coulomb interaction}
\begin{figure}[t]
\begin{center}
	\includegraphics[width=1\linewidth]{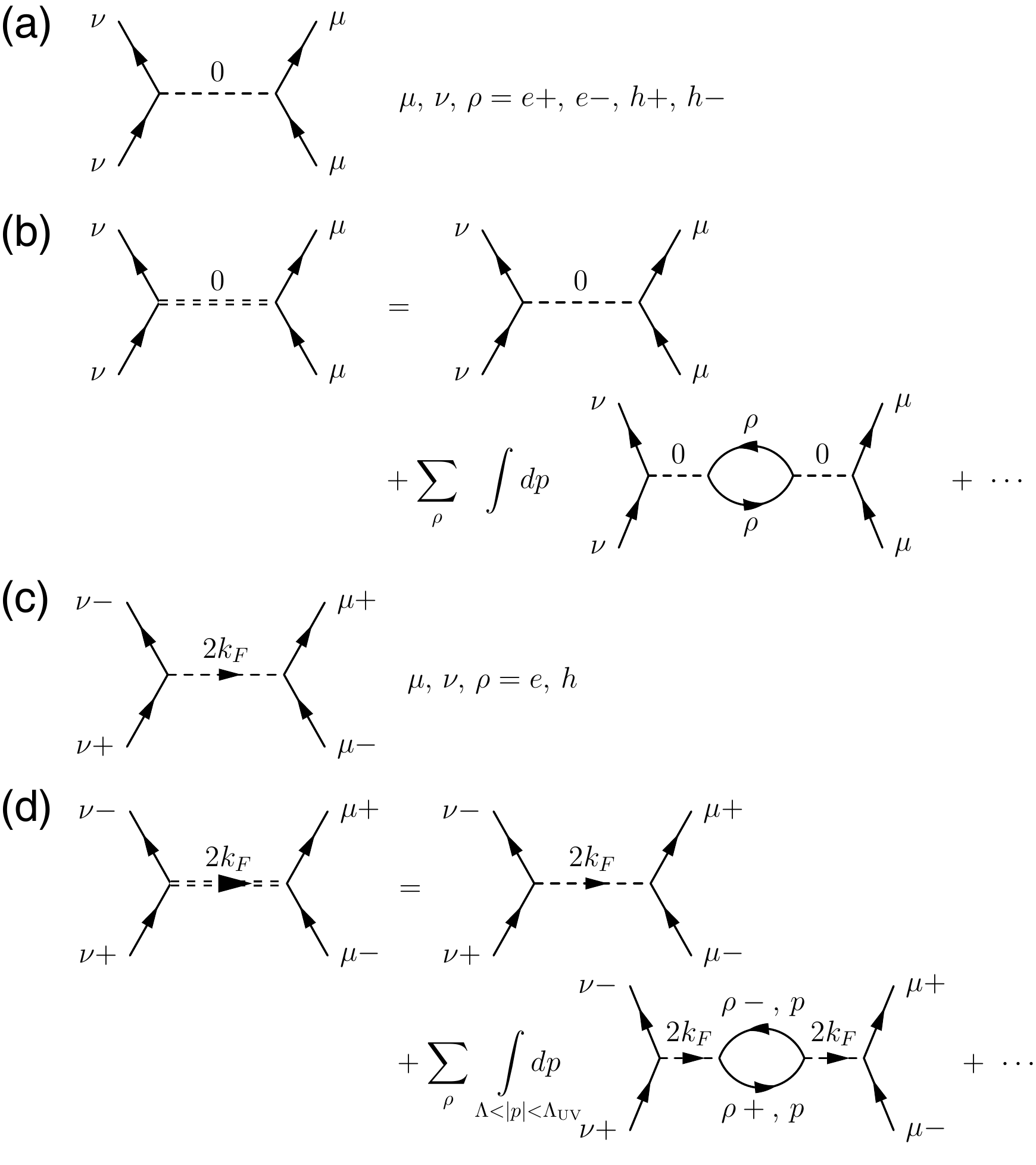}
	\caption{(a) interaction potentials in $H_1$;   
$\Gamma_{\mu\nu}$ in Eq.~(\ref{int-h1}) ($\mu,\nu=e_{+},e_{-},h_{+},h_{-}$). 
(b) screened form of $\Gamma_{\mu\nu}$ in Eq.~(\ref{int-h1}). (c)
interaction potentials in $H_2$; $\Phi^{+-}_{\mu\nu}$ in Eq.~(\ref{int-h2}) ($\mu,\nu=e,h$). 
(d) screened form of $\Phi^{+-}_{\mu\nu}$ in Eq.~(\ref{int-h2}), where 
we only include the fast mode of the $2k_F$ density fluctuation. The slow 
$2k_F$ density fluctuation shall be included in the parquet RG equations; see Fig.~\ref{fig:6}, e.g. 
the last two Feynman diagrams in the right-hand side of the first and second lines as well as the first two 
Feynman diagrams in the right-hand side of the fourth line.}
	\label{fig:4}
\end{center}
\end{figure}

The interaction potentials in $H_1$ are screened by low-energy density fluctuations 
at $p_z=0$, 
while the interaction potentials in $H_2$ are screened by the $2k_F$ density 
fluctuations. The respective screened interaction comprises of a sum of the bare 
interaction part and an effective interaction mediated by the density fluctuations. Using the 
random phase approximation (Fig.~\ref{fig:4}(b,d)) with a help of the homomorphic nature of 
the interaction potentials, Eqs.~(\ref{homomorphic-0},\ref{homomorphic-1}), we 
can show that the screened forms for the interaction potentials $\Gamma_{\mu\nu}$ and 
$\Phi^{+-}_{\mu\nu}$ take exactly the same forms as their respectively bare forms 
in Eqs.~(\ref{int-h1},\ref{int-h2}), 
except that their arguments, $I_0(q_x,k_y)$ and $I_{2k_F}(q_x,k_y)$, are replaced by 
their screened counterparts, $\overline{I}_0(q_x,k_y)$ and $\overline{I}_{2k_F}(q_x,k_y)$, 
respectively,  
\begin{align}
& \overline{I}_{0}(q_x,k_y) = \frac{I_{0}(q_x,k_y)}{1-\frac{1}{\hbar}\sum_{\lambda} 
\Pi_{0,\lambda}(\omega=0) I_{0}(q_x,k_y)}, 
\label{screen-1b}  \\
&\overline{I}_{2k_{F}}(q_x,k_y) = 
\frac{I_{2k_F}(q_x,k_y)}{1-\frac{1}{\hbar}\sum_{\lambda} \Pi^{-+}_{0,\lambda}(\omega=0) I_{2k_F}(q_x,k_y)}.
\label{screen-2b}  
\end{align} 
Here $\Pi_{0,\lambda}(\omega)$ denotes a bare polarization function for the 
$p_z=0$ density fluctuation in the right/left-mover electron band ($\lambda=e_{+/-}$) or 
left/right-mover hole band ($\lambda=h_{+/-}$). It is given by  
\begin{align}
\Pi_{0,\lambda}(\omega=0) &= \nonumber \\
&\hspace{-0.8cm} \frac{1}{2\pi l^2} \int \frac{dp_1}{2\pi} 
\bigg\{ \frac{\theta(\varepsilon_F - \varepsilon_{\lambda,p_1}) \theta(
\varepsilon_{\lambda,p_1+p}-
 \varepsilon_F)}{\varepsilon_{\lambda,p_1}-\varepsilon_{\lambda,p_1+p}} \nonumber \\
& \hspace{0.5cm} - \frac{\theta(\varepsilon_{\lambda,p_1}-\varepsilon_F) \theta(\varepsilon_F -
\varepsilon_{\lambda,p_1+p})}{\varepsilon_{\lambda,p_1}-\varepsilon_{\lambda,p_1+p}}
\bigg\}, \label{bare-po-0}
\end{align}
with the Heaviside step function $\theta(x)$.  From Eq.~(\ref{kin}), 
$\varepsilon_{\lambda,p}-\varepsilon_F \equiv v_F p$ for $\lambda=e_{+},h_{-}$ 
and $\varepsilon_{\lambda,p}-\varepsilon_F \equiv -v_F p$ for $\lambda=e_{-},h_{+}$. Noting that 
$p$ is much smaller than $k_F$, one obtains the bare polarization function as 
\begin{align}
\Pi_{0,\lambda}(\omega=0) = -\frac{1}{(2\pi l)^2} \frac{1}{v_F},  \label{bare-p1}
\end{align}   
for $\lambda=e_{+},e_{-},h_{+},h_{-}$. $\Pi^{-+}_{0,\lambda}(\omega)$ 
is the bare polarization function for the 
$2k_F$ density fluctuation within the electron band ($\lambda=e$)
or hole band ($\lambda=h$),  
\begin{align}
\Pi^{-+}_{0,\lambda}(\omega=0) &= \nonumber \\
&\hspace{-0.8cm} 
\frac{1}{2\pi l^2} \int \frac{dp_1}{2\pi} 
\bigg\{ \frac{\theta(\varepsilon_{F}-\varepsilon_{\lambda_{-},p_1})
\theta(\varepsilon_{\lambda_{+},p_1+p}-\varepsilon_F)}
{\varepsilon_{\lambda_{-},p_1}-\varepsilon_{\lambda_{+},p_1+p}} \nonumber \\
& \hspace{-0.3cm} - \frac{\theta(\varepsilon_{\lambda_{-},p_1}-\varepsilon_{F})
\theta(\varepsilon_{F}-\varepsilon_{\lambda_{+},p_1+p})}
{\varepsilon_{\lambda_{-},p_1}-\varepsilon_{\lambda_{+},p_1+p}}  \bigg\}.  \label{2kf-bare-p}
\end{align}
From Eq.~(\ref{kin}), $\varepsilon_{\lambda_{\pm},p}-\varepsilon_{F}= \pm v_{F} p $ 
for $\lambda=e$ and $\varepsilon_{\lambda_{\pm},p}-\varepsilon_{F}= \mp v_{F} p $ for 
$\lambda=h$.

\begin{figure}[t]
\begin{center}
	\includegraphics[width=1\linewidth]{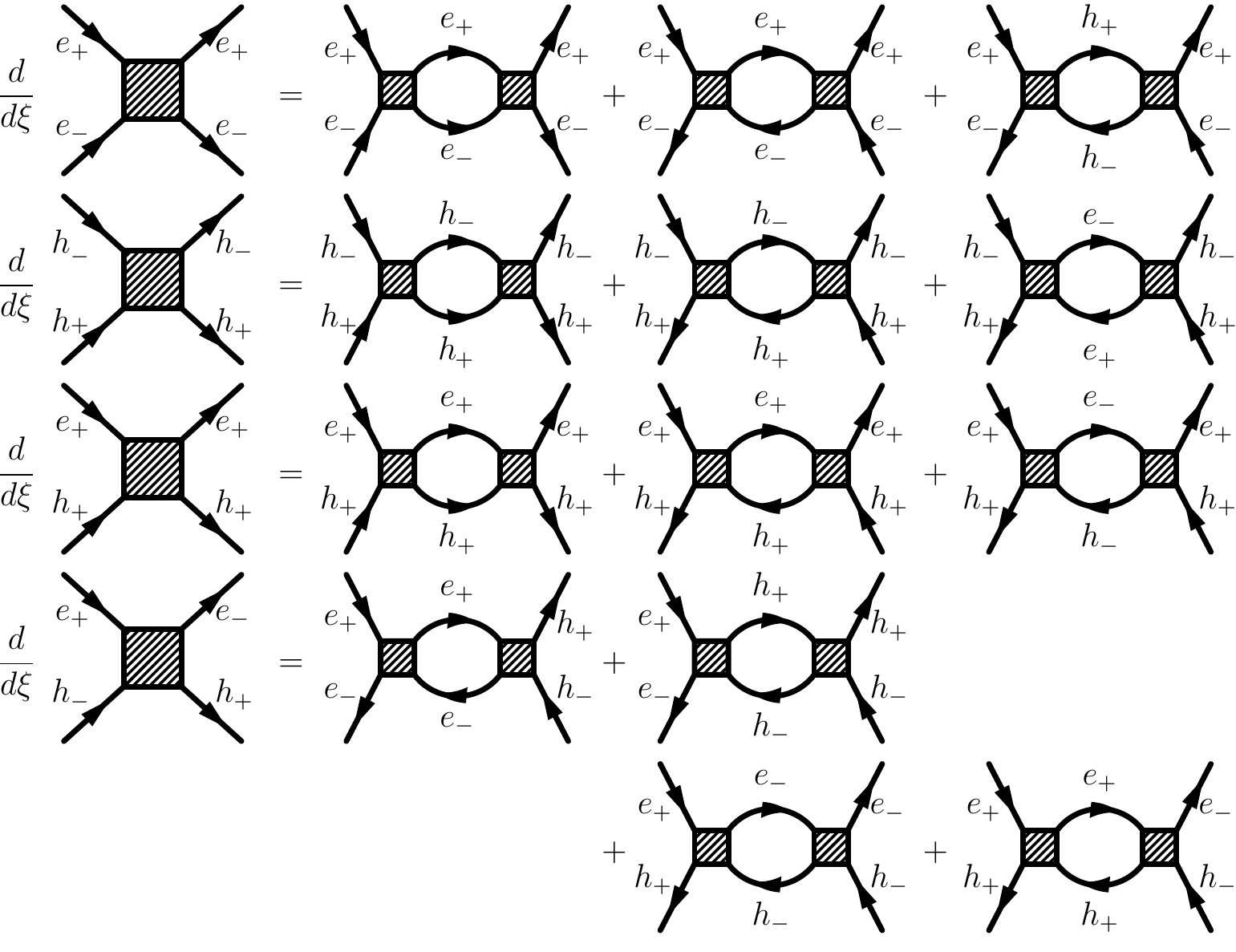}
	\caption{parquet RG equations in terms of Feynman diagrams. From the top to the bottom, each line of 
equations correspond to Eqs.~(\ref{wb-rg2},\ref{wd-rg2},\ref{we-rg2},\ref{wg-rg2}) respectively}
	\label{fig:6}
\end{center}
\end{figure}

The polarization function for $2k_F$ density fluctuation can be divided into low-energy (slow) $2k_F$ density 
fluctuation part and high-energy (fast) $2k_F$ density fluctuation part,
\begin{align}
&\Pi^{-+}_{0,\lambda}(\omega=0) = \Pi^{-+,{\rm s}}_{0,\lambda}(\omega=0) 
+ \Pi^{-+,{\rm f}}_{0,\lambda}(\omega=0), \label{decomp-1} \\
&\Pi^{-+,{\rm s}}_{0,\lambda}(\omega=0)  \equiv \frac{1}{2\pi l^2} \int_{|p_1|<\Lambda} 
\frac{dp_1}{2\pi} \bigg\{\cdots \bigg\}, \label{decomp-2} \\
& \Pi^{-+,{\rm f}}_{0,\lambda}(\omega=0)  \equiv \frac{1}{2\pi l^2} 
\int_{\Lambda<|p_1|<\Lambda_{\rm UV}} \frac{dp_1}{2\pi} \bigg\{\cdots \bigg\}.  \label{decomp} 
\end{align}  
Here $\Lambda_{\rm UV}$ is a momentum cut off associated with the Brillouin zone boundary, while 
$\Lambda$ separates the slow mode from the fast mode, $\Lambda<\Lambda_{\rm UV}$.   
The slow $2k_{F}$ density fluctuation part ($|p_1|<\Lambda$) leads to the logarithmic 
singularity at $p=0$~\cite{zheleznyak97}, 
\begin{align}
 \Pi^{-+,{\rm s}}_{0,\lambda}(\omega=0)  \simeq 
-\frac{1}{(2\pi l)^2} \frac{1}{v_F} \ln \bigg(\frac{2\Lambda}{|p|}\bigg), \label{2kf-sin}
\end{align}
for any $\lambda=e,h$. The singularity is a seed of the $2k_F$ DW instability in 
each pocket~\cite{gruner00,abrikosov70,brazovskii72,yakovenko93}. More generally, `bubble' Feynman diagrams 
composed of two single-particle Green functions 
with opposite sign of the Fermi velocities have the same logarithmic singularity both in particle-hole 
and in particle-particle channels~\cite{abrikosov70,brazovskii72}. 
These logarithmic singularities suggest that the ground state  
of the electron-hole model at the charge neutrality point has several competing 
instabilities at lower temperature. To clarify the most dominant instability precisely, we 
thus take into account the slow $2k_F$ density fluctuation in the framework of parquet 
RG equations and include it as well as the other low-energy fluctuations 
with the logarithmic singularity on the equal footing (see the next section). Therefore, to avoid the double counting of the slow $2k_F$ density fluctuation part, we include in Eq.~(\ref{screen-2b}) only the high-energy (fast) $2k_F$ density fluctuation part first. This determines a form of the screened interaction. The screened interaction thus obtained is then used for an `initial' 
interaction potential for the RG equations (Fig.~\ref{fig:5}). 
Finally, the low-energy (slow) $2k_F$ density fluctuation as well as other 
dominant low-energy fluctuations shall be included sequentially in the 
framework of the RG procedure (Fig.~\ref{fig:6}). 
For example, in Fig.~\ref{fig:6}, the last two Feynman diagrams in the right-hand side of the first and 
second lines as well as the first two Feynman diagrams in the right-hand side of the fourth line 
represents the inclusions of the slow $2k_F$ density fluctuations. The fast $2k_F$ density 
fluctuation part, Eq.~(\ref{decomp}), takes a constant finite value at $p=0$,     
\begin{align}
\Pi^{-+,{\rm f}}_{0,\lambda}(\omega=0) 
& \simeq  -\frac{1}{(2\pi l)^2} \frac{1}{v_F} 
\ln\bigg(\frac{\Lambda_{\rm UV}}{\Lambda}\bigg), \label{high}
\end{align} 
for $\lambda=e,h$. 

To summarize, we will use the following form of the screened interaction potentials   
as the initial interaction forms for the later parquet RG studies; 
\begin{align}
H_{\rm int} & \equiv \overline{H}_{1} + \overline{H}_{2}, \nonumber \\
\overline{H}_1 & = \frac{1}{2} \sum_{\mu,\nu=e_{+},e_{-},h_{+},h_{-}}   \int \frac{dp \!\ dp_1 \!\ dp_2}{(2\pi)^3}
\nonumber \\
&\hspace{-0.7cm}\int d(lQ_1) d(lQ^{\prime}_1) d(lQ^{\prime}_2) d(dQ_2) \!\  
\Gamma_{\mu\nu}(Q_1,Q^{\prime}_1,Q^{\prime}_{2},Q_{2};\overline{I}_{0})  \nonumber \\
& \hspace{-0.4cm} a^{\dagger}_{\mu}(Q^{\prime}_1,p_2+p) a_{\mu}(Q^{\prime}_2,p_2) 
a^{\dagger}_{\nu}(Q_1,p_1-p) a_{\nu}(Q_2,p_1) 
, \label{screen-1c} \\
\overline{H}_{2}& =  \sum_{\mu,\nu=e,h} \int \frac{dp \!\ dp_1\!\ dp_2}{(2\pi)^3} \nonumber \\
& \hspace{-0.7cm} \int d(lQ_1)  d(lQ^{\prime}_1)  d(lQ^{\prime}_2) d(lQ_2) 
\Phi^{+-}_{\mu\nu}(Q_1,Q^{\prime}_1,Q^{\prime}_{2},Q_{2};\overline{I}_{2k_{F}})  
 \nonumber \\ 
& \hspace{-0.5cm}  a^{\dagger}_{\mu_{+}}(Q^{\prime}_1,p_2+p) a_{\mu_{-}}(Q^{\prime}_2,p_2)  
a^{\dagger}_{\nu_{-}}(Q_1,p_1-p) a_{\nu_{+}}(Q_2,p_1),  \label{screen-2c}\\
& \overline{I}_{0}(q_x,k_y) = \frac{4\pi e^2 l^2}{\varepsilon} 
\frac{e^{-\frac{1}{2}(q^2_x+k^2_y)l^2}}{(q^2_x+k^2_y) l^2 + \frac{1}{A} 
e^{-\frac{1}{2}(q^2_x+k^2_y)l^2}}, \label{screen-1d}  \\
&\overline{I}_{2k_F}(q_x,k_y) = \frac{4\pi e^2 l^2}{\varepsilon} 
\frac{e^{-\frac{1}{2}(q^2_x+k^2_y)l^2}}{(q^2_x+k^2_y) l^2 +B + \frac{1}{A^{\prime}} 
e^{-\frac{1}{2}(q^2_x+k^2_y)l^2}} 
\label{screen-2d} 
\end{align}  
with $B \equiv 4 k^2_F l^2$ and 
\begin{align}
\frac{1}{A} &\equiv - \frac{4\pi e^2 l^2}{\hbar \varepsilon}   
\sum_{\lambda=e_{\pm},h_{\pm}}\Pi_{0,\lambda}(\omega=0) = \frac{4e^2}{\hbar \pi v_F \varepsilon}, \label{0-pi} \\
\frac{1}{A^{\prime}} & \equiv - \frac{4\pi e^2 l^2}{\hbar \varepsilon}  
\sum_{\lambda=e,h} \Pi^{-+,{\rm f}}_{0,\lambda}(\omega=0) = 
\frac{2 e^2 \ln\big[\frac{\Lambda_{\rm UV}}{\Lambda}\big]}{\hbar \pi  v_F \varepsilon}. \label{2kf-pi}
\end{align}
For simplicity, we take $\ln(\Lambda_{\rm UV}/\Lambda) = 2$ henceforth and identify 
$A^{\prime}$ with $A$. For reminder, the 
functionals $\Gamma_{\mu\nu}$ 
($\mu,\nu=e_{+},e_{-},h_{+},h_{-}$) and 
$\Phi^{+-}_{\mu\nu}$ ($\mu,\nu=e,h$) in Eqs.~(\ref{screen-1c},\ref{screen-2c}) 
are defined in Eqs.~(\ref{e+e+},\ref{e-e-},\ref{e+e-},\ref{e-e+},\ref{2kF-1},
\ref{2kF-2},\ref{2kF-3}) respectively.

\section{parquet renormalization group equation}
The polarization function for the slow $2k_F$ density fluctuation has 
the logarithmic singularity [Eq.~(\ref{2kf-sin})]. More generally, all the 
`bubble' diagrams composed of the two Green functions with opposite sign of the Fermi velocities 
have the same logarithmic singularity in both particle-hole and particle-particle channels. The presence of 
the logarithmic singularities in several distinct channels means {\it competing} ground-state 
instabilities in the semimetal model. To reveal the ground-state phase diagram of the model precisely,  
we thus include all the relevant fluctuations with the logarithmic singularity on the equal footing. 

To this end, we derive in this section the parquet renormalization group (RG) 
equations~\cite{abrikosov70,brazovskii72,yakovenko93}, 
where consecutive integration of the higher-energy fermionic degree of freedom 
renormalizes the interaction potentials among the lower-energy fermions. The renormalization 
gives rise to either enhancement, suppression or convergence of the interaction potentials. By identifying 
the most divergent potentials among the others, we shall tell 
the dominant ground-state instability in the model. 

\begin{figure}[htbp]
\begin{center}
	\includegraphics[width=1\linewidth]{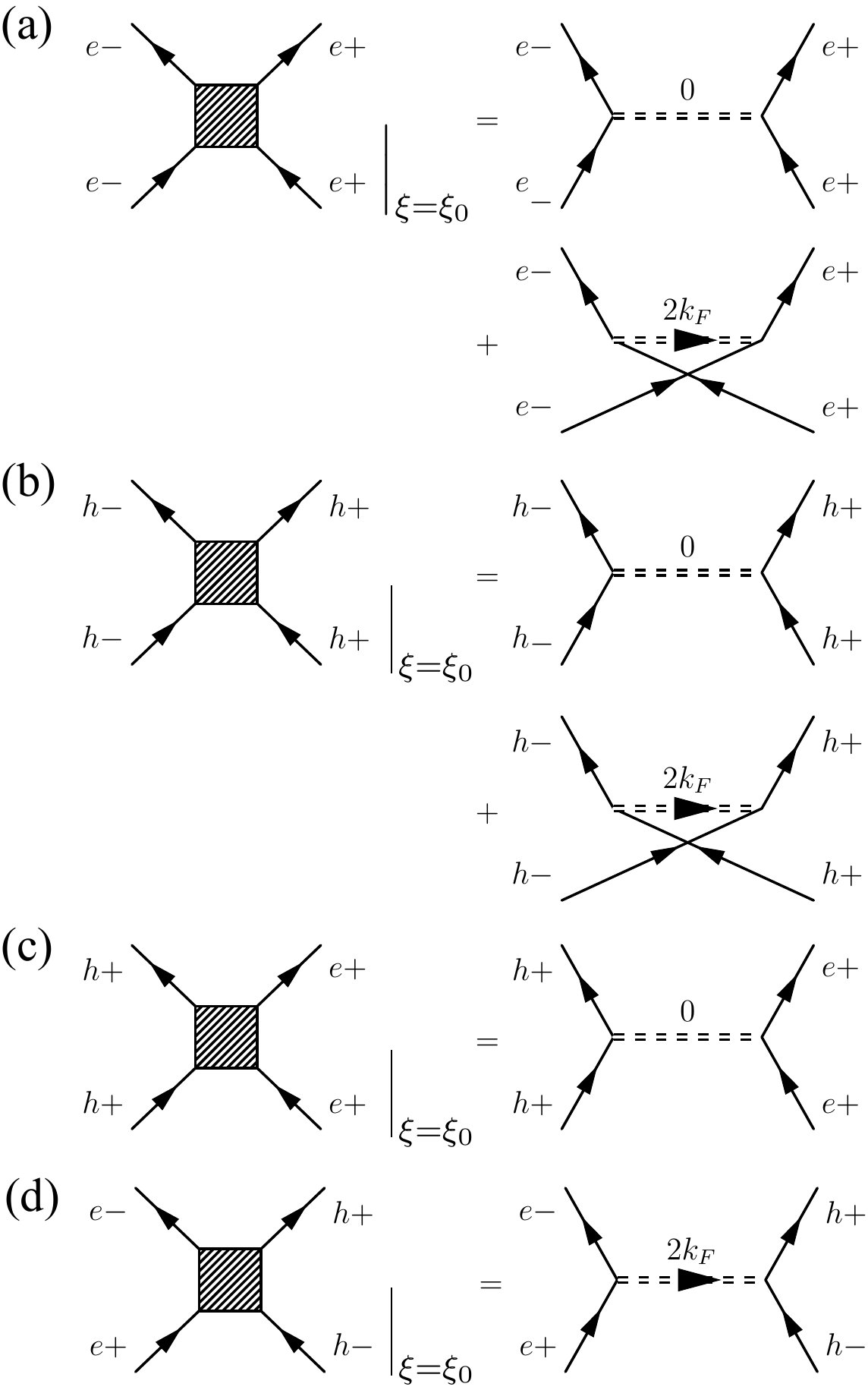}
	\caption{A set of initial forms of the interaction potentials for the parquet RG equations 
are given by the RPA screened Coulomb interactions, 
Eqs.~(\ref{screen-1c},\ref{screen-2c},\ref{screen-1d},\ref{screen-2d}). 
(a) $W_{b}({\bm k},\xi)$ at the initial RG scale ($\xi=0$) that corresponds 
to Eq.~(\ref{wbd}), (b) $W_{d}({\bm k})$ at $\xi=0$, corresponding to 
 Eq.~(\ref{wbd}) (c) $W_{e}({\bm k})$ at $\xi=0$, corresponding to Eq.~(\ref{we}) 
(d) $W_{g}({\bm k})$ at $\xi=0$, corresponding to Eq.~(\ref{wg}).}
	\label{fig:5}
\end{center}
\end{figure}

The one-loop parquet RG equations can be derived by a standard momentum 
shell renormalization. Thereby, we begin with a partition function of the interacting 
fermion model;
\begin{align}
Z &= \int {\cal D}e^{\dagger}_{\pm} {\cal D} e_{\pm} {\cal D}h^{\dagger}_{\pm} {\cal D} h_{\pm} 
e^{-S_0-S_1} \nonumber \\
S_0 =& \sum_{\sigma=\pm} \int \frac{d(l\omega)}{2\pi} \int_{|p|<\Lambda} dp
\int dQ \!\ \Big\{ \big(- i\omega + \sigma v_F p\big) \nonumber \\ 
&\times e^{\dagger}_{\sigma}(Q,p,\omega) e_{\sigma}(Q,p,\omega) + 
\big(- i\omega - \sigma v_F p\big) \nonumber \\
 & \times h^{\dagger}_{\sigma}(Q,p,\omega) h_{\sigma}(Q,p,\omega) \Big\}, 
\label{s0} 
\end{align}
\begin{align}
S_1& = \int_{1,2,3} \int dk_1 dq_1  dk_2 dq_2 \!\ \!\ e^{i{\bm k}_1 \wedge {\bm k}_2} 
W_{b}({\bm k}_1-{\bm k}_2) \nonumber \\
&\hspace{0.3cm} e^{\dagger}_{+}(k_1,1) e^{\dagger}_{-}(q_1,2) 
e_{-}(q_2,3) e_{+}(k_2,1+2-3) \nonumber \\
& +  \int_{1,2,3} \int dk_1 dq_1  dk_2 dq_2 \!\ \!\ e^{i{\bm k}_1 \wedge {\bm k}_2} 
W_{d}({\bm k}_1-{\bm k}_2) \nonumber \\
&\hspace{0.3cm} h^{\dagger}_{-}(k_1,1) h^{\dagger}_{+}(q_1,2) 
h_{+}(q_2,3) h_{-}(k_2,1+2-3) \nonumber \\ 
&+ \int_{1,2,3} \int dk_1 dq_1  dk_2 dq_2 \!\ \!\ e^{i{\bm k}_1 \wedge {\bm k}_2} 
W_{e}({\bm k}_1-{\bm k}_2) \nonumber \\
&\hspace{0.3cm} e^{\dagger}_{+}(k_1,1) h^{\dagger}_{+}(q_1,2) 
h_{+}(q_2,3) e_{+}(k_2,1+2-3) \nonumber \\ 
&+ \int_{1,2,3} \int dk_1 dq_1  dk_2 dq_2 \!\ \!\ e^{i{\bm k}_1 \wedge {\bm k}_2} 
W_{e}({\bm k}_1-{\bm k}_2) \nonumber \\
&\hspace{0.3cm} h^{\dagger}_{-}(k_1,1) e^{\dagger}_{-}(q_1,2) 
e_{-}(q_2,3) h_{-}(k_2,1+2-3) \nonumber \\  
&+ \int_{1,2,3} \int dk_1 dq_1  dk_2 dq_2 \!\ \!\ e^{i(k_1q_1+k_2q_2)} 
W_{g}({\bm k}_1-{\bm k}_2) \nonumber \\
&\hspace{0.3cm} e^{\dagger}_{+}(k_1,1) h^{\dagger}_{-}(k_2,2) 
h_{+}(q_2,3) e_{-}(q_1,1+2-3) \nonumber \\  
&+ \int_{1,2,3} \int dk_1 dq_1  dk_2 dq_2 \!\ \!\ e^{-i(k_1q_1+k_2q_2)} 
W^{*}_{g}({\bm k}_1-{\bm k}_2) \nonumber \\
&\hspace{-0.3cm} h^{\dagger}_{+}(q_1,1) e^{\dagger}_{-}(q_2,2) 
e_{+}(k_2,3) h_{-}(k_1,1+2-3)  + \cdots . \label{s1}
\end{align}
Here $Q=k_y$ or $q_x$ is rescaled by the magnetic length $l$; $Q_{\rm new} \equiv Q_{\rm old} l$. 
Besides, we used the following notations, 
\begin{align}
&1 \equiv (p_1,\omega_1), \ \  2 \equiv (p_2,\omega_2), \ \ 
3 \equiv (p_3,\omega_3) \nonumber \\ 
&\int_{1,2,3} \equiv \int \frac{d\omega_1 \!\ d\omega_2  \!\ d\omega_3}{(2\pi)^3}
\int \frac{dp_1 \!\ dp_2 \!\ dp_3}{(2\pi)^3} \nonumber \\
& {\bm k}_1 \equiv (k_1,q_1), \ \ {\bm k}_2 \equiv (k_2,q_2), \nonumber \\  
&{\bm k}_1 \wedge {\bm k}_2 \equiv k_1 q_2 - k_2 q_1. \nonumber 
\end{align}  
For the repulsive Coulomb interaction case, the interaction potentials in $S_1$ are given by either 
some of $\overline{H}_1$ or $\overline{H}_2$ or their combination from 
Eqs.~(\ref{screen-1c},\ref{screen-2c},\ref{screen-1d},\ref{screen-2d}) (see Fig.~\ref{fig:5}). 
Namely, $W_{b}({\bm k})$ and $W_{d}({\bm k})$ are given by a sum of Eq.~(\ref{e+e-}) 
and Eq.~(\ref{2kF-1}) with $I_0$ and $I_{2k_F}$ replaced  by $\overline{I}_0$ and $\overline{I}_{2k_F}$. 
$W_e({\bm k})$ and $W_{g}({\bm k})$ are given by Eq.~(\ref{e+e-}) and by Eq.~(\ref{2kF-3}) 
respectively with $\overline{I}_0$ and $\overline{I}_{2k_F}$. To be more specific, we consider 
the following set of the screened interaction as the initial interaction forms of the RG equations, 
\begin{align}
W_b({\bm k}) &= W_{d}({\bm k}) \nonumber \\
&\hspace{-1.2cm} = \overline{I}_0(q_x,k_y) - \int \frac{dq^{\prime}_x \!\ dk^{\prime}_y}{2\pi} 
e^{ik_y q^{\prime}_x - iq_x k^{\prime}_y} \overline{I}_{2k_F}(q^{\prime}_x,k^{\prime}_y), \label{wbd} \\
W_{e}({\bm k}) &= \overline{I}_0(q_x,k_y), \label{we} \\
W_{g}({\bm k})  &= \int \frac{dq^{\prime}_x \!\ d k^{\prime}_y }{2\pi} e^{i q^{\prime}_x k^{\prime}_y + 
iq^{\prime}_x k_y + i k^{\prime}_y q_x} 
\overline{I}_{2k_F}(q^{\prime}_x,k^{\prime}_y), \label{wg}
\end{align}
with ${\bm k}\equiv (k_y,q_x)$ and Eqs.~(\ref{screen-1d},\ref{screen-2d}). 
$``\cdots"$ in Eq.~(\ref{s1}) denotes those interaction parts that are not renormalized by others and 
do not renormalize others at the level of the one-loop RG equations. Such interaction parts 
are irrelevant in the framework of the one-loop RG analyses; we thus omit them henceforth.

Following the standard momentum shell renormalization process (Appendix B), we first decompose 
the fermionic field into fast mode ($\Lambda-d\Lambda<|p|<\Lambda$) 
and slow mode ($|p|<\Lambda-d\Lambda$) in the momentum along the field. The integration 
of the fast mode in the partition function leads to the renormalizations of the interaction potentials among 
the slow modes. This gives out a set of coupled RG equations for the 
interaction potentials, 
\begin{align}
\frac{dW_{b}({\bm k})}{d\xi} &= \int d{\bm k}^{\prime} \!\ W_{b}({\bm k}^{\prime}) 
W_{b}({\bm k}-{\bm k}^{\prime}) \big(1-e^{-i{\bm k}\wedge {\bm k}^{\prime}}\big)  \nonumber \\
& \hspace{1.2cm} +  \int d{\bm k}^{\prime} \!\ W_{g}({\bm k}^{\prime}) 
W^{*}_{g}({\bm k}-{\bm k}^{\prime}) \label{wb-rg2} \\ 
\frac{dW_{d}({\bm k})}{d\xi} &= \int d{\bm k}^{\prime} \!\ W_{d}({\bm k}^{\prime}) 
W_{d}({\bm k}-{\bm k}^{\prime}) \big(1-e^{-i{\bm k}\wedge {\bm k}^{\prime}}\big)  \nonumber \\
& \hspace{1.2cm} +  \int d{\bm k}^{\prime} \!\ W_{g}({\bm k}^{\prime}) 
W^{*}_{g}({\bm k}-{\bm k}^{\prime}) \label{wd-rg2} \\ 
\frac{dW_{e}({\bm k})}{d\xi} &= \int d{\bm k}^{\prime} \!\ W_{e}({\bm k}^{\prime}) 
W_{e}({\bm k}-{\bm k}^{\prime}) \big(1-e^{-i{\bm k}\wedge {\bm k}^{\prime}}\big)  \nonumber \\
& \hspace{-0.4cm} +  \int d{\bm k}^{\prime} \!\ e^{-ikq+ik^{\prime}q+ikq^{\prime}} W_{g}({\bm k}^{\prime}) 
W^{*}_{g}({\bm k}-{\bm k}^{\prime}) \label{we-rg2} \\ 
\frac{dW_{g}({\bm k})}{d\xi} &= \int d{\bm k}^{\prime} \!\ W_{g}({\bm k}-{\bm k}^{\prime})  
\Big\{ W_{b}({\bm k}^{\prime}) + W_{d}({\bm k}^{\prime}) \nonumber \\
& \hspace{2.0cm} +2\!\ e^{-ikq^{\prime}-ik^{\prime}q+ik^{\prime}q^{\prime}} 
W_{e}({\bm k}^{\prime}) \Big\} 
 \label{wg-rg2}  
\end{align} 
with ${\bm k}\equiv (k,q)$, ${\bm k}^{\prime} \equiv (k^{\prime},q^{\prime})$, and 
$d{\bm k}^{\prime} \equiv dk^{\prime}dq^{\prime}$. $d\xi$ denotes a RG scale,  
\begin{align}
d\xi \equiv \frac{1}{(2\pi)^3 l^2} \frac{d\Lambda}{v_F \Lambda}. \label{d-xi}
\end{align}

In order to solve the coupled RG equations numerically, we put them in the dual-space representation 
by the Fourier transform of $W_{\mu}({\bm k})$,
\begin{align}
&F_{\mu}({\bm r}) \equiv \int d{\bm k} e^{-i{\bm k}{\bm r}} W_{\mu}({\bm k}),  \label{f1} \\
&W_{\mu}({\bm k}) \equiv \int \frac{d{\bm r}}{(2\pi)^2} e^{i{\bm k}{\bm r}} F_{\mu}({\bm r}), \label{f2} 
\end{align} 
for $\mu=b,d,e,g$ with 
\begin{align}
\tilde{F}_{g}({\bm r}) \equiv e^{-ir_x r_y} F_{g}({\bm r}), \label{f3}
\end{align}
${\bm r}\equiv (r_x,r_y)$ and ${\bm k}\equiv (k,q)$. It turns out that parquet RG equations for 
$F_{\mu}({\bm r})$ ($\mu=b,d,e$) and $\tilde{F}_g({\bm r})$ 
as well as their initial forms respect the following O(2) symmetry and real-valued-ness;
\begin{align}
&F_{\mu}(\hat{R}_{\theta} {\bm r}) = F_{\mu}({\bm r}) = 
F^{*}_{\mu}({\bm r}) \equiv \Gamma_{\mu}(r), \label{Gamma1} \\ 
&\tilde{F}_{g}(\hat{R}_{\theta} {\bm r}) = \tilde{F}_{g}({\bm r}) = 
\tilde{F}^{*}_g({\bm r}) \equiv \Gamma_{g}(r),  \label{Gamma2} \\
&\hat{R}_{\theta} \equiv \left(\begin{array}{cc} 
\cos \theta & \sin\theta \\
-\sin \theta & \cos \theta \\
\end{array}\right), \label{u1}
\end{align}
with $r\equiv |{\bm r}|$ for arbitrary $\theta \in (0,2\pi]$. Using this 
symmetry, we can finally reach O(2)-symmetric parquet RG equations for $\Gamma_{\mu}(r)$ 
($\mu=b,d,e$) and $\Gamma_{g}(r)$ as follows~\cite{brazovskii72},  
\begin{align}
\frac{d\Gamma_{b/d}(r)}{d\xi} = &\Gamma^2_{b/d}(r) + \Gamma^2_g(r) \nonumber \\
& \hspace{-1.5cm}  - \int^{\infty}_{0} dr^{\prime} \int^{\infty}_{0} dr^{\prime\prime} \Gamma_{b/d}(r^{\prime}) 
\Gamma_{b/d}(r^{\prime\prime}) K(r,r^{\prime},r^{\prime\prime}), \label{gamma-bd} \\
\frac{d\Gamma_{e}(r)}{d\xi} = &\Gamma^2_{e}(r) + \overline{\Gamma}^2_g(r) \nonumber \\
& \hspace{-1.cm} - \int^{\infty}_{0} dr^{\prime} \int^{\infty}_{0} dr^{\prime\prime} \Gamma_{e}(r^{\prime}) 
\Gamma_{e}(r^{\prime\prime}) K(r,r^{\prime},r^{\prime\prime}), \label{gamma-e} \\
\frac{d\Gamma_{g}(r)}{d\xi} = & \Gamma_{g}(r) \big(\Gamma_{b}(r) + \Gamma_{d}(r)\big) 
\nonumber \\
& + 2 \int^{\infty}_{0} r^{\prime} dr^{\prime} \Gamma_{e}(r^{\prime}) 
\overline{\Gamma}_{g}(r^{\prime}) J_{0}(rr^{\prime}), \label{gamma-g} \\
\frac{d\overline{\Gamma}_{g}(r)}{d\xi} = & 2 \Gamma_{e}(r) \overline{\Gamma}_{g}(r) 
\nonumber \\ 
& \hspace{-0.6cm} + \int^{\infty}_{0} r^{\prime} dr^{\prime} 
\Gamma_{g}(r^{\prime}) \big(\Gamma_{b}(r^{\prime}) + \Gamma_{d}(r^{\prime})\big) 
J_{0}(r^{\prime}r), \label{over-gamma-g} 
\end{align}  
with 
\begin{align}
K(r,r^{\prime},r^{\prime\prime}) \equiv r^{\prime} r^{\prime\prime} 
\sum^{\infty}_{m=-\infty} J_{2m}(rr^{\prime}) J_{2m}(rr^{\prime\prime}) J_{2m}(r^{\prime}r^{\prime\prime}), 
\end{align}
and Bessel function $J_{2m}(r)$ (integer $m$).  
$\overline{\Gamma}_{g}(r)$ is a Hankel transform of $\Gamma_{g}(r)$;
\begin{align}
&\overline{\Gamma}_g(r) = \int^{\infty}_{0} 
r^{\prime}dr^{\prime} \Gamma_{g}(r^{\prime}) J_{0}(rr^{\prime}), \label{hankel1}\\ 
&\Gamma_g(r) = \int^{\infty}_{0} 
r^{\prime}dr^{\prime} \overline{\Gamma}_{g}(r^{\prime}) J_{0}(rr^{\prime}). \label{hankel2}  
\end{align}
The initial forms of $\Gamma_{\mu}(r)$ ($\mu=b,d,e,g$) and $\overline{\Gamma}_{g}(r)$ are 
obtained from Eqs.~(\ref{wbd},\ref{we},\ref{wg}) as follows,
\begin{align}
&\Gamma_{b/d}(r) \equiv 2\pi \Big\{ \int^{\infty}_{0} r^{\prime} dr^{\prime} J_0(rr^{\prime}) 
\overline{I}_0(r^{\prime}) - \overline{I}_{2k_F}(r)\Big\},  \label{gamma-bd=in} \\ 
&\Gamma_{e}(r) \equiv 2\pi \int^{\infty}_{0} r^{\prime} dr^{\prime} J_0(rr^{\prime}) 
\overline{I}_0(r^{\prime}),   \label{gamma-e=in} \\ 
&\Gamma_{g}(r) \equiv 2\pi \overline{I}_{2k_F}(r),  \label{gamma-g=in} \\  
&\overline{\Gamma}_{g}(r) \equiv 2\pi \int^{\infty}_{0} r^{\prime} dr^{\prime} J_0(rr^{\prime}) 
\overline{I}_{2k_F}(r^{\prime}), \label{over-gamma-g=in} 
\end{align}
with 
\begin{align}
\overline{I}_{0}(r) &\equiv \frac{4\pi e^2 l^2}{\varepsilon} \frac{1}{r^2 e^{\frac{1}{2}r^2}+A^{-1}}, \label{I0-a} \\
\overline{I}_{2k_F}(r) &\equiv \frac{4\pi e^2 l^2}{\varepsilon} \frac{1}{(r^2 + B) 
e^{\frac{1}{2}r^2}+A^{-1}}. \label{I2kf-a}
\end{align}
Note that the overall factor of $\overline{I}_0(r)$ and $\overline{I}_{2k_F}(r)$, $4\pi e^2 l^2/\varepsilon$, 
can be absorbed into a redefinition of the RG scale change $\xi$; it does not alter the ground-state 
phase diagram. Only two dimensionless parameters, $A$ and $B \equiv 4 k^2_F l^2$ in 
Eqs.~(\ref{I0-a},\ref{I2kf-a}), play vital role in a determination of the ground-state phase diagram. 

We solved Eqs.~(\ref{gamma-bd},\ref{gamma-e},\ref{gamma-g},\ref{over-gamma-g}) numerically, 
with $\Gamma_{\mu}(r)$ ($\mu=b,d,e,g$) and $\overline{\Gamma}_{g}(r)$ at the initial 
RG scale $(\xi=0)$ being given 
by Eqs.~(\ref{gamma-bd=in},\ref{gamma-e=in},\ref{gamma-g=in},
\ref{over-gamma-g=in},\ref{I0-a},\ref{I2kf-a}). By doing 
so, we numerically observed that in the two-dimensional $A$-$B$ space, either a set of 
$\Gamma_{b}(r,\xi)=\Gamma_{d}(r,\xi)$ and 
$\Gamma_{g}(r,\xi)$ or a set of 
$\Gamma_{e}(r,\xi)$ and $\overline{\Gamma}_{g}(r,\xi)$ show divergences 
at certain values of $r$ and $\xi$ (see the next section). The divergence indicates 
a certain type of pairing instabilities in the ground state. To identify the favored pairings 
and natures of resulting symmetry-broken phases, we rewrite 
the interaction potentials in Eq.~(\ref{s1}) in the {\it same} 
basis of the Landau gauge. For example, we put $e_{-}$ and $h_{+}$ as well as $e_{+}$ and
 $h_{-}$ in the basis of the $x$-gauge eigenstates by using Eqs.~(\ref{ytox},\ref{xtoy}). This leads to 
\begin{widetext}
\begin{align}
S_1 =& \int_{1,2,3} \int dk_1 d\overline{k}_1 d\overline{k}_2 \!\ 
\Phi_b(\overline{k}_2-\overline{k}_1,k_1-\overline{k}_2) \!\ \!\ 
\Big\{ \!\ 
e^{\dagger}_{+}(k_1,1) e^{\dagger}_{-}(\overline{k}_1,2) 
e_{-}(\overline{k}_2,3) e_{+}(k_1+\overline{k}_{1}-\overline{k}_2,1+2-3) \nonumber \\
&\hspace{3.3cm} + \!\ 
h^{\dagger}_{-}(k_1,1) h^{\dagger}_{+}(\overline{k}_1,2) 
h_{+}(\overline{k}_2,3) h_{-}(k_1+\overline{k}_{1}-\overline{k}_2,1+2-3) \Big\} \nonumber \\ 
& \!\ + \int_{1,2,3} \int dk_1 d\overline{k}_1 d\overline{k}_2 \!\ 
\Phi_e(\overline{k}_2-\overline{k}_1,k_1-\overline{k}_2) \!\ \Big\{ \!\ 
e^{\dagger}_{+}(k_1,1) h^{\dagger}_{+}(\overline{k}_1,2) 
h_{+}(\overline{k}_2,3) e_{+}(k_1+\overline{k}_{1}-\overline{k}_2,1+2-3) \nonumber \\
&\hspace{3.3cm} + \!\ 
h^{\dagger}_{-}(k_1,1) e^{\dagger}_{-}(\overline{k}_1,2) 
e_{-}(\overline{k}_2,3) h_{-}(k_1+\overline{k}_{1}-\overline{k}_2,1+2-3) \Big\} \nonumber \\ 
&+ \int_{1,2,3} \int dk_1 dk_2 d\overline{k}_2 \!\  \Phi_g(k_1-\overline{k}_2,k_2-\overline{k}_2) \!\ 
e^{\dagger}_{+}(k_1,1) h^{\dagger}_{-}(k_2,2) 
h_{+}(\overline{k}_2,3) e_{-}(k_1+k_{2}-\overline{k}_2,1+2-3) + {\rm h.c.}, \label{int-3} 
\end{align}
\end{widetext}
where $k_1$, $\overline{k}_1$, $k_2$ and $\overline{k}_2$ are the momentum along the $y$-direction. 
Note that due to the translational symmetry along the $y$ direction in the $x$-gauge, 
all the interaction potentials preserve a center of mass in the momentum. From 
Eqs.~(\ref{f2},\ref{Gamma1},\ref{Gamma2}), 
one can readily see that $\Phi_b(k,k^{\prime})$, $\Phi_e(k,k^{\prime})$ and $\Phi_g(k,k^{\prime})$ 
in Eq.~(\ref{int-3}) are given by $\Gamma_b(r)=\Gamma_d(r)$, $\Gamma_e(r)$, $\Gamma_{g}(r)$ 
and $\overline{\Gamma}_g(r)$ as follows,    
\begin{align}
\Phi_b(k,k^{\prime}) 
&\equiv \int \frac{dr_x}{2\pi} e^{ik r_x} 
\Gamma_{b}\Big( \sqrt{r^2_x+{k^{\prime}}^2} \Big), \label{phib} \\ 
\Phi_e(k,k^{\prime}) 
&\equiv \int \frac{dr_x}{2\pi} e^{i k r_x} 
\Gamma_{e}\Big(\sqrt{r^2_x+{k^{\prime}}^2}\Big), \label{phie} \\ 
%
\Phi_g(k,k^{\prime}) 
&\equiv  \int \frac{dr_x}{2\pi} e^{i k r_x} 
\Gamma_{g}\Big(\sqrt{r^2_x+{k^{\prime}}^2}\Big), \label{phig} \\
& \equiv \int \frac{dr_y}{2\pi} e^{ik^{\prime}r_y} 
\overline{\Gamma}_{g}\Big(\sqrt{k^2+r^2_y}\Big). \label{phigb}
\end{align}
One could also rewrite $e_{+}$ and $h_{-}$ as well as $e_{-}$ and
 $h_{+}$ in the basis of the $y$-gauge eigenstates. Of course, this leads to  
the same conclusions as we will reach in the $x$-gauge 
eigenstates (see the following two sections).

\section{Ground-state phase diagram in the presence of repulsive Coulomb interaction}
The parquet RG equations have a dual structure; $\Gamma_{b}(r)=\Gamma_{d}(r)$ and 
$\Gamma_{g}(r)$ couple with each other exactly in the same way as 
$\Gamma_{e}(r)$ and $\overline{\Gamma}_{g}(r)$ do, and $\Gamma_{g}(r)$ 
and $\overline{\Gamma}_{g}(r)$ are Fourier transforms of the other 
[Eqs.~(\ref{hankel1},\ref{hankel2})]. In the case of the repulsive interaction, 
this dual structure in the RG equations leads to a ground-state competition between 
the excitonic insulator phase~\cite{brazovskii72} 
and Ising-type spin density wave phase (Fig.~\ref{fig:1}). 
The numerical solution of the RG equations shows that in the two-dimensional $A$-$B$ parameter 
space, either a set of $\Gamma_{e}(r)$ and $\overline{\Gamma}_{g}(r)$ or a set of 
$\Gamma_{b}(r)$ and $\Gamma_{g}(r)$ diverge at a certain critical RG scale, $\xi=\xi_c$.  

\begin{figure}[t]
\begin{center}
	\includegraphics[width=1\linewidth]{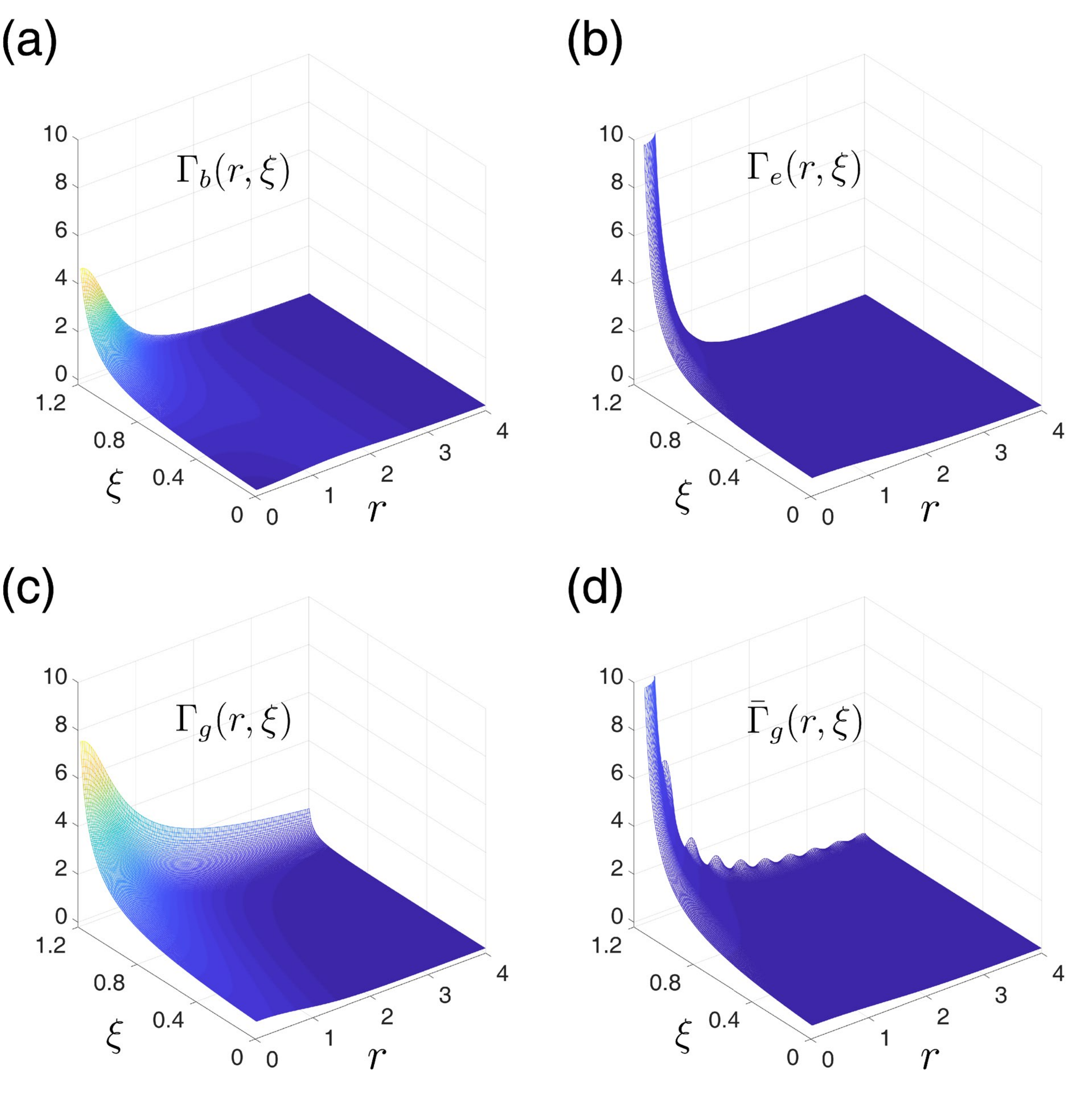}
	\caption{Numerical solution of the parquet RG equations, 
Eqs.~(\ref{gamma-bd},\ref{gamma-e},\ref{gamma-g},\ref{over-gamma-g}), with 
the initial interaction forms given by Eqs.~(\ref{gamma-bd=in},\ref{gamma-e=in},\ref{gamma-g=in},\ref{over-gamma-g=in}) 
in the strong screening regime ($\log_{10} A=0$ and $\log_{10} B = 0$). The solution tells how  
the interaction potentials of $r$,  $\Gamma_{b}(r,\xi)=\Gamma_d(r,\xi)$, 
$\Gamma_e(r,\xi)$, $\Gamma_g(r,\xi)$ and $\overline{\Gamma}_g(r,\xi)$, grow 
as a function of the RG scale $\xi$. $\Gamma_{e}(r,\xi)$ and $\overline{\Gamma}_{g}(r,\xi)$ 
help each other and show the diverge at $r=0$ around $\xi=1.2$.}
	\label{fig:6b}
\end{center}
\end{figure}

\subsection{strong screening region}
To understand the phase diagram qualitatively, let us keep only those terms in 
the parquet RG equations that couple the functions locally in the radial 
coordinate $r$;~\cite{brazovskii72,yakovenko93,tsai02a,alicea09} 
\begin{align}
\frac{d\Gamma_{b/d}(r,\xi)}{d\xi} &= \Gamma^2_{b/d}(r,\xi) + \Gamma^2_{g}(r,\xi), \label{gbd-app} \\
\frac{d\Gamma_{g}(r,\xi)}{d\xi} &= \Gamma^2_{g}(r,\xi)\big(\Gamma_{b}(r,\xi) + \Gamma_{d}(r,\xi)\big), \label{gg-app} 
\end{align}    
and 
\begin{align}
\frac{d\Gamma_{e}(r,\xi)}{d\xi} &= \Gamma^2_{e}(r,\xi) + \overline{\Gamma}^2_{e}(r,\xi), \label{ge-app} \\
 \frac{d\overline{\Gamma}_{g}(r,\xi)}{d\xi} &= 2\overline{\Gamma}^2_{g}(r,\xi) \Gamma_{e}(r,\xi).  \label{ogb-app}  
\end{align}
When the RG scale is near (but below) the critical RG scale, $\xi\lesssim \xi_c$, 
the local terms become leading order than those terms neglected, and the approximation and 
solutions below are justified. Without the constraint between $\Gamma_g(r)$ and 
$\overline{\Gamma}_{g}(r)$ (Eqs.~(\ref{hankel1}),(\ref{hankel2})), 
the approximate RG equations can be solved and the solutions are determined only by the 
initial forms of the interaction potentials, 
\begin{align}
\Gamma_{b/d}(r,\xi) &= \frac{1}{2} \bigg\{\frac{1}{\frac{1}{\Gamma_{b/d}(r,0)+\Gamma_{g}(r,0)}-\xi} 
+ \frac{1}{\frac{1}{\Gamma_{b/d}(r,0)-\Gamma_{g}(r,0)}-\xi}\bigg\}, \label{gbd-app-s} \\
\Gamma_{g}(r,\xi) & = \frac{1}{2} \bigg\{\frac{1}{\frac{1}{\Gamma_{b/d}(r,0)+\Gamma_{g}(r,0)}-\xi} 
- \frac{1}{\frac{1}{\Gamma_{b/d}(r,0)-\Gamma_{g}(r,0)}-\xi}\bigg\}, \label{gg-app-s} 
\end{align} 
and 
\begin{align}
\Gamma_{e}(r,\xi) &= \frac{1}{2} \bigg\{\frac{1}{\frac{1}{\Gamma_{e}(r,0)+\overline{\Gamma}_{g}(r,0)}-\xi} 
+ \frac{1}{\frac{1}{\Gamma_{e}(r,0)-\overline{\Gamma}_{g}(r,0)}-\xi}\bigg\}, \label{ge-app-s} \\
\overline{\Gamma}_{g}(r,\xi) & = \frac{1}{2} \bigg\{\frac{1}{\frac{1}{\Gamma_{e}(r,0)
+\overline{\Gamma}_{g}(r,0)}-\xi} 
- \frac{1}{\frac{1}{\Gamma_{e}(r,0)-\overline{\Gamma}_{g}(r,0)}-\xi}\bigg\}. \label{over-gg-app-s} 
\end{align} 
With Eqs.~(\ref{gamma-bd=in},\ref{gamma-e=in},\ref{gamma-g=in},\ref{over-gamma-g=in},\ref{I0-a},\ref{I2kf-a}) 
at the initial RG scale ($\xi=0$), $\Gamma_{e}(r,0)+\overline{\Gamma}_{g}(r,0)$ 
takes the largest positive value at $r=0$ among the other three at any $r$,
\begin{align} 
&\Gamma_e(r=0,0)+\overline{\Gamma}_g(r=0,0)  \nonumber \\
 & \ge \Gamma_{e}(r,0)\pm \overline{\Gamma}_{g}(r,0), \Gamma_b(r,0)\pm \Gamma_g(r,0). \nonumber 
\end{align} 
Thus, the approximate solution dictates that positive 
$\Gamma_{e}(r,\xi)$ and positive $\overline{\Gamma}_g(r,\xi)$ diverge at $r=0$ 
simultaneously on the renormalization as~\cite{brazovskii72},
\begin{align}
\Gamma_{e}(r,\xi_c) = \overline{\Gamma}_{g}(r,\xi_c) = \frac{A^{\prime}}{r^2} + \cdots \!\  
(A^{\prime}>0). \label{aprime}
\end{align}   
Fig.~\ref{fig:6b} demonstrates how the four interaction 
potentials, $\Gamma_{b}(r,\xi)=\Gamma_d(r,\xi)$, $\Gamma_e(r,\xi)$, $\Gamma_g(r,\xi)$ and 
$\overline{\Gamma}_g(r,\xi)$, change their forms under the parquet RG equations,  
Eqs.(\ref{gamma-bd},\ref{gamma-e},\ref{gamma-g},\ref{over-gamma-g}), in 
the strong screening regime.  
 

When $\Gamma_{e}(r=0)$ and $\overline{\Gamma}_g(r=0)$ dominate over the others, 
an excitonic pairing is formed between electron and hole bands 
at the same Fermi point and at the same spatial coordinate within the $xy$ plane,
\begin{align}
&\langle e^{\dagger}_{+}(k_y) h_{+}(k_y) \rangle = \langle e^{\dagger}_{-}(k_y) h_{-}(k_y) \rangle \ne 0, \label{eq1}  \\
&\langle e^{\dagger}_{+}(q_x) h_{+}(q_x) \rangle = \langle e^{\dagger}_{-}(q_x) h_{-}(q_x) \rangle \ne 0. \label{eq2}
\end{align}
Namely, the asymptotic forms of $\Gamma_{e}(r,\xi_c)$ and $\overline{\Gamma}_{g}(r,\xi_c)$ 
make the following scattering channels in Eq.~(\ref{int-3}) to be dominant among the others,
\begin{align}
&\Phi_{e}(\overline{k}_2-\overline{k}_1,k_1-\overline{k}_2=0) \rightarrow +\infty,   \label{Pe} \\  
&\Phi_{g}(k_1-\overline{k}_2=0,k_2-\overline{k}_2) \rightarrow +\infty,   \label{Pg}
\end{align} 
for arbitrary $\overline{k}_2-\overline{k}_1$ (Eq.~(\ref{Pe})) and arbitrary 
$k_2-\overline{k}_2$ (Eq.~(\ref{Pg})) respectively. 
These scatterings favor electron-hole pairings at the same Fermi points and 
at the same two-dimensional space coordinates within the $xy$ plane;
\begin{align}
S_1 = - &\int_{1,2,3} \int_{k_1-\overline{k}_2=0} d k_1 d\overline{k}_1 d\overline{k}_2 \!\ 
\Phi_{e}(\overline{k}_2-\overline{k}_1,k_1-\overline{k}_2) \nonumber \\
&\hspace{-1.2cm} 
\Big\{ \!\ \big\langle e^{\dagger}_{+}(\overline{k}_2,1) h_{+}(\overline{k}_2,3) \big\rangle 
 \big\langle h^{\dagger}_{+}(\overline{k}_1,2) e_{+}(\overline{k}_1,1+2-3) \big\rangle \nonumber \\ 
& \hspace{-1.0cm} + \big\langle h^{\dagger}_{-}(\overline{k}_2,1) e_{-}(\overline{k}_2,3) \big\rangle 
\big\langle e^{\dagger}_{-}(\overline{k}_1,2) h_{-}(\overline{k}_1,1+2-3) \big\rangle \Big\} \nonumber \\
& 
 - \int_{1,2,3} \int_{k_1-\overline{k}_2=0} d k_1 d\overline{k}_1 d\overline{k}_2 \!\ 
\Phi_{g}(\overline{k}_1-\overline{k}_2,k_2-\overline{k}_2) \nonumber \\
&\hspace{-1.2cm} \Big\{ \!\ 
\big\langle e^{\dagger}_{+}(\overline{k}_2,1) h_{+}(\overline{k}_2,3) \big\rangle 
 \big\langle h^{\dagger}_{-}(k_2,2) e_{-}(k_2,1+2-3) \big\rangle \nonumber \\ 
& \hspace{4cm} +{\rm c.c.} \Big\}  + \cdots.  \label{119}  
\end{align}
Note that the relative U(1) phase between the excitonic pairing field at the right Fermi point 
and that at the left Fermi point is locked to be zero by the positively large 
$\Phi_{g}(k_1-\overline{k}_2=0,k_2-\overline{k}_2)$. 

The excitonic pairing between the electron band with $\uparrow$ spin and the hole band 
with $\downarrow$ spin results in a ferro-type order of an $XY$ component of the spin-1 moment;
\begin{align}
\langle \Psi^{\dagger}_{e,\uparrow}({\bm r}) \Psi_{h,\downarrow}({\bm r}) \rangle 
\equiv X({\bm r}) + i Y({\bm r}) \propto e^{i\theta}.  
\end{align} 
The ferro-type order breaks the U(1) spin rotational symmetry around the magnetic field. 
However, detailed microscopic magnetism of the excitonic phase depends on atomic 
orbitals (localized Wannier orbitals) that form the electron band and the hole band. 

\begin{figure}[t]
\begin{center}
	\includegraphics[width=1\linewidth]{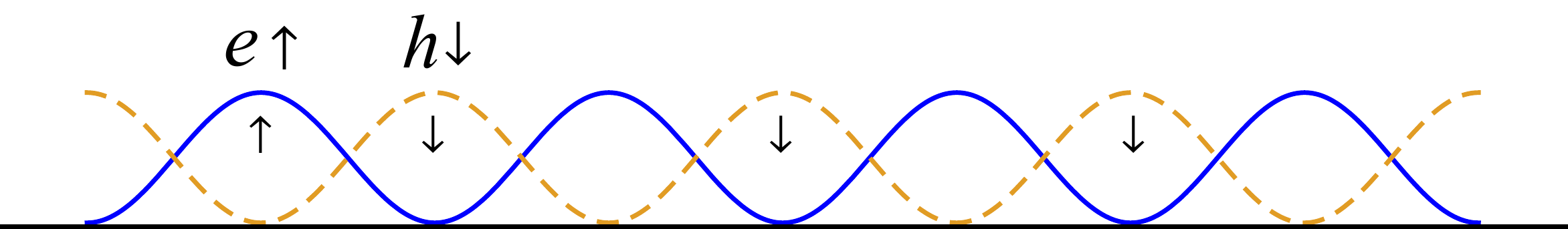}
	\caption{Schematic picture of Ising-type spin density wave. The $\uparrow$ and 
$\downarrow$ arrows are spins along the field. The horizontal axis is along the field direction.}
	\label{fig:7}
\end{center}
\end{figure}

\subsection{weak screening region}
When the screening length is longer than the magnetic length [$A\ge 3$ for $2k_Fl \simeq 1$], 
the numerical solution shows that $\Gamma_{b}(r)$ and $\Gamma_{g}(r)$ diverges at $r=0$ as;
\begin{align}
\Gamma_{b}(r,\xi_c) = \Gamma_{g}(r,\xi_c) = \frac{A^{\prime}}{r^2} + \cdots \!\ \!\ 
 (A^{\prime}>0). \nonumber  
\end{align}
The divergence identifies the relevant scattering channels in Eq.~(\ref{int-3}) as,
\begin{align}
&\Phi_{b}(\overline{k}_2-\overline{k}_1,k_1-\overline{k}_2=0) \rightarrow +\infty,   \nonumber 
\end{align} 
for any $\overline{k}_2-\overline{k}_1$, and  
\begin{align}
&\Phi_{g}(k_1-\overline{k}_2,k_2-\overline{k}_2=0) \rightarrow +\infty,   \nonumber 
\end{align} 
for any $k_1-\overline{k}_2$. These scattering channels cause 
an instability to a charge density wave of the electron band and that of the hole band,
\begin{align}
\langle e^{\dagger}_{+}(k) e_{-}(k) \rangle = e^{i\pi} 
\langle h^{\dagger}_{+}(k) h_{-}(k) \rangle. \label{eq3}
\end{align}
Both density waves share the same spatial pitch ($2\pi/2k_{F}$) along the field direction. 
The relative U(1) phase between the electron-band density wave and hole-hand density wave 
is locked to be $\pi$ by the positively large $\Phi_{g}(k_1-\overline{k}_2,k_2-\overline{k}_2=0)$. 
Due to the $\pi$ phase shift, the ground state in the weak screening region is accompanied by 
Ising-type spin density wave that preserves the U(1) spin rotational symmetry around the 
magnetic field (Fig.~\ref{fig:7}).

\section{Ground-state phase diagram in the presence of effective attractive interaction}
In the previous section, we have studied how the repulsive Coulomb interaction leads to the 
low-temperature instability in the semimetal under high magnetic field. As the complimentary aspect, 
we consider in this section an effect of another relevant many-body interaction; electron-(acoustic) phonon interaction. 
We employ an argument based on an equivalence between an electron-phonon coupled 
system and a system with an electron-electron interaction, and adopt the following effective 
attractive electron-electron interaction;
\begin{align}
H_{\rm eff} &= \frac{1}{2} \int d^3{\bm r} d^3{\bm r}^{\prime} 
\rho({\bm r})\rho({\bm r}^{\prime}) V_{\rm eff}({\bm r}-{\bm r}^{\prime}), \label{veff0} \\
V_{\rm eff}({\bm r}) &= \int \frac{d^3{\bm q}}{(2\pi)^3} V_{\rm eff}({\bm q}) \!\ e^{i{\bm q}{\bm r}}, \label{veff1} \\
V_{\rm eff}({\bm q}) &\equiv - U^2_{\rm eff}({\bm q}), \label{veff2} \\
U_{\rm eff}({\bm q}) &\equiv \Big(\frac{\rho_0}{M}\Big)^{\frac12} 
\frac{4\pi Z e^2 l^2}{\varepsilon c} 
\frac{1}{(q^2_z+q^2_{\perp}) l^2 + A^{-1} e^{-\frac{1}{2}q^2_{\perp} l^2}}. \label{uq} 
\end{align}   
Here $U_{\rm eff}({\bm q})$ is the Fourier transform of a screened Coulomb potential between electron 
and (longitudinal acoustic) phonon. $Z$ and $c$ are an electron valence of positively 
charged nucleus ion and a sound velocity of the acoustic phonon respectively, $\rho_0$ and 
$M$ is the density of the charged nuclei, and a mass of the charged nucleus ion. 
Within the random phase approximation, (the square of) 
the screening length `$A$' in $U_{\rm eff}({\bm q})$ was calculated in the previous section 
[Eq.~(\ref{0-pi},\ref{2kf-pi})]. Here we consider the case with $A=A^{\prime}$ for simplicity.   

Using Eqs.~(\ref{veff0},\ref{veff1},\ref{veff2},\ref{uq}) as the effective electron-electron interaction, 
we study low-temperature instabilities in semimetal under high magnetic field in the presence of the  
electron-phonon coupling. To this end, we solve numerically the same parquet RG equations as in the previous 
section, while we use the following set as the interaction forms at the initial RG scale ($\xi=0$);
\begin{align}
&\Gamma_{b/d}(r) \equiv -2\pi \Big\{ \int^{\infty}_{0} r^{\prime} dr^{\prime} J_0(rr^{\prime}) 
\overline{I}^{\prime}_0(r^{\prime}) - \overline{I}^{\prime}_{2k_F}(r)\Big\},  \label{gamma-bd=in-a} \\ 
&\Gamma_{e}(r) \equiv -2\pi \int^{\infty}_{0} r^{\prime} dr^{\prime} J_0(rr^{\prime}) 
\overline{I}^{\prime}_0(r^{\prime}),   \label{gamma-e=in-a} \\ 
&\Gamma_{g}(r) \equiv -2\pi \overline{I}^{\prime}_{2k_F}(r),  \label{gamma-g=in-a} \\  
&\overline{\Gamma}_{g}(r) \equiv -2\pi \int^{\infty}_{0} r^{\prime} dr^{\prime} J_0(rr^{\prime}) 
\overline{I}^{\prime}_{2k_F}(r^{\prime}), \label{over-gamma-g=in-a} 
\end{align}
with 
\begin{align}
\overline{I}^{\prime}_{0}(r) &\equiv \Big(\frac{4\pi Z e^2 l^2}{\varepsilon c}\Big)^2 \frac{\rho_0}{M}  
\frac{e^{\frac{1}{2}r^2}}{(r^2 e^{\frac{1}{2}r^2}+A^{-1})^2}, \label{I0-a-prime} \\
\overline{I}^{\prime}_{2k_F}(r) &\equiv \Big(\frac{4\pi Z e^2 l^2}{\varepsilon c}\Big)^2 \frac{\rho_0}{M} 
\frac{e^{\frac{1}{2}r^2}}{\big((r^2 + B) 
e^{\frac{1}{2}r^2}+A^{-1}\big)^2}. \label{I2kf-a-prime}
\end{align}
Again, the overall factor of $\overline{I}^{\prime}_{0}(r)$ and $\overline{I}^{\prime}_{2k_F}(r)$ does  
not play any role in a determination of the phase diagram within the one-loop RG analyses. Only the two 
dimensionless parameters $A$ and $B$ play the crucial role.   

\subsection{intermediate screening region}
Fig.~\ref{fig:2} is a phase diagram obtained by the numerical solutions. In an intermediate screening 
region ($A\simeq 10^{-1}$), the ground state shows an instability toward a charge Wigner crystal phase, where 
$\Gamma_{b}(r)$ and $\Gamma_g(r)$ diverges at nonzero $r$ ($r= r_c \ne 0$) at a certain 
critical RG scale ($\xi=\xi_c$) as, 
\begin{align}
\Gamma_{b}(r,\xi_c) = -\Gamma_{g}(r,\xi_c) = \frac{A^{\prime\prime}}{|r-r_c|^2} + \cdots \!\ \!\ 
(A^{\prime\prime}>0).  \label{a-pp}
\end{align}  
Substituting this into Eqs.~(\ref{phib},\ref{phig}), one can see that dominant scattering channels in Eq.~(\ref{int-3}) 
take the following asymptotic forms,  
\begin{align}
&\Phi_{b}(\overline{k}_2-\overline{k}_1,k_1-\overline{k}_2= r_c \cos\theta)  \nonumber \\
& \hspace{2.0cm} \rightarrow \cos\big((\overline{k}_2-\overline{k}_1) r_c \sin\theta\big) \times (+ \infty), \label{a-pp-1} \\
&\Phi_{g}(k_1-\overline{k}_2,\overline{k}_2-k_2= r_c \cos\theta) \nonumber \\
&\hspace{2.0cm} \rightarrow \cos\big((k_1-\overline{k}_2) r_c \sin\theta\big) \times (- \infty), \label{a-pp-2}
\end{align}
for any $\theta \in [0,\pi)$, and for any $\overline{k}_2-\overline{k}_1$ (Eq.~(\ref{a-pp-1}))
and any $k_1-\overline{k}_2$ (Eq.~(\ref{a-pp-2})), respectively. The scattering channels induce $2k_F$ density-wave pairings 
within the electron band and hole hand. The induced density-wave pairings generally 
connect different two-dimensional coordinates within the $xy$ plane, 
\begin{widetext}
\begin{align}
S_1 \equiv &\int^{\pi}_0 d\theta \!\ r_c \sin\theta \!\  S_1(\theta) + \cdots, \nonumber \\
S_1(\theta) = - &\int_{1,2,3} \int 
d\overline{k}_1 d\overline{k}_2 \!\ 
\Phi_{b}(\overline{k}_2-\overline{k}_1,r_c\cos\theta) 
\Big\{ \!\ \big\langle e^{\dagger}_{+}(\overline{k}_2+r_c\cos\theta,1) e_{-}(\overline{k}_2,3) \big\rangle 
 \big\langle e^{\dagger}_{-}(\overline{k}_1,2) 
e_{+}(\overline{k}_1+r_c\cos\theta,\cdots) \big\rangle \nonumber \\
&\hspace{0.5cm} 
+ \big\langle h^{\dagger}_{-}(\overline{k}_2-r_c\cos\theta,1) h_{+}(\overline{k}_2,3) \big\rangle 
\big\langle h^{\dagger}_{+}(\overline{k}_1,2) h_{-}(\overline{k}_1-r_c\cos\theta,\cdots) \big\rangle \Big\} \nonumber \\
& \hspace{-1.5cm} 
+ \int_{1,2,3} \int d k_1 d\overline{k}_2 \!\ 
\Phi_{g}(k_1-\overline{k}_2,r_c\cos\theta) \!\ 
\big\langle e^{\dagger}_{+}(k_1,1) e_{-}(k_1-r_c\cos\theta,\cdots) \big\rangle 
 \big\langle h^{\dagger}_{-}(\overline{k}_2-r_c\cos\theta,2) h_{+}(\overline{k}_2,3) \big\rangle +{\rm c.c.} .  \label{s1-ep-a}
\end{align}
\end{widetext}
Due to the coordinate-dependent ($k$-dependent) cosine functions in Eqs.~(\ref{a-pp-1},\ref{a-pp-2}), 
the action $S_1(\theta)$ is fully minimized by the pairing fields that have coordinate-dependent phases, 
\begin{align}
\langle e^{\dagger}_{+}(k) e_{-}(k-r_c\cos\theta) \rangle 
&= \langle h^{\dagger}_{+}(k) h_{-}(k-r_c\cos\theta) \rangle \nonumber \\
&= B e^{-i\lambda\mp i k r_c \sin\theta}. 
\end{align} 
Such pairings lead to the density waves in the electron and hole bands, 
that break the translational symmetries within the $xy$ plane,
\begin{align}
&\langle \Psi^{\dagger}_{e,\uparrow}({\bm r}) \Psi_{e,\uparrow}({\bm r}) \rangle 
= \langle \Psi^{\dagger}_{h,\downarrow}({\bm r}) \Psi_{h,\downarrow}({\bm r}) \rangle \nonumber \\
&= B^{\prime} \cos \big(2k_F z + r_c (y \!\  \cos\theta  \pm  x\!\ \sin\theta) + \lambda^{\prime} \big).  \nonumber 
\end{align} 
The density wave of the electron band with $\uparrow$ spin 
and the density wave of the hole band with the $\downarrow$ spin have the same phase; the 
superpose of these two is nothing but the charge density wave without any spin texture. 
The spatial pitches within the $xy$ plane and along the field direction is 
$2\pi l/r_c$ and $2\pi/(2k_F)$ respectively. 

The `propagation' direction of the density wave within the $xy$ plane is specified by $\theta$,  
that can take any value in $[0,\pi)$ according to Eqs.~(\ref{a-pp-1},\ref{a-pp-2},\ref{s1-ep-a}). 
The ground state is generally a superposition of 
the density waves with different propagation directions within the $xy$ plane. One of the most 
plausible superposition is a symmetric superposition, 
\begin{align}
&\langle \Psi^{\dagger}_{e,\uparrow}({\bm r}) \Psi_{e,\uparrow}({\bm r}) \rangle = 
\langle \Psi^{\dagger}_{h,\downarrow}({\bm r}) \Psi_{h,\downarrow}({\bm r}) \rangle  \nonumber \\
&\ \ \ \propto \sum_{j=1,2,3}\cos\big(2k_Fz + r_c{\bm n}_j\cdot{\bm r}_{\perp}+\theta\big) + {\rm const}, \label{wigner}
\end{align} 
with ${\bm r}_{\perp}=(x,y)$, ${\bm n}_1=(1,0)$, ${\bm n}_2=(-\frac{1}{2},\frac{\sqrt{3}}{2})$, 
and ${\bm n}_3=(-\frac{1}{2},-\frac{\sqrt{3}}{2})$ [or its O(2) rotation within the $xy$ plane]. 
This leads to a triangle lattice of the charge density within the $xy$ plane 
(charge Wigner crystal; Fig.~\ref{fig:8}(a)).  

\subsection{strong screening region}
In a strong screening region ($A < 10^{-2}$), the ground state exhibits 
an instability to an excitonic phase, where $\Gamma_{e}(r)$ and $\overline{\Gamma}_g(r)$ diverge at 
nonzero $r$ ($r=r_c\ne 0$) at the critical RG scale ($\xi=\xi_c$) as;
\begin{align}
\Gamma_{e}(r,\xi_c) = \overline{\Gamma}_{g}(r,\xi_c) = \frac{A^{\prime\prime}}{|r-r_c|^2} + \cdots, \!\ \!\ 
(A^{\prime\prime}>0).
\end{align}   
The divergence gives rise to the following forms of the dominant scattering channels in Eq.~(\ref{int-3}),
\begin{align}
&\Phi_e(\overline{k}_2-\overline{k}_1,k_1-\overline{k}_2=r_c \cos\theta) \nonumber \\
& \hspace{2cm} \rightarrow \cos\big((\overline{k}_2-\overline{k}_1) r_c \sin\theta\big) \times (+\infty), \label{a-pp-3} \\
&\Phi_g(k_1-\overline{k}_2=r_c \cos\theta, k_2-\overline{k}_2) \nonumber \\
& \hspace{2cm} \rightarrow \cos\big( (k_2-\overline{k}_2) r_c \sin\theta \big) \times (+\infty), \label{a-pp-4}
\end{align}
for any $\theta$, and for any $\overline{k}_2-\overline{k}_1$ (Eq.~(\ref{a-pp-3})) 
and any $k_2-\overline{k}_2$ (Eq.~(\ref{a-pp-4})) respectively. These 
scattering channels mediate the excitonic pairings between different spatial coordinate within the $xy$ plane;
\begin{widetext}
\begin{align}
S_1 \equiv &  \int^{\pi}_{0} d\theta  \!\ r_c \sin\theta \!\ S_{1}(\theta) + \cdots  \nonumber \\
S_1(\theta) = - &\int_{1,2,3} \int d\overline{k}_1 d\overline{k}_2 \!\ 
\Phi_{e}(\overline{k}_2-\overline{k}_1,r_c \cos\theta) 
\Big\{ \!\ \big\langle e^{\dagger}_{+}(\overline{k}_2+r_c\cos\theta,1) h_{+}(\overline{k}_2,3) \big\rangle 
 \big\langle h^{\dagger}_{+}(\overline{k}_1,2) 
e_{+}(\overline{k}_1+r_c\cos\theta,\cdots) \big\rangle \nonumber \\ 
& \hspace{1cm} + \big\langle h^{\dagger}_{-}(\overline{k}_2-r_c\cos\theta,1) e_{-}(\overline{k}_2,3) \big\rangle 
\big\langle e^{\dagger}_{-}(\overline{k}_1,2) h_{-}(\overline{k}_1-r_c\cos\theta,\cdots) \big\rangle \Big\} \nonumber \\
& \hspace{-1.6cm} - \int_{1,2,3} \int dk_2 d\overline{k}_2 \!\ 
\Phi_{g}(r_c\cos\theta,k_2-\overline{k}_2) 
\big\langle e^{\dagger}_{+}(\overline{k}_2+r_c\cos\theta,1) h_{+}(\overline{k}_2,3) \big\rangle 
 \big\langle h^{\dagger}_{-}(k_2,2) e_{-}(k_2+r_c\cos\theta,\cdots) \big\rangle  +{\rm c.c.}. \label{s1-5}
\end{align}
\end{widetext}
Namely, the action with Eqs.~(\ref{a-pp-3},\ref{a-pp-4}) 
is minimized by the excitonic pairing within the same Fermi points 
but between different spatial coordinates within the $xy$ plane. The pairing fields thus determined 
have the coordinate-dependent phase factors, 
\begin{align}
\langle e^{\dagger}_{+}(\overline{k}_2+r_c \cos\theta) h_{+}(\overline{k}_2) \rangle 
&= \langle e^{\dagger}_{-}(\overline{k}_2+r_c \cos\theta) h_{-}(\overline{k}_2) \rangle \nonumber \\ 
&= C e^{i\psi \pm i \overline{k}_2 r_c \sin\theta}. \label{vortex-lattice}
\end{align} 
Such excitonic pairings leads to a density wave of the $XY$ component of the spin-1 moment, 
that breaks the translational symmetry within the $xy$ plane; 
\begin{align}
\langle \Psi^{\dagger}_{e,\uparrow}({\bm r}) \Psi_{h,\downarrow}({\bm r}) \rangle 
\equiv X({\bm r}) + i Y({\bm r}) = e^{i\psi + ir_c (y \cos\theta \pm x \sin \theta)}. \label{dw-xy}
\end{align}

\begin{figure}[t]
\begin{center}
 \includegraphics[width=0.9\linewidth]{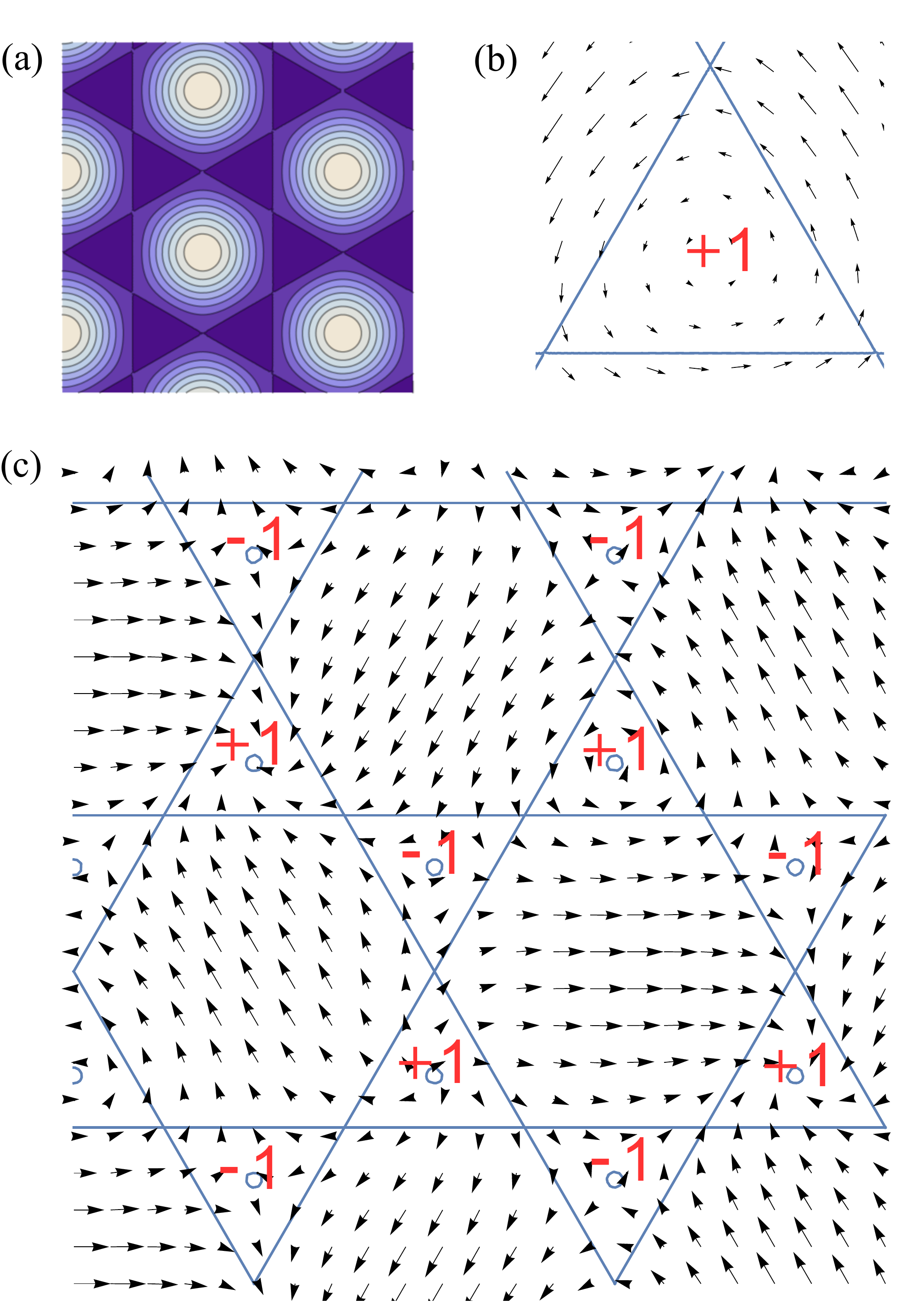}
	\caption{Schematic picture of charge Wigner crystal (a), and a vortex lattice of the $XY$ `spin' moment (b,c). 
The $XY$ spin moment forms vortices with $\pm$ chirality. The vortices with $+/-$ chirality enters $A/B$ sublattice 
of the two-dimensional honeycomb lattice. Note that the lattice constant of these two-dimensional textures 
is given by $l/r_c$, where $l$ is the magnetic length.}
	\label{fig:8}
\end{center}
\end{figure}

The propagation direction of the $XY$ spin density wave is characterized by the  
arbitrary phase $\theta$; the ground state takes a form of the superposition of the density 
waves over different propagation directions within the $xy$ plane. 
From an analogy of the charge Wigner crystal phase, one of the possible
spatial structures of the $XY$ spin moment is the symmetric superposition,  
\begin{align}
 X({\bm r}) + i Y({\bm r}) & \propto  
\sum_{j=1,2,3} e^{i\theta_j + ir_c {\bm n}_j \cdot {\bm r}_{\perp}} \nonumber \\
& = e^{i\theta_T}\sum_{j=1,2,3} e^{ir_c {\bm n}_j \cdot ({\bm r}_{\perp}-{\bm r}_{\perp,0})} \label{vortex} 
\end{align}
with ${\bm n}_1=(1,0)$, ${\bm n}_2=(-\frac{1}{2},\frac{\sqrt{3}}{2})$, 
and ${\bm n}_3=(-\frac{1}{2},-\frac{\sqrt{3}}{2})$. This results in a two-dimensional 
vortex lattice, where vortices of the $XY$ spin with $\pm 1$ charges form a two-dimensional 
honeycomb structure (Fig.~\ref{fig:8}(b,c)). 

In actual semimetal compounds, the emergent two-dimensional lattice structure of the $XY$ spin moment 
as well as the charge density wave must be extremely sensitive to actual 
crystal symmetry of underlying lattice structure in each compound. Especially, $k$-$p$ expansion around 
a high symmetric momentum line (parallel to the magnetic field $\parallel z$) often gives rise to an anisotropy 
in the effective mass or effective velocity within a plane perpendicular to the field ($xy$ plane). The anisotropy 
reduces the in-plane O(2) symmetry in the model dictated by Eqs.~(\ref{HT},\ref{h0},\ref{landau-gauge}) 
down to a discrete rotational symmetry around the field. For example, in the case of the graphite, the relevant 
electron and hole pockets around the zone boundary lines ($H$-$K$-$H$ and  
$H^{\prime}$-$K^{\prime}$-$H^{\prime}$) respect a Z$_3$ discrete rotational symmetry, 
reflecting the graphite crystal structure. Speaking symmetry, the triangle lattice structure 
of the charge density (Fig.~\ref{fig:8}(a)) as well as the two-dimensional vortex lattice 
structure of the $XY$ spin moment (Fig.~\ref{fig:8}c) is compatible with this Z$_3$ discrete 
rotational symmetry.  

\subsection{weak screening region}
In the weak screening region ($A \ge 1$), the phase diagram 
is covered by either charge density wave (smaller $k_F l$ region) or possible non-Fermi liquid 
(larger $k_Fl$ region). In the charge density wave phase, 
$\Gamma_{b}(r)$ and $\Gamma_{g}(r)$ diverges at $r=0$ at a certain critical RG scale as,
\begin{align}
\Gamma_b(r,\xi_c) = -\Gamma_g(r,\xi_c) = \frac{A^{\prime\prime}}{r^2} + \cdots \!\ 
(A^{\prime\prime}>0). \nonumber 
\end{align}
Equivalently, the effective pontentials in Eq.~(\ref{int-3}) will be dominated by the following 
scattering channels, 
\begin{align}
&\Phi_b(\overline{k}_2-\overline{k}_1,k_1-\overline{k}_2=0) \rightarrow + \infty, \nonumber \\ 
&\Phi_g(k_1-\overline{k}_2,k_2-\overline{k}_2=0) \rightarrow  - \infty, \nonumber
\end{align}
for any $\overline{k}_2-\overline{k}_1$ and any $k_1-\overline{k}_2$ respectively. 
As in the previous section, the scatterings give rise to the $2k_F$ 
density wave of the electron band and that of the hole band. The relative U(1) phase between 
the two density waves is locked to be zero by the negatively large 
$\Phi_g(k_1-\overline{k}_2,k_2-\overline{k}_2=0)$,
\begin{align}
\langle e^{\dagger}_{+}(k) e_{-}(k) \rangle = \langle h^{\dagger}_{+}(k) h_{-}(k) \rangle \ne 0.  
\end{align} 
The resulting ground state has a simple charge density modulation along the field direction 
(without any spin texture), whose spatial pitch is $2\pi/(2k_F)$.

\begin{figure}[t]
\begin{center}
 \includegraphics[width=0.9\linewidth]{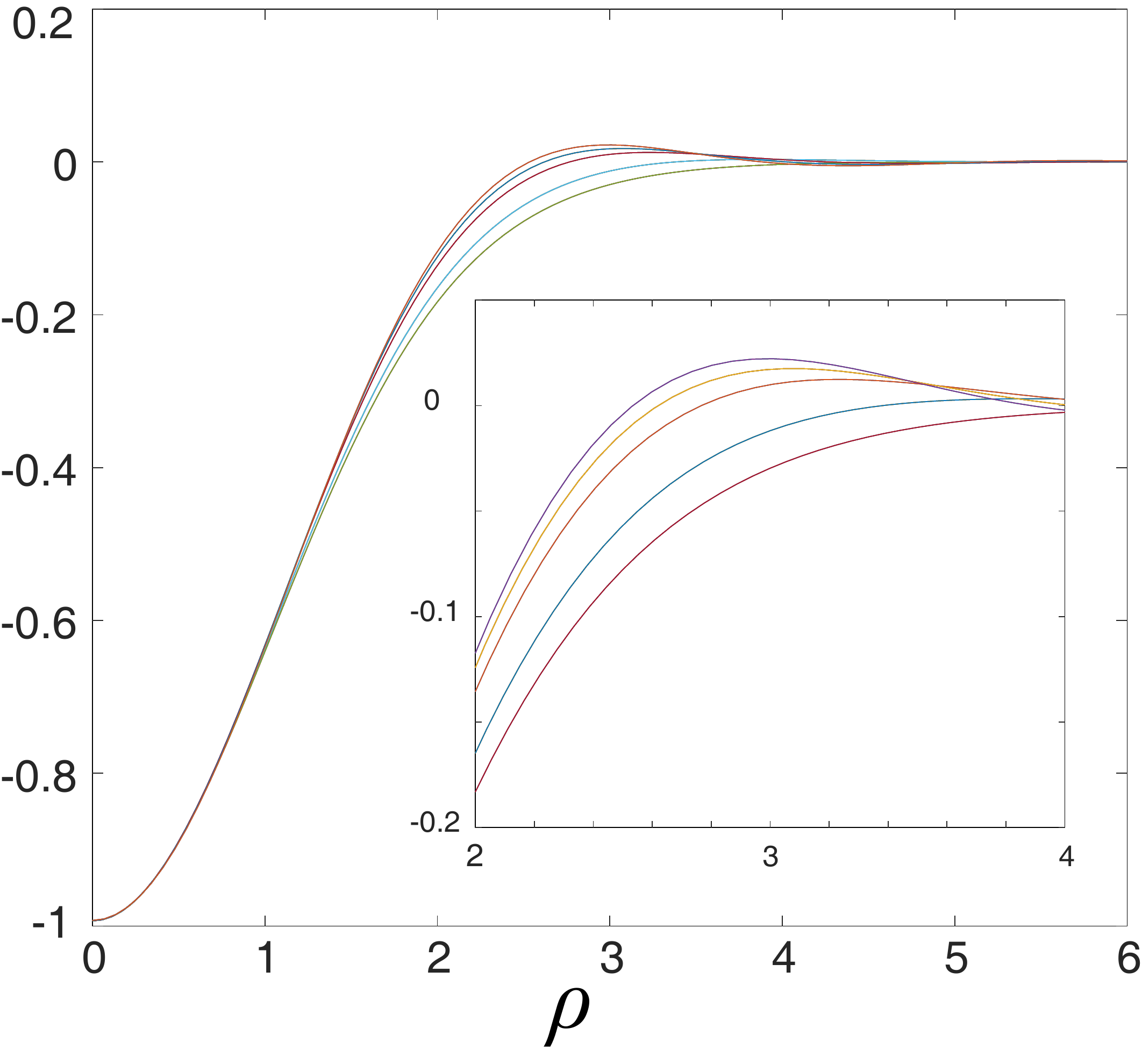}
	\caption{Numerical solution of 
$\Gamma_{b}(r,\xi)$ as a function of $\rho \equiv (\xi-\xi_{1})^{\frac{1}{6}} r)$ 
near for larger $\xi=0,\!\ 5, \!\ 25, \!\ 50, \!\ 100$ 
(from bottom to top) in the possible non-Fermi liquid phase with 
$\xi_1 = -40.2$.}
	\label{fig:9}
\end{center}
\end{figure}

When the spatial pitch of the charge density modulation becomes shorter than the magnetic length 
($1/(2k_Fl) \lesssim 1$), the density wave undergoes a phase transition, and the ground state 
becomes a critical phase. 
In the critical phase, $\Gamma_{g}(r)$ as well as $\overline{\Gamma}_{g}(r)$ get renormalized to the zero 
at any $r$. Since $\Gamma_{g}(r) \equiv \overline{\Gamma}_{g}(r)\equiv 0$, 
the coupled parquet RG equations are decoupled into two RG equations,
\begin{align}
\frac{d\Gamma_{\mu}(r)}{d\xi} = \Gamma^2_{\mu}(r) - \int dr^{\prime} dr^{\prime\prime} 
\Gamma_\mu(r^{\prime})\Gamma_\mu(r^{\prime\prime}) K(r,r^{\prime},r^{\prime\prime}), 
\end{align}    
for $\mu=e,b$. Being attractive, both $\Gamma_{b}(r)$ and $\Gamma_{e}(r)$ converge to  
universal functions of $r$. The universal functions are solutions of the decoupled RG equation 
at larger RG scale, $\xi\gg \xi_1$ where $\xi_1$ is a certain short-range cutoff of the RG scale. 
The functions have a `self-similar' structure (Fig.~\ref{fig:9})~\cite{yakovenko93},
\begin{align} 
\Gamma_b(r,\xi\gg \xi_1)=W_{b,*}((\xi-\xi_1)^{\frac{1}{6}}r), \label{yako1} \\
\Gamma_{e}(r,\xi \gg \xi^{\prime}_1) = W_{e,*}((\xi-\xi^{\prime}_1)^{\frac{1}{6}}r). \label{yako2} 
\end{align}
Yakovenko previously discovered this critical phase in single band model under the magnetic field 
and characterized this critical phase as  marginal Fermi liquid phase, where the renormalization factor of 
the electron Green function vanishes in the large $\xi$ limit~\cite{yakovenko93}.

\subsection{topological excitonic insulator} 
The numerical solutions also found a three-dimensional topological excitonic insulator phase between 
the charge Wigner crystal phase and the excitonic insulator phase with the $XY$-spin vortex lattice. 
Thereby, $\Gamma_{e}(r)$ and $\overline{\Gamma}_{g}(r)$ diverge at $r=0$ as 
\begin{align}
\Gamma_{e}(r) = - \overline{\Gamma}_{g}(r) = \frac{A^{\prime\prime}}{r^2} + \cdots \!\ (A^{\prime\prime}>0). 
\end{align}
The divergence chooses Eqs.~(\ref{Pe},\ref{Pg}) as the dominant scattering channels in Eq.~(\ref{int-3}), 
while the sign of $\Phi_{g}(k_1-\overline{k}_2=0,k_2-\overline{k}_2)$ is negative. Such scattering 
channels lead to a formation of a `(spatially) odd-parity' excitonic pairing that connects the same spatial 
coordinate within the $xy$ plane;
\begin{align}
\langle e^{\dagger}_{+}(k_y) h_{+}(k_y) \rangle = - \langle e^{\dagger}_{-}(k_y) h_{-}(k_y) \rangle 
= |\Delta| e^{i\theta}. \label{odd-parity}
\end{align}
Due to the opposite sign between the two pairings at the right and left Fermi points, the $XY$ components 
that come from these two Fermi points cancel each other. The phase has no local $XY$ component 
of the spin-1 moment. 

As shown by the author previously, the excitonic insulator phase can be regarded as a topological 
band insulator that has a single copy of $(2+1)$D massless Dirac surface fermion at its side 
surface [side surface is along the field direction; $zx$ plane or $yz$ plane]~\cite{pan18}. The emergence of the 
surface state results from the odd-parity excitonic pairing in the bulk and is a direct consequence 
of a ${\bm Z_2}$ topological integer defined in a bulk mean-field electronic Hamiltonian. 

To explain this, note first that the bulk mean-field Hamiltonian takes a form of a sum of `one-dimensional' Hamiltonian, 
as the excitonic pairing connects the same two-dimensional spatial coordinate within the $xy$ plane, 
$H_{\rm mf} \equiv \int dk_y H_{\rm 1D}(k_y)$ with;
\begin{widetext}
\begin{align}
H_{\rm 1D}(k_y) &\equiv \int dk_z \left(\begin{array}{cc} 
e^{\dagger}(k_z,k_y) & h^{\dagger}(k_z,k_y) \\
\end{array}\right) \left(\begin{array}{cc} 
M(k_z,k_y) & \Delta(k_z) e^{-i\theta} \\
\Delta(k_z) e^{i\theta} & -M(k_z,k_y) \\
\end{array}\right) \left(\begin{array}{c}
e(k_z,k_y) \\
 h(k_z,k_y) \\
\end{array}\right),  
\end{align}
\end{widetext} 
\begin{align}
M(k_z,k_y) \equiv \frac{\hbar^2 k^2_z}{2m} - \mu_0 + V_{c}(k_y l^2),  
\end{align}
and $\mu_0\equiv E_g + H_z -\frac{1}{2} \hbar \omega$. 
Here we went back to Eq.~(\ref{LL-bulk}) and 
wrote down explicitly the whole $k_z$-dependence of the kinetic 
energy along the field. Besides, we put a confining potential 
$V_c(x)$ in the $x$ gauge (Landau gauge) with $x=k_y l^2$. $V_c(x)$ is zero in the bulk region 
($|x|<L/2$) and it becomes positively 
 large in the vacuum region ($|x|>L/2$). $\Delta(k_z)$ is the excitonic pairing potential. From 
Eq.~(\ref{119}) and Eq.~(\ref{odd-parity}), the potential 
is an odd function of the momentum along the field,
\begin{align}
\Delta(k_z=\pm k_F) = \mp |\Delta| \int dk \big(\Phi_{e}(k,0) - \Phi_{g}(0,k)\big). \label{odd-parity-2}
\end{align}

When the U(1) phase in Eq.~(\ref{odd-parity}) is spatially uniform, one can absorb the phase  
 into a relative gauge between the electron and hole bands. For a fixed $\theta$, one defines a  
winding number for the one-dimensional mean-field Hamiltonian~\cite{heeger88,wen89},
\begin{align}
Z \equiv \int \frac{dk_z}{2\pi} \big(N_3 \partial_{k_z} N_1 - N_1 \partial_{k_z} N_3\big),  
\end{align}
with
\begin{align}
h_{\rm 1D}(k_z,k_y) &\equiv \left(\begin{array}{cc}
M(k_z,k_y) & \Delta(k_z) \\
\Delta(k_z) & -M(k_z,k_y) \\
\end{array}\right), \nonumber \\
&\equiv X_3(k_z,k_y) \sigma_3 + X_1(k_z,k_y)\sigma_1,  \nonumber \\
(X_3,X_1) &\equiv \sqrt{X^2_3+X^2_1} (N_3,N_1).  
\end{align}
Note that in the bulk region ($V_{c}(k_y l^2)=0$), $N_{3}(k_z,k_y)$ is negative 
for $|k_z|<k_F$ and positive otherwise. Thus, the winding number must be 
an odd integer ($\pm 1$), since $N_1(k_z,k_y)$ is an odd function in $k_z$. Meanwhile, 
in the vacuum region ($V_c(k_y l^2)=+\infty$), the confining potential 
becomes positively large, so that the winding number is always zero; the electron band 
and the hole band are `re-inverted' and 
$N_3$ is always positive for all $k_z$. 

The odd-even difference in the winding number in the one-dimensional 
mean-field Hamiltonians causes the emergence of the surface state in a boundary 
region between bulk and vacuum.  Namely, by regarding $k_{y}l^2$ as 
a `parameter' of one-dimensional electronic system, one can expect that the one-dimensional 
topological band insulator `phase' with the odd integer winding ($|k_y l^2|<L/2$) and 
one-dimensional trivial band insulator `phase' with the zero winding ($|k_y l^2| >L/2$) must be 
intervened by a one-dimensional topological `critical point', that comes at the 
boundaries ($|k_y l^2| \simeq L/2$). In fact, since the excitonic pairing is spatially odd, 
the critical point is generally described by the (1+1)D 
massless Dirac fermion with a linear dispersion along the momentum $k_z$ at $k_z=0$. 
Besides, $M(k_z=0,k_y)$ changes its sign at $|k_y l^2|\simeq L/2$. Thus, the mean-field Hamiltonian 
forms the $(2+1)$D massless Dirac Hamiltonian in the $k_z$-$k_y$ plane around $k_z=0$ and 
$|k_y l^2|\simeq L/2$;
\begin{align}
h_{\rm 1D}(k_z,k_y) = C k_z \sigma_1 \pm D \Big(k_y l^2 \mp \frac{L}{2}\Big) \sigma_3 + 
{\cal O}\big(k^2_z,(\delta k_y)^2\big). \nonumber 
\end{align}  
Note that the massless surface Dirac fermion has helical velocities in any directions 
within the side surface ($yz$ plane 
in the $x$-gauge). It has a helical velocity not only along the field direction ($\parallel z$) 
but also along the perpendicular direction ($\parallel y$). From these observations, 
the excitonic insulator phase with the odd-parity excitonic pairing can be regarded 
as a three-dimensional topological band insulator in the quantum limit. 

In the next section, we will describe the effect of 
the tilted magnetic field on the topological surface state on the side surface. 

\section{effect of titled magnetic field on topological surface states}
\begin{figure}[t]
\begin{center}
 \includegraphics[width=1.0\linewidth]{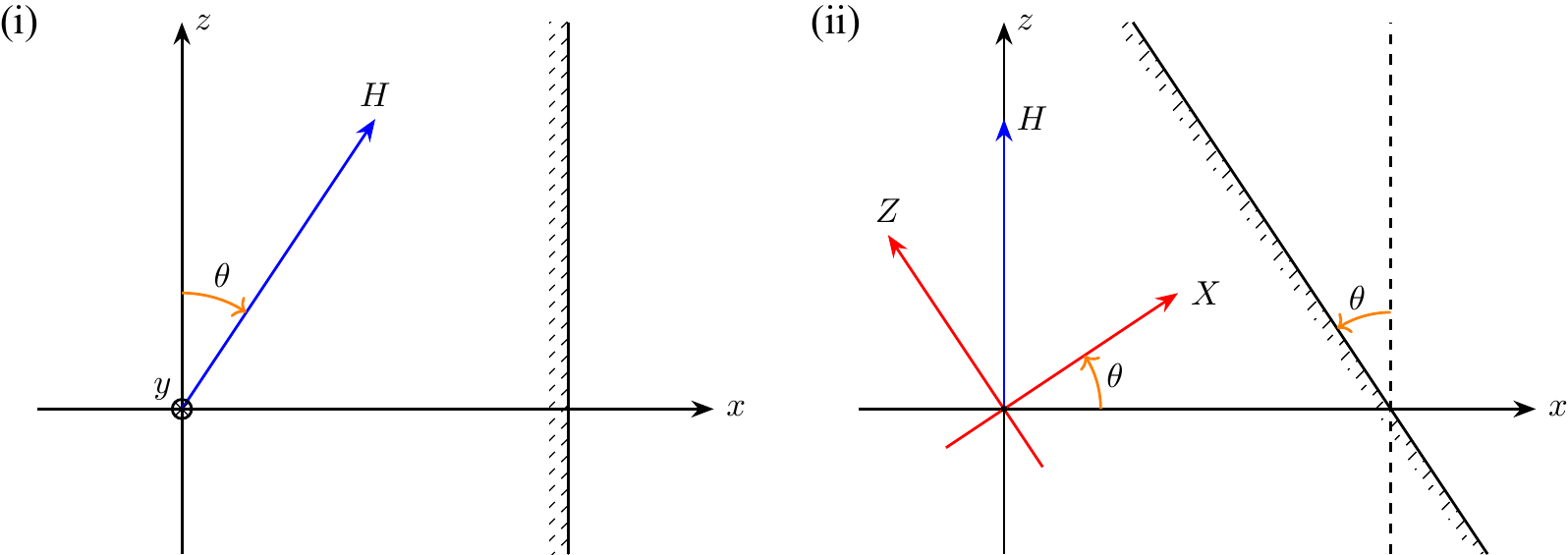}
	\caption{Geometry of the system (i) under tilted magnetic field along $Z\equiv -x\cos\theta+z\sin\theta$ 
with a side surface ($yz$ plane) and (ii) under the magnetic field along $z$ with 
a tilted `side' surface ($yZ$ plane). The system has the translational symmetry along the $y$ direction.}
	\label{fig:10}
\end{center}
\end{figure}

When the magnetic field is tilted from the $z$-axis to $Z$-axis with 
$Z \equiv -x \sin\theta + z \cos\theta$ ($0\ge \theta < \pi$), 
the excitonic pairing in the {\it bulk} remains intact; the three-dimensional semimetal model has 
a spatially isotropic effective mass [Eqs.~(\ref{HT},\ref{h0})]. Meanwhile, the $(2+1)$D massless surface 
Dirac fermion on the side surface ($yz$ plane) forms Landau levels due to a finite out-of-surface 
component of the magnetic field. Equivalently, we can consider the same situation by tilting the `side'  
surface from the $yz$ plane to the $yZ$ plane, and keep the field along the $z$ axis (Fig.~\ref{fig:10}).

Specifically, we add in Eqs.~(\ref{h0},\ref{landau-gauge}) 
a confining potential $V_c(X)$ that depends only on $X \equiv x \cos\theta+z\sin\theta$. 
For simplicity, we take the system is translationally symmetric 
along the $y$-direction, so that $-i\hbar \partial_y$ in 
Eqs.~(\ref{h0},\ref{landau-gauge}) is replaced by $\hbar k_y$. This gives out 
\begin{widetext}
\begin{align}
H_{1D}(k_y) = \int dx \int dz & \left(\begin{array}{cc} 
e^{\dagger}(x,z,k_y) & h^{\dagger}(x,z,k_y) \\
\end{array}\right) \!\ 
\hat{h}_{\rm 1D}(k_y,x,z,\nabla_x,\nabla_z) \!\ 
\left(\begin{array}{c} 
e(x,z,k_y) \\
h(x,z,k_y) \\
\end{array}\right), \label{h1D-a} \\
\hat{h}_{\rm 1D}(k_y,x,z,\nabla_x,\nabla_z) &\equiv 
\left(\begin{array}{cc} 
M(k_y,x,z,\nabla_x,\nabla_z) & i\Delta_0 \nabla_z \\
i\Delta_0 \nabla_z & - M(k_y,x,z,\nabla_x,\nabla_z) \\
\end{array}\right),\label{h1D-b} \\
M(k_y,x,z,\nabla_x,\nabla_z) &\equiv -\frac{\hbar^2 \nabla^2_z}{2m} 
- \frac{\hbar^2 \nabla^2_x}{2m} + \frac{1}{2}m\omega^2 (k_y l^2 + x)^2 + V_{c}(X).  
\label{h1D-c}
\end{align}
\end{widetext}
Here we assume that the odd-parity excitonic pairing is linear in $k_z$, $\Delta(k_z) \equiv \Delta_0 k_z$. The confining 
potential $V_c(X)$ takes a constant value, $V_c(X)=-E_g-H_z$,  
for those $X$ in the bulk region. $V_c(X)$ becomes increasingly large for 
those $X$ in the vacuum region. In the following, we obtain the eigenstates and eigenvalues of
this mean-field Hamiltonian, that are localized at the boundary region. To this end,  
we Taylor-expand $V_{c}(X)$ around the boundary and keep only up to the linear term in the spatial 
coordinate,
\begin{align}
V_{c}(X) &= V_{c}\big(X=\frac{L}{2}\big) + \big(X-\frac{L}{2} \big) \partial_X V_{c}(X)_{|X=\frac{L}{2}} 
+ \cdots \nonumber \\ 
&= V_0 + V_{1} X + {\cal O}\Big(\big(X-\frac{L}{2}\big)^2\Big) \label{taylor-expansion}
\end{align} 
with positive $V_1$. We define the boundary, $L$, such that for $\theta=0$, $M(k_y,k_z=0)$ 
changes its sign at $X=x=-k_y l^2 - \frac{V_1}{m\omega^2}=\frac{L}{2}$. This definition of $L$ 
gives out,   
\begin{align}
\frac{\hbar \omega}{2} + V_0 + \frac{1}{2} V_1 L + \frac{V^2_1}{2m\omega^2} =0. \label{Ldef}
\end{align}
The Taylor expansion will be a priori justified, provided that the potential varies in space much slower than the 
magnetic length; $l \partial_X V_c \ll \hbar \omega$ (see below).

Under a proper basis change of the 2 by 2 Pauli matrices, $\hat{h}_{\rm 1D}(k_y)$ thus given 
can be expressed in terms of raising and lower operators,
\begin{align}
\hat{h}_{\rm 1D}(k_y) \equiv \left(\begin{array}{cc} 
0 & \beta a^{\dagger} + \hbar \omega \!\ b^{\dagger} b \\ 
\beta a + \hbar \!\  \omega b^{\dagger} b & 0 \\
\end{array}\right),  
\end{align} 
with $\beta \equiv \sqrt{2\Delta_0 V_1 \sin\theta}$. 
The raising and lower operators, $[a,a^{\dagger}]=[b,b^{\dagger}]=1$, are defined in the 
following way,
\begin{align}
a^{\dagger} &= -\frac{\hbar^2 \nabla^2_{\tilde{z}}}{2m\beta} + \frac{1}{\sqrt{2}} \Big( \frac{1}{l_{\perp}} \tilde{z}
- l_{\perp} \nabla_{\tilde{z}} \Big), \label{a-dagger} \\
a &= -\frac{\hbar^2 \nabla^2_{\tilde{z}}}{2m\beta} + \frac{1}{\sqrt{2}} \Big( \frac{1}{l_{\perp}} \tilde{z}
+ l_{\perp} \nabla_{\tilde{z}} \Big), \label{a-non-dagger} \\ 
b^{\dagger} &= \frac{1}{\sqrt{2}} \Big( \frac{1}{l} \tilde{x}
- l \nabla_{\tilde{x}} \Big), \ \ \ b =  \frac{1}{\sqrt{2}} \Big( \frac{1}{l} \tilde{x}
+ l \nabla_{\tilde{x}} \Big). \label{b-dagger-non-dagger}  
\end{align} 
with $l_{\perp} \equiv \sqrt{\frac{\Delta_0}{V_1 \sin\theta}}$, 
$\tilde{x} \equiv x - x_0$, $\tilde{z} \equiv z-z_0$ and 
\begin{align}
x_0 &\equiv - \Big(k_y l^2 + \frac{V_1\cos\theta}{m\omega^2}\Big), \nonumber \\
z_0 &\equiv \frac{\cos\theta}{\sin\theta} \Big(k_y l^2 + \frac{V_1\cos\theta}{2m\omega^2}\Big) 
- \frac{1}{V_1 \sin\theta} \Big(\frac{\hbar \omega}{2} + V_0\Big). \nonumber 
\end{align}
The raising (lowering) operators, $a^{\dagger}$ ($a$) and $b^{\dagger}$ ($b$), 
have ladders of number states, $|n\rangle_a$, $|n\rangle_b$, 
\begin{align}
a|0\rangle_a =0, \!\ \!\ \!\  a^{\dagger}|n-1\rangle_a = \sqrt{n}|n\rangle_a, \nonumber \\
b|0\rangle_b =0,  \!\ \! \!\ b^{\dagger}|n-1\rangle_b = \sqrt{n}|n\rangle_b. \nonumber   
\end{align}
These number states are functions only of $z$ and $x$. They are localized 
around $z=z_0$ and $x=x_0$ with localization length $l_{\perp}$ and $l$ respectively. 
Especially, $|0\rangle_a$ is given by the Airy function. 

$\hat{h}_{\rm 1D}(k_y)$ thus given has following set of eigenstates and eigenvalues; 
\begin{align}
\phi_0(x,z) \equiv &\left(\begin{array}{c}
|0\rangle_a |0 \rangle_{b} \\
0 \\
\end{array}\right) \hspace{1.5cm} \big(E=0 \big), \label{0-LL} \\
\phi_{\pm |n|}(x,z) \equiv &\frac{1}{\sqrt{2}} 
\left(\begin{array}{c}
|n\rangle_a |0 \rangle_{b} \\
\pm |n-1\rangle_a |0\rangle_b \\
\end{array}\right) \hspace{0.0cm} \!\  \big(E = \pm |E_n|\big),  \label{n-LL}  
\end{align}
with $n\ge 1$ and $E_n \equiv \sqrt{2\Delta_0 \sin\theta H |n|}$.  The $k_y$ dependence 
is encoded into $x_0$ and $z_0$ in the number states. Irrespective of $k_y$, 
the eigenstates are localized around $X = \frac{L}{2}$ along the $X$-direction;
\begin{align}
X_0 &\equiv x_0 \cos\theta + z_0 \sin \theta \nonumber \\
&= -\frac{1}{V_1}\Big(\frac{\hbar \omega}{2} 
+ V_0\Big) - \frac{1}{2} \frac{V_1\cos^2\theta}{m\omega^2} = \frac{L}{2} + {\cal O}\Big(\frac{l^2}{\lambda}\Big). 
\label{X0}
\end{align} 
Here $\lambda$ is a characteristic length scale with which the confining potential varies in space 
around the boundary, $\lambda V_1 \equiv \hbar \omega$. Provided that $\lambda\gg l$, the eigenstates 
with different $k_y$ are all localized at $X=L/2$. The localized feature of the eigenstates a priori 
justifies the Taylor expansion of $V_{c}(X)$ around $X=L/2$ in $\hat{h}_{\rm 1D}(k_y)$. 

The eigenstates with different $k_y$ are energetically degenerate in each sLL and 
they are localized at different locations along the $Z$-axis, 
\begin{align} 
Z_{0} &\equiv -x_0 \sin\theta + z_0 \cos\theta \nonumber \\
&= \frac{\cos\theta}{\sin\theta} \frac{L}{2} 
+ \frac{k_y l^2}{\sin\theta} + {\cal O}\Big(\frac{l^2}{\lambda}\Big).  \label{Z0}
\end{align} 
Accordingly, the degeneracy at each surface Landau level is proportional to an 
area of the side surface and the out-of-surface component of the magnetic field,
\begin{align}
k_{y} = \frac{2\pi m}{L_y} \!\ \!\ \!\ \!\ \!\ 
 \bigg(m=1,2,\cdots,\frac{L_yL_Z \sin\theta}{2\pi l^2}\bigg). 
\end{align}

In conclusion, the $(2+1)$D massless surface Dirac state in the topological excitonic 
insulator under the tilted magnetic field forms a sequence of the surface Landau levels, 
$E_{n}={\rm sgn} (n) \sqrt{2\Delta_{0}H_{\perp} |n|}$ ($n=\cdots,-1,0,1,\cdots$).  
Each surface Landau level has an degeneracy of $L_{y}L_{Z}\sin\theta/(2\pi l^2)$, where 
$H_{\perp}$ is the out-of-surface component of the magnetic field, 
$H_{\perp} \equiv H \sin\theta$. 

\section{summary and discussion on experiment}  
In this paper, we clarify comprehensive ground-state phase diagrams of 
a three-dimensional semimetal model in the quantum limit. The semimetal model 
has a pair of electron and hole pocket. We study two limiting cases 
at the charge neutrality point, (i) the model with screened Coulomb 
interaction and (ii) the model with an effective attractive interaction 
mediated by the screened electron-phonon interaction. The results show rich phase diagram 
structures as a function of the Fermi wavelength and the screening 
length (normalized by the magnetic length). In the repulsive 
interaction case, we found that an Ising-type spin density wave phase / excitonic 
insulator phase with ferro-type order of $XY$ spin moment is stabilized in the 
weak / strong screening regime respectively. In the attractive interaction case, we 
found that the plain charge density phase or a possible non-Fermi liquid phase is 
stabilized for weak screening regime, while from the intermediate to strong 
screening regimes, the ground state is dominated by charge Wigner crystal phase, 
topological excitonic insulator phase, and excitonic insulator with a 
two-dimensional vortex lattice of the $XY$ component of the spin-$1$ moment.  

The topological excitonic insulator phase in the attractive interaction case is an three-dimensional 
interaction-driven topological band insulator in the quantum limit. Thereby, 
the odd-parity excitonic pairing in the bulk gives rise to a single copy of 
the $(2+1)$D massless surface Dirac 
fermion state at those surfaces parallel to the magnetic field. We show that when 
an in-plane transport is dominated by the surface transport through the $(2+1)$D 
massless Dirac state, the in-plane resistivity must show a $\sqrt{H_{\perp}}$-type 
surface SdH oscillation under canted magnetic field $H_{\perp}$. 

Recently, a comprehensive resistivity measurement in graphite under 
high magnetic field has been carried out up to 90 T~\cite{zhu19}. The graphite 
under the high magnetic field (the field $\perp$ the graphene plane) exhibits 
consecutive metal-insulator transitions as well as insulator-metal re-entrant 
transition at low temperature in an electric resistivity 
along the out-of-plane (field) direction~\cite{yaguchi09,yaguchi98a,fauque13,akiba15,arnold17,zhu19}. 
Experimentally, there exists two insulating phases, 
one insulating phase in the range of 30 T $< \!\ H \!\ <$ 53 T and 
the other in the range of 53 T $< \!\ H \!\ < $ 75 T. The recent experiment 
shows that the resistivity within the in-plane direction shows unusual `metallic' behaviour 
in the second `insulating' phase at 53 T $< \!\ H \!\ < $ 75 T~\cite{yaguchi09,fauque13,akiba15,zhu19}. 

The semimetal model studied in this paper can be applicable to the latter field 
regime (53 T $< \!\ H \!\ < $ 75 T), where an electron pocket locates around the $K$ point and a hole pocket 
locates around the $H$ point in the first Brillouin zone of the graphite under $H$. Previously, 
the authors argued that the low-$T$ insulator phase in the range of 53 T $< \!\ H \!\ < $ 75 T  
could be the topological EI phase, where the unusual metallic in-plane resistivity was attributed 
to the surface transport through the $(2+1)$D massless Dirac states~\cite{pan18}.

A relevant electronic energy band in graphite under the high magnetic field has a band width 
of $40$ meV, the lattice constant along the out-of-plane direction $c_0$ is 6.7 $\AA$, and the 
relative permittivity $\varepsilon$ in graphite is from 9 to 16.  We assume that 
$k_F=\pi/(6c_0)$ for $H=64$ T.   
For $H=64$ T, this gives out $\hbar v_F = -\partial t\cos(kc_0)/\partial k|_{k=k_F} = tc_0/2= 10$ 
meV $\times 6.7 \AA$  with $t=20$ meV, $\log_{10} B =2\log_{10} (2k_F l) \simeq 1.4$. From 
Eq.~(\ref{0-pi}), $1/A = (4e^2)/(\hbar \pi v_F \varepsilon) 
\simeq 22$ and $\log_{10} A \simeq -1.34$. The sound velocity in graphite along 
the $c_0$ axis is around 500 m/s. Carbon is 12 amu heavy ($M=12$ amu), and 
the density of carbon atom in graphite is $\rho_0=0.12 ... \AA^{-3}$. From low-carrier 
density feature in graphite in the zero field (at most $n=10^{18}$ cm$^{-3}$), 
we take $Z=10^{-5}$. For $H=64$ T with $1/A=22$, this set of 
parameters give a ratio between the overall factor of the effective 
attractive interaction mediated by the screened electron-phonon interaction 
and that of the screened repulsive Coulomb interaction. It turns out that the 
effective attractive interaction is much smaller than the screened repulsive 
Coulomb interaction, 
\begin{align} 
\frac{4\pi e^2 A l^2}{\varepsilon} : \frac{\rho_0}{Mc^2} 
\bigg(\frac{4\pi e^2 A l^2 Z}{\varepsilon}\bigg)^2 = 1:2.5\times 10^{-7}. \label{ratio}
\end{align}
The small value of the effective attractive interaction is mainly because of tiny electron valence 
of charged nucleus ion (carbon atom), $Z$. The tiny $Z$ even overcomes very large screening length, e.g. 
$l_{\rm scr} \equiv \sqrt{A} l = 6.8 \AA$ for $H=64$ T. Thereby, if we simply add 
these two interactions with the opposite signs at the initial RG scale, 
the repulsive interaction clearly dominates over the attractive interaction. 
This would be the case even if we used 100 times larger value of $Z$ 
than the value given above. From this observation, we consider in the following 
the case with only the repulsive Coulomb interaction.

In Fig.~\ref{fig:1}, the parameter point with $(\log_{10} A, \log_{10} B)\simeq (-1.34,1.4)$ 
corresponds to the EI phase with the broken U(1) spin rotational symmetry around 
the field direction. Please note that the excitonic pairing in the EI phase in Fig.~\ref{fig:1} has 
the spatially even parity [see Eqs.~(\ref{eq1},\ref{eq2})], and thereby it is non-topological EI instead 
of the topological EI. In fact, the non-topological 
EI phase seems to be consistent with the second `insulating' phase in a recent graphite 
experiment. The recent in-plane resistivity experiment under the canted magnetic 
field does {\it not} show any SdH oscillation as a function of the canted component 
of the magnetic field~\cite{zhu19}, unlike the expectation from the surface 
transport in the topological EI phase.

Depending on other factors, the excitonic pairing in the non-topological EI phase in the repulsive 
interaction case could be formed between electron band and hole band at {\it different} spatial coordinates 
within the $xy$ plane, as was the case in Sec.~VIB and Fig.~\ref{fig:8}(c). 
For example, a ratio between $A$ and 
$A^{\prime}$ may not be $1$. $A$ itself could be smaller by several factors than the value 
given above, due to additional screening from the higher LLs and from the other valley in graphite.

The excitonic pairings between different spatial coordinates within the $xy$ plane could 
induce coherent carrier transports within the plane. Since the excitonic pairing is 
between $\uparrow$ spin electron-type band and $\downarrow$ spin 
hole-type band, the transport must be free from pinning effect due to charged 
impurities~\cite{shayegan97}. Thereby, we can expect that such EI phase with broken 
translational symmetries within the $xy$ plane may give a simple theory explanation 
for the in-plane metallic bulk-transport behaviour in the second `insulating' phase 
of 53 T $< H <$ 75 T in the graphite experiment~\cite{yaguchi09,fauque13,akiba15,zhu19}. 
In fact, the recent transport experiment up to 90 T shows that the in-plane  
resistivity in the second `insulating' phase is nearly {\it constant} in the field~\cite{zhu19}. 
This observation is consistent with the two-dimensional $XY$-spin vortex lattice shown in 
Fig.~\ref{fig:8}(c) whose lattice constant is proportional to the magnetic length $l$. 
Since the lattice constant within the $xy$ plane is scaled by $l$, an Aharonov-Bohm 
(AB) flux that penetrates through a unit cell of the two-dimensional spin vortex lattice 
is {\it independent} of the field. This results in an {\it absence} of any 
SdH-like oscillation in the in-plane transport inside the second `insulating' phase. 
Nonetheless, for further understandings of the unusual transport in graphite 
as well as the re-entrant insulator-metal transition, we need further theoretical 
studies and relevant results will be discussed elsewhere.    

\section*{ACKNOWLEDGMENTS}
RS acknowledge helpful and enlighting discussions with Zengwei Zhu, 
Miguel A. Cazalilla, Tomi Ohtsuki, Alexei Tsvelik, and Masatoshi Sato. 
The work was supported by NBRP of China (Grant No. 2014CB920901, 
Grant No. 2015CB921104 and Grant No. 2017A040215). 

\appendix 
\section{RPA screening}
The interaction potentials that carry the zero momentum along the field, 
i.e. $\Gamma_{\mu\nu}$ in Eq.~(\ref{int-h1}), are screened by low-energy density 
fluctuations within each branch (`right-mover' or `left-mover' branch) of the 
electron-type band or hole-type band. The screened interaction comprises of a sum of the bare 
interaction part and an effective interaction mediated 
by the density fluctuations~\cite{fetter03}. According to the linear response theory, the effective interaction part 
is given by a retarded correlation functions between the density fluctuations;
\begin{widetext}
\begin{align}
\overline{H}_1 &\equiv \frac{1}{2} \int \frac{dp}{2\pi} \int d(lQ_1) d(lQ^{\prime}_1) d(lQ^{\prime}_2)  d(lQ_2) 
\nonumber \\
& \hspace{0.4cm} 
\Gamma_{\mu\nu}(Q_1,Q^{\prime}_1,Q^{\prime}_{2},Q_{2};I_0) 
\int \frac{dp_2}{2\pi}
a^{\dagger}_{\mu}(Q^{\prime}_1,p_2+p) a_{\mu}(Q^{\prime}_2,p_2)  \!\ 
\int \frac{dp_1}{2\pi}
a^{\dagger}_{\nu}(Q_1,p_1-p) a_{\nu}(Q_2,p_1)  \nonumber \\  
& \hspace{-0.2cm} +  
\frac{1}{2 \hbar} \sum_{\mu,\nu,\lambda,\psi} \int \frac{dp^{\prime}}{2\pi} \frac{dp}{2\pi}
\int d(lQ_1) d(lQ^{\prime}_1) d(lQ^{\prime}_2)  d(lQ_2)  
 \int \frac{dp_2}{2\pi}
a^{\dagger}_{\mu}(Q^{\prime}_1,p_2+p^{\prime}) a_{\mu}(Q^{\prime}_2,p_2)  \!\ 
\int \frac{dp_1}{2\pi} a^{\dagger}_{\nu}(Q_1,p_1-p) a_{\nu}(Q_2,p_1) \nonumber \\  
&\hspace{0.2cm} 
\int d(lQ^{\prime\prime}_1) d(lQ^{\prime\prime\prime}_1)  d(lQ^{\prime\prime\prime}_2) 
 d(lQ^{\prime\prime}_2) \!\ \!\ 
\Gamma_{\mu\lambda}(Q^{\prime\prime}_1,Q^{\prime}_1,Q^{\prime}_2,Q^{\prime\prime}_2;I_0) 
\!\ D^{R}_{\lambda\psi}(-p^{\prime},p,\omega=0) \!\  
\Gamma_{\psi\nu}(Q_1,Q^{\prime\prime\prime}_1,Q^{\prime\prime\prime}_2,Q_2;I_0).  
 \label{int-1}
\end{align}    
\end{widetext}
with $\mu,\nu,\lambda,\psi=e_{+},e_{-},h_{+},h_{-}$ and Eq.~(\ref{notation-a}).
The first term in the right-hand side is the bare interaction part and second term 
is the effective interaction part.  
The retarded correlation function $D^R_{\lambda\psi}(-p^{\prime},p,\omega)$ 
is obtained from a time-ordered correlation function in the 
static limit, 
$D^{R}_{\lambda\psi}(-p^{\prime},p,\omega=0) = D^{T}_{\lambda\psi}(-p^{\prime},p,\omega=0)$ with;
\begin{align}
&i D^{T}_{\lambda\psi}(-p^{\prime},p,t-t^{\prime}) \equiv  \nonumber \\
&\hspace{0.1cm} 
\frac{\langle \Psi_0 | T\big\{\delta \hat{T}_{\lambda,H}(Q^{\prime\prime}_1,Q^{\prime\prime}_2,-p^{\prime},t) 
\delta \hat{T}_{\psi,H}(Q^{\prime\prime\prime}_1,Q^{\prime\prime\prime}_2,p,t^{\prime}) \big\} | \Psi_0 \rangle}{\langle \Psi_0 |\Psi_0\rangle},  \nonumber \\
& D^{T}_{\lambda\psi}(-p^{\prime},p,\omega) \equiv \int^{\infty}_{-\infty} dt e^{i\omega t} 
D^{T}_{\lambda\psi}(-p^{\prime},p,t). \nonumber 
\end{align}
Here $|\Psi_0\rangle$ is a many-body ground-state wavefunction and (real-)time dependence of 
the operator is in the Heisenberg picture. 
$\delta \hat{T}_{\mu}(Q_1,Q_2,q)$ is the density fluctuation operator within 
every branch $\mu=e_{+},e_{-},h_{+},h_{-}$, 
\begin{align}
&\hat{T}_{\mu}(Q_1,Q_2,p) \equiv \int \frac{dp_1}{2\pi} 
a^{\dagger}_{\mu}(Q_1,p_1+p) a_{\mu}(Q_2,p_1), \nonumber \\
& \delta \hat{T}_{\mu}(Q_1,Q_2,p) \equiv \hat{T}_{\mu}(Q_1,Q_2,p) - 
\frac{\langle \Psi_0 | \hat{T}_{\mu}(Q_1,Q_2,p) | \Psi_0 \rangle}{\langle \Psi_0 | \Psi_0 \rangle}. \nonumber 
\end{align}
According to the Feynman-Dyson perturbation theory~\cite{fetter03,mahan00}, 
the time-ordered correlation function is given by a proper part of the 
polarization function. The RPA approximates the proper part 
by its lowest order in the electron correlation. This gives out  
\begin{align}
&D^{T,{\rm RPA}}_{\lambda\psi}(-p^{\prime},p,\omega) = \nonumber \\
& \hspace{0.2cm}\delta(p^{\prime}-p)  \Big\{ 
\delta\big(l(Q^{\prime\prime}_2-Q^{\prime\prime\prime}_1)\big) 
\delta\big(l(Q^{\prime\prime\prime}_2-Q^{\prime\prime}_1)\big) 
\Pi_{0,\lambda}(\omega) \delta_{\lambda\psi} \nonumber \\
& \hspace{0.6cm} 
+ \frac{1}{2\pi \hbar} \Gamma_{\lambda\psi}(Q^{\prime\prime\prime}_2,Q^{\prime\prime}_2,
Q^{\prime\prime}_1,Q^{\prime\prime\prime}_1;I_0) 
\!\ \Pi_{0,\lambda}(\omega) \!\ \Pi_{0,\psi}(\omega) \nonumber \\ 
&\hspace{-0.2cm} + \frac{1}{(2\pi \hbar)^2} 
\int d(l\tilde{Q}_{1}) d(l\tilde{Q}_2)  \!\
\Pi_{0,\lambda}(\omega) \!\ \Gamma_{\lambda\rho}(\tilde{Q}_2,Q^{\prime\prime}_2,
Q^{\prime\prime}_1,\tilde{Q}_1;I_0)  \nonumber \\ 
&\hspace{0.4cm} 
\Pi_{0,\rho}(\omega) \!\ \Gamma_{\rho\psi}(Q^{\prime\prime\prime}_{2},
\tilde{Q}_1,\tilde{Q}_2,Q^{\prime\prime\prime}_1;I_0) \!\ 
 \Pi_{0,\psi}(\omega) + \cdots \Big\}, \label{0-rpa}
\end{align}
where the summation over $\rho=e_{+},e_{-},h_{+},h_-$ is omitted in the right hand side. 
In the static limit, the 
bare polarization function $\Pi_{0,\lambda}(\omega)$ for $\lambda=e_{+},e_{-},h_{+},h_{-}$  
is given by Eqs.~(\ref{bare-po-0},\ref{bare-p1}). In terms of the homomorphic 
nature of the interaction potential functional, Eq.~(\ref{homomorphic-0}), 
Eq.~(\ref{int-1}) with the RPA correlation function Eq.~(\ref{0-rpa}) reduces to 
\begin{align}
\overline{H}_1 &= \frac{1}{2} \sum_{\mu,\nu} \int \frac{dp \!\ dp_1 \!\ dp_2}{(2\pi)^3} 
\int d(lQ_1) \cdots d(dQ_2) \nonumber \\
&\hspace{0.5cm} 
a^{\dagger}_{\mu}(Q^{\prime}_1,p_2+p) a_{\mu}(Q^{\prime}_2,p_2)  \!\ 
a^{\dagger}_{\nu}(Q_1,p_1-p) a_{\nu}(Q_2,p_1) \nonumber \\
& \hspace{2.5cm} \times \Gamma_{\mu\nu}
(Q_1,Q^{\prime}_1,Q^{\prime}_{2},Q_{2};\overline{I}_{0}), \label{screen-1a} 
\end{align}
where $\overline{I}_{0}(q_x,k_y)$ is given by Eq.~(\ref{screen-1b}). 

The interaction potentials that carry $2k_F$ momentum along the field, 
$\Phi^{+-}_{\mu\nu}$ in Eq.~(\ref{int-h2}), are also screened by $2k_F$ density fluctuations. 
As above, the screened interaction is characterized by the retarded density 
correlation function between the $2k_F$ density fluctuation operators;
\begin{widetext}
\begin{align}
\overline{H}_2 &=  \sum_{\mu,\nu=e,h} \int \frac{dp}{2\pi} 
\int d(lQ_1) \int d(lQ^{\prime}_1) \int d(lQ^{\prime}_2) 
\int d(lQ_2)  \nonumber \\
& \hspace{0.4cm} 
\Phi^{+-}_{\mu\nu}(Q_1,Q^{\prime}_1,Q^{\prime}_{2},Q_{2};I_{2k_{F}}) 
\int \frac{dp_2}{2\pi}
a^{\dagger}_{\mu_{+}}(Q^{\prime}_1,p_2+p) a_{\mu_{-}}(Q^{\prime}_2,p_2)  \!\ 
\int \frac{dp_1}{2\pi}
a^{\dagger}_{\nu_{-}}(Q_1,p_1-p) a_{\nu_{+}}(Q_2,p_1)  \nonumber \\   
&\hspace{-0.4cm} + \frac{1}{\hbar} \sum_{\mu,\nu,\lambda,\psi} \int \frac{dp^{\prime}}{2\pi} \frac{dp}{2\pi}
\int d(lQ_1) d(lQ^{\prime}_1) d(lQ^{\prime}_2)  d(lQ_2)  
 \int \frac{dp_2}{2\pi}
a^{\dagger}_{\mu_{+}}(Q^{\prime}_1,p_2+p^{\prime}) a_{\mu_{-}}(Q^{\prime}_2,p_2)  \!\ 
\int \frac{dp_1}{2\pi} a^{\dagger}_{\nu_{-}}(Q_1,p_1-p) a_{\nu_{+}}(Q_2,p_1) \nonumber \\  
&\hspace{-0.3cm} 
\int d(lQ^{\prime\prime}_1) d(lQ^{\prime\prime\prime}_1)  d(lQ^{\prime\prime\prime}_2) 
 d(lQ^{\prime\prime}_2) \!\ \!\ 
\Phi^{+-}_{\mu\lambda}(Q^{\prime\prime}_1,Q^{\prime}_1,Q^{\prime}_2,Q^{\prime\prime}_2;I_{2k_F}) 
\!\ D^{R,-+}_{\lambda\psi}(-p^{\prime},p,\omega=0) \!\  
\Phi^{+-}_{\psi\nu}(Q_1,Q^{\prime\prime\prime}_1,Q^{\prime\prime\prime}_2,Q_2;I_{2k_F}).  
\label{int-2}
\end{align} 
\end{widetext}
In the static limit ($\omega=0$), the retarded correlation function 
$D^{R,-+}_{\lambda\psi}(-p^{\prime},p,\omega)$ is identical to the corresponding 
time-ordered correlation function;
\begin{align}
&iD^{T,-+}_{\lambda\psi}(-p^{\prime},p,t-t^{\prime}) \equiv \nonumber \\
& \ \frac{\langle\Psi_0|T\{ \delta \hat{S}^{-}_{\lambda,H}(Q^{\prime\prime}_1,Q^{\prime\prime}_2,t) 
\delta \hat{S}^{+}_{\psi,H}(Q^{\prime\prime\prime}_1,Q^{\prime\prime\prime}_2,t^{\prime}) \} 
|\Psi_0\rangle}{\langle \Psi_0|\Psi_0\rangle}, \nonumber \\
& D^{T,-+}_{\lambda\psi}(-p^{\prime},p,\omega) = \int^{\infty}_{-\infty} dt e^{i\omega t} 
 D^{T,-+}_{\lambda\psi}(-p^{\prime},p,t). \nonumber 
\nonumber 
\end{align}
$\delta \hat{S}^{\pm}_{\mu}$ is the $ \pm 2k_F$ density fluctuation operator within electron pocket ($\mu=e$) 
or hole pocket ($\mu=h$), 
\begin{align}
&\hat{S}^{\pm}_{\mu}(Q_1,Q_2,p) \equiv \int \frac{dp_1}{2\pi} a^{\dagger}_{\mu_{\pm}}
(Q_1,p_1+p) a_{\mu_{\mp}}(Q_2,p_1), \nonumber \\
& \delta \hat{S}^{\pm}_{\mu}(Q_1,Q_2,p) \equiv \hat{S}^{\pm}_{\mu}(Q_1,Q_2,p)
- \frac{\langle \Psi_0 |  \hat{S}^{\pm}_{\mu}(Q_1,Q_2,p) | \Psi_0 \rangle}
{\langle \Psi_0 | \Psi_0 \rangle}. \nonumber 
\end{align}
Within the RPA, the time-ordered correlation function is given by a bare polarization function that carries 
$2k_F$ momentum;
\begin{align}
& D^{T,-+}_{\lambda\psi}(-p^{\prime},p,\omega) = \nonumber \\
& \hspace{0.2cm}\delta(p^{\prime}-p)  \Big\{ 
\delta\big(l(Q^{\prime\prime}_2-Q^{\prime\prime\prime}_1)\big) 
\delta\big(l(Q^{\prime\prime\prime}_2-Q^{\prime\prime}_1)\big) 
\Pi^{-+}_{0,\lambda}(\omega) \delta_{\lambda\psi} \nonumber \\
& \hspace{0.6cm} 
+ \frac{1}{2\pi \hbar} \Psi^{+-}_{\lambda\psi}(Q^{\prime\prime\prime}_2,Q^{\prime\prime}_2,
Q^{\prime\prime}_1,Q^{\prime\prime\prime}_1;I_{2k_F}) 
\!\ \Pi^{-+}_{0,\lambda}(\omega) \!\ \Pi^{-+}_{0,\psi}(\omega) \nonumber \\ 
&\hspace{-0.2cm} + \frac{1}{(2\pi \hbar)^2} 
\int d(l\tilde{Q}_{1}) d(l\tilde{Q}_2)  \!\
\Pi^{-+}_{0,\lambda}(\omega) \!\ \Psi^{+-}_{\lambda\rho}(\tilde{Q}_2,Q^{\prime\prime}_2,
Q^{\prime\prime}_1,\tilde{Q}_1;I_{2k_F})  \nonumber \\ 
&\hspace{0.4cm} 
\Pi^{-+}_{0,\rho}(\omega) \!\ \Psi^{+-}_{\rho\psi}(Q^{\prime\prime\prime}_{2},\tilde{Q}_1,
\tilde{Q}_2,Q^{\prime\prime\prime}_1;I_{2k_F}) \!\ 
 \Pi^{-+}_{0,\psi}(\omega) + \cdots \Big\}, \label{2k_F-rpa}
\end{align}
where the polarization function at $p_z=2k_F$, 
$\Pi^{-+}_{0,\lambda}(\omega=0)$, is given by Eq.~(\ref{2kf-bare-p}).
In terms of the homomorphic relation, Eq.~(\ref{homomorphic-1}), 
Eq.~(\ref{int-2}) with Eq.~(\ref{2k_F-rpa})
reduces to 
\begin{align}
\overline{H}_2 &=  \sum_{\mu,\nu} \int \frac{dp \!\ dp_1\!\ dp_2}{(2\pi)^3} 
\int d(lQ_1)  \cdots d(lQ_2)  \nonumber \\
& \hspace{-1.5cm} 
a^{\dagger}_{\mu_{+}}(Q^{\prime}_1,p_2+p) a_{\mu_{-}}(Q^{\prime}_2,p_2)  
a^{\dagger}_{\nu_{-}}(Q_1,p_1-p) a_{\nu_{+}}(Q_2,p_1)  \nonumber \\ 
& \hspace{0.2cm} 
\times \Phi^{+-}_{\mu\nu}(Q_1,Q^{\prime}_1,Q^{\prime}_{2},Q_{2};\overline{I}_{2k_{F}}),   
\label{screen-2a}
\end{align}
where $\overline{I}_{2k_F}(q_x,k_y)$ is given by Eq.~(\ref{screen-2b}).

\section{derivation of parquet RG equation}  
A derivation of the one-loop parquet RG equation can be implemented by a standard momentum 
shell renormalization. Thereby, we begin with a partition function of the interacting fermion 
model, Eqs.~(\ref{s0},\ref{s1}), and decompose the fermionic field into fast mode 
($e_{\pm,>}$, $h_{\pm,>}$) and slow mode ($e_{\pm,<}$, $h_{\pm,<}$) in the 
momentum space 
\begin{align}
e_{\pm}(Q,p,\omega) &= \left\{\begin{array}{cc} 
e_{\pm,<}(Q,p,\omega) & (|p|<\Lambda^{\prime}) \\
e_{\pm,>}(Q,p,\omega) & (\Lambda^{\prime} <|p|<\Lambda) \\
\end{array} \right.  \\
h_{\pm}(Q,p,\omega) &= \left\{\begin{array}{cc} 
h_{\pm,<}(Q,p,\omega) & (|p|<\Lambda^{\prime}) \\
h_{\pm,>}(Q,p,\omega) & (\Lambda^{\prime} <|p|<\Lambda) \\
\end{array} \right.  
\end{align}
with $\Lambda^{\prime} \equiv \Lambda e^{-\ln b}$. 
The integration of the fast mode in the partition function leads to a 
renormalization of the effective action for the slow mode,
\begin{align}
Z &= \int {\cal D}e_{<} {\cal D} h_{<} e^{-S_{0,<}-S_{1,<}} 
\int {\cal D}e_{>} {\cal D} h_{>} e^{-S_{0,>}-S_{1,>}} \nonumber \\
&= Z_{0,>} \int {\cal D}e_{<} {\cal D} h_{<} \!\ e^{-S_{0,<}-S_{1,<}} \nonumber \\ 
& \hspace{1.5cm} 
e^{-\langle S_{1,>}\rangle_{0,>} + \frac{1}{2} \big(\langle S^2_{1,>} \rangle_{0,>}
 - \langle S_{1,>}\rangle^2_{0,>}\big) + \cdots }, \label{total}
\end{align}
where 
\begin{align}
&\langle \cdots \rangle_{0,>} = \frac{1}{Z_{0,>}} \int {\cal D}e_{>} {\cal D} h_{>} e^{-S_{0,>}} \cdots, \nonumber \\
& Z_{0,>} \equiv  \int {\cal D}e_{>} {\cal D} h_{>} e^{-S_{0,>}}, \nonumber  
\end{align} 
and 
\begin{align}
S_{0,<(>)} = &\sum_{\sigma} \int \frac{dl\omega}{2\pi} \int_{|p|<\Lambda^{\prime}
(\Lambda^{\prime}<|p|<\Lambda)} dp \int dQ \nonumber \\
&  \ \ \big\{ (-i\omega + \sigma v_F p) 
e^{\dagger}_{\sigma,<(>)} e_{\sigma,<(>)} \nonumber \\
& \ \ + (-i\omega - \sigma v_F p)
 h^{\dagger}_{\sigma,<(>)} h_{\sigma,<(>)}\big\}. \nonumber 
\end{align} 
$S_{1,<}$ is the interaction part that is comprised only of the slow modes. 
$S_{1,>}$ is the other part of the interaction term that contains the fast modes. 
$\langle S_{1,>} \rangle_{0,>}$ in Eq.~(\ref{total}) renormalizes the Fermi velocity of 
the electron and hole pocket. Due to a particle-hole symmetry that exchanges the electron ahd hole 
bands ($m_e=m_h$), the renormalization of the Fermi velocity of the electron band and that of the 
hole band are identical to each other at the charge neutrality point. At the one-loop level of the 
renormalization group (RG), the Fermi velocity renormalization can be always absorbed into a scale 
change of the RG (see Eq.~(\ref{rg-scale})). Thereby, we do not keep track of the Fermi velocity 
renormalization from $\langle S_{1,>}\rangle_{0,>}$ in Eq.~(\ref{total}).  
 
$\langle S^2_{1,>} \rangle_{0,>} - \langle S_{1,>}\rangle^2_{0,>}$ gives rise to 
a renormalization of the interaction potentials. To calculate the renormalization, we have only 
to consider the following part of $S_{1,>}$,
\begin{align}
S_{1,>} &= \int_{1,2,3} \int dk_1 dq_1 dk_2 dq_2 \!\ e^{i{\bm k}_1 \wedge {\bm k}_2} 
W_{b}({\bm k}_1-{\bm k}_2) \nonumber \\ 
& \big\{ \!\ e^{\dagger}_{+,>} e^{\dagger}_{-,>}  e_{-,<} e_{+,<} 
+ e^{\dagger}_{+,<} e^{\dagger}_{-,<} e_{-,>} e_{+,>} \nonumber \\
& \ \ + e^{\dagger}_{+,>} e^{\dagger}_{-,<}  e_{-,>} e_{+,<} 
+ e^{\dagger}_{+,<} e^{\dagger}_{-,>}  e_{-,<} e_{+,>} \big\} \nonumber \\
& \!\ + \int_{1,2,3} \int dk_1 dq_1 dk_2 dq_2 \!\ e^{i{\bm k}_1 \wedge {\bm k}_2} 
W_{d}({\bm k}_1-{\bm k}_2) \nonumber \\ 
& \big\{ \!\ h^{\dagger}_{-,>} h^{\dagger}_{+,>}  h_{+,<} h_{-,<} 
+ h^{\dagger}_{-,<} h^{\dagger}_{+,<} h_{+,>} h_{-,>} \nonumber \\
& \ \ + h^{\dagger}_{-,>} h^{\dagger}_{+,<}  h_{+,>} h_{-,<} 
+ h^{\dagger}_{-,<} h^{\dagger}_{+,>} h_{+,<} h_{-,>} \big\} \nonumber \\
&\!\ + \int_{1,2,3} \int dk_1 dq_1 dk_2 dq_2 \!\ e^{i{\bm k}_1 \wedge {\bm k}_2} 
W_{e}({\bm k}_1-{\bm k}_2) \nonumber \\ 
& \big\{ \!\ e^{\dagger}_{+,>} h^{\dagger}_{+,>}  h_{+,<} e_{+,<} 
+ e^{\dagger}_{+,<} h^{\dagger}_{+,<} h_{+,>} e_{+,>} \nonumber \\
& \ \ + e^{\dagger}_{+,>} h^{\dagger}_{+,<}  h_{+,>} e_{+,<} 
+ e^{\dagger}_{+,<} h^{\dagger}_{+,>}  h_{+,<} e_{+,>} \big\} \nonumber \\
& \!\ + \int_{1,2,3} \int dk_1 dq_1 dk_2 dq_2 \!\ e^{i{\bm k}_1 \wedge {\bm k}_2} 
W_{e}({\bm k}_1-{\bm k}_2) \nonumber \\ 
& \big\{ \!\ h^{\dagger}_{-,>} e^{\dagger}_{-,>}  e_{-,<} h_{-,<} 
+ h^{\dagger}_{-,<} e^{\dagger}_{-,<} e_{-,>} h_{-,>} \nonumber \\
& \ \ + h^{\dagger}_{-,>} e^{\dagger}_{-,<}  e_{-,>} h_{-,<} 
+ h^{\dagger}_{-,<} e^{\dagger}_{-,>} e_{-,<} h_{-,>} \big\} \nonumber \\
&\!\ + \int_{1,2,3} \int dk_1 dq_1 dk_2 dq_2 \!\ e^{i(k_1 q_1+k_2 q_2)} 
W_{g}({\bm k}_1-{\bm k}_2) \nonumber \\ 
& \big\{ \!\ e^{\dagger}_{+,<} h^{\dagger}_{-,>}  h_{+,>} e_{-,<} 
+ e^{\dagger}_{+,<} h^{\dagger}_{-,>} h_{+,<} e_{-,>} \nonumber \\
& \ \ + e^{\dagger}_{+,>} h^{\dagger}_{-,<}  h_{+,>} e_{-,<} 
+ e^{\dagger}_{+,>} h^{\dagger}_{-,<}  h_{+,<} e_{-,>} \big\} \nonumber \\
& \!\ + \int_{1,2,3} \int dk_1 dq_1 dk_2 dq_2 \!\ e^{-i(k_1q_1+k_2q_2)} 
W^{*}_{g}({\bm k}_1-{\bm k}_2) \nonumber \\ 
& \big\{ \!\ h^{\dagger}_{+,<} e^{\dagger}_{-,>}  e_{+,>} h_{-,<} 
+ h^{\dagger}_{+,<} e^{\dagger}_{-,>} e_{+,<} h_{-,>} \nonumber \\
& \hspace{-0.1cm} + h^{\dagger}_{+,>} e^{\dagger}_{-,<}  e_{+,>} h_{-,<} 
+ h^{\dagger}_{+,>} e^{\dagger}_{-,<} e_{+,<} h_{-,>} \big\}, \label{s1>}
\end{align}  
(the others do not contribute to the renormalization of the 
interaction potentials at the one-loop level RG).

$\langle S^2_{1,>} \rangle_{0,>} - \langle S_{1,>}\rangle^2_{0,>}$ in Eq.~(\ref{total}) 
gives out the following one-loop renormalization to the interaction potentials in Eqs.~(\ref{s1});
\begin{align}
&dW_b({\bm k}) = \frac{1}{(2\pi)^3 l^2} \frac{d\Lambda}{v_F\Lambda} 
\int d{\bm k}^{\prime} \!\ \nonumber \\
& \hspace{1.7cm} \Big\{ 
 \!\ W_{b}({\bm k}^{\prime}) W_{b}({\bm k}-{\bm k}^{\prime}) \big( 1 - 
e^{-i{\bm k} \wedge {\bm k}^{\prime}} \big)  \nonumber \\
& \hspace{1.6cm} + \!\ W_{g}(k^{\prime},-q^{\prime}) 
W^{*}_{g}(k-k^{\prime},-q+q^{\prime}) \Big\},  \label{wb-re} \\
&dW_d({\bm k}) = \frac{1}{(2\pi)^3 l^2} \frac{d\Lambda}{v_F\Lambda} 
\int d{\bm k}^{\prime} \!\  \nonumber \\
& \hspace{1.7cm} 
\Big\{ W_{d}({\bm k}^{\prime}) W_{d}({\bm k}-{\bm k}^{\prime}) \Big( 1 - 
e^{-i{\bm k} \wedge {\bm k}^{\prime}} \Big)  \nonumber \\
& \hspace{1.6cm} + \!\ W_{g}(k^{\prime},-q^{\prime}) 
W^{*}_{g}(-k+k^{\prime},q-q^{\prime}) \Big\},  \label{wd-re} \\
&dW_e({\bm k}) = \frac{1}{(2\pi)^3 l^2} \frac{d\Lambda}{v_F\Lambda}  
\int d{\bm k}^{\prime} \!\ \nonumber \\
& \hspace{1.7cm}  
\Big\{ W_{e}({\bm k}^{\prime}) W_{e}({\bm k}-{\bm k}^{\prime}) \Big( 1 - 
e^{-i{\bm k} \wedge {\bm k}^{\prime}} \Big)  \nonumber \\
& \hspace{1.5cm} + \!\ e^{-ikq+ikq^{\prime}+ik^{\prime}q} 
W_{g}({\bm k}^{\prime}) 
W^{*}_{g}({\bm k}-{\bm k}^{\prime}) \Big\},  \label{we-re} \\
&dW_g({\bm k}) = \frac{1}{(2\pi)^3 l^2} \frac{d\Lambda}{v_F\Lambda} \int d{\bm k}^{\prime} \!\ 
 W_{g}({\bm k}-{\bm k}^{\prime}) \nonumber \\
& \hspace{1.2cm} 
\Big\{ W_{b}(k^{\prime},-q^{\prime}) + W_{d}(-k^{\prime},q^{\prime})  \nonumber \\
& \hspace{1.2cm} +e^{-ikq^{\prime}-ik^{\prime}q+ik^{\prime}q^{\prime}}
\big(W_{e}({\bm k}^{\prime})+W_{e}(-{\bm k}^{\prime})\big)  
\Big\},  \label{wg-re} 
\end{align}
with $d\Lambda \equiv \Lambda \ln b$ ${\bm k}\equiv (k,q)$,  
${\bm k}^{\prime} \equiv (k^{\prime},q^{\prime})$ and $d{\bm k}^{\prime} \equiv dk^{\prime} dq^{\prime}$.  
After the integration of the fast modes, we scale the momentum along the field ($p$), single-particle 
frequency ($\omega$) and the field operators ($e_{\sigma}$ and $h_{\sigma}$) as 
\begin{align}
&p =  p^{\prime}/b, \ \omega = \omega^{\prime}/b,  \nonumber \\ 
&e_{\sigma}(Q,p,\omega) = e^{\frac{3}{2} \ln b} e^{\prime}_{\sigma}(Q,p^{\prime},\omega^{\prime}), \nonumber \\ 
&h_{\sigma}(Q,p,\omega) = e^{\frac{3}{2} \ln b} h^{\prime}_{\sigma}(Q,p^{\prime},\omega^{\prime}). \label{scale} 
\end{align}
This (tree-level) scale change keeps $S_{0,<}$ as well as $S_{1,<}$ 
to be invariant, while putting $\Lambda^{\prime}$ in $S_{0,<}$ and $S_{1,<}$
back to $\Lambda$. Accordingly, Eqs.~(\ref{wb-re},\ref{wd-re},\ref{we-re},\ref{wg-re}) 
lead to the following one-loop renormalization group equations for the interaction potentials,
\begin{align} 
\frac{dW_{b}({\bm k})}{d\xi} &= \int d{\bm k}^{\prime} \!\ W_{b}({\bm k}^{\prime}) 
W_{b}({\bm k}-{\bm k}^{\prime}) \big(1-e^{-i{\bm k}\wedge {\bm k}^{\prime}}\big)  \nonumber \\
& \hspace{-0.2cm} +  \int d{\bm k}^{\prime} \!\ W_{g}(k^{\prime},-q^{\prime}) 
W^{*}_{g}(k-k^{\prime},-q+q^{\prime}) \label{wb-rg1} \\ 
\frac{dW_{d}({\bm k})}{d\xi} &= \int d{\bm k}^{\prime} \!\ W_{d}({\bm k}^{\prime}) 
W_{d}({\bm k}-{\bm k}^{\prime}) \big(1-e^{-i{\bm k}\wedge {\bm k}^{\prime}}\big)  \nonumber \\
& \hspace{-0.2cm} +  \int d{\bm k}^{\prime} \!\ W_{g}(k^{\prime},-q^{\prime}) 
W^{*}_{g}(-k+k^{\prime},q-q^{\prime}) \label{wd-rg1} \\ 
\frac{dW_{e}({\bm k})}{d\xi} &= \int d{\bm k}^{\prime} \!\ W_{e}({\bm k}^{\prime}) 
W_{e}({\bm k}-{\bm k}^{\prime}) \big(1-e^{-i{\bm k}\wedge {\bm k}^{\prime}}\big)  \nonumber \\
& \hspace{-0.4cm} +  \int d{\bm k}^{\prime} \!\ e^{-ikq+ik^{\prime}q+ikq^{\prime}} W_{g}({\bm k}^{\prime}) 
W^{*}_{g}({\bm k}-{\bm k}^{\prime}) \label{we-rg1} \\ 
\frac{dW_{g}({\bm k})}{d\xi} &= \int d{\bm k}^{\prime} 
\Big\{ W_{b}(k^{\prime},-q^{\prime}) + W_{d}(-k^{\prime},q^{\prime}) \nonumber \\
& \hspace{-0.9cm} +e^{-ikq^{\prime}-ik^{\prime}q+ik^{\prime}q^{\prime}}
\big(W_{e}({\bm k}^{\prime})+W_{e}(-{\bm k}^{\prime})\big) \Big\} W_{g}({\bm k}-{\bm k}^{\prime})  
 \label{wg-rg1}  
\end{align}  
with 
\begin{align}
d\xi \equiv \frac{1}{(2\pi)^3 l^2} \frac{d\Lambda} {v_F \Lambda}. \label{rg-scale}
\end{align}
Note that the above one-loop RG equations 
as well as the initial forms of the interaction potentials, Eqs.~(\ref{wbd},\ref{we},\ref{wg}),  
respect the following symmetries, 
\begin{align}
&W^{*}_{b}(k,q) = W_{b}(k,q) = W_{b}(k,-q)=W_{b}(-k,q), \nonumber \\
&W^{*}_{d}(k,q) = W_d(k,q) = W_d(k,-q) = W_{d}(-k,q), \nonumber \\
&W^{*}_e(k,q) = W_e(k,-q) = W_e(-k,q), \nonumber \\
&W^{*}_g(k,q) = W_g(k,-q) = W_{g}(-k,q). \nonumber 
\end{align} 
Using these symmetries, the RG equations 
can be also written in Eqs.~(\ref{wb-rg2},\ref{wd-rg2},\ref{we-rg2},\ref{wg-rg2}).

Consider the Fourier transform of $W_{\mu}({\bm k})$,
\begin{align}
&F_{\mu}({\bm r}) \equiv \int d{\bm k} e^{-i{\bm k}{\bm r}} W_{\mu}({\bm k}),  \label{f1} \\
&W_{\mu}({\bm k}) \equiv \int \frac{d{\bm r}}{(2\pi)^2} e^{i{\bm k}{\bm r}} F_{\mu}({\bm r}), \label{f2-s} 
\end{align} 
for $\mu=b,d,e,g$ with 
\begin{align}
\tilde{F}_{g}({\bm r}) \equiv e^{-ir_x r_y} F_{g}({\bm r}), \label{f3}
\end{align}
and ${\bm r}\equiv (r_x,r_y)$, ${\bm k}\equiv (k,q)$. In terms of this dual representation, 
Eqs.~(\ref{wb-rg2},\ref{wd-rg2},\ref{we-rg2},\ref{wg-rg2}) reduce to 
\begin{align}
\frac{dF_{b/d}({\bm r})}{d\xi} &= F^2_{b/d}({\bm r}) + \tilde{F}_g({\bm r}) \tilde{F}_{g}(-{\bm r}) \nonumber \\
& \hspace{-1.6cm} 
- \int\frac{d{\bm r}^{\prime}d{\bm r}^{\prime\prime}}{(2\pi)^2} F_{b/d}({\bm r}^{\prime}) 
F_{b/d}({\bm r}^{\prime\prime}) e^{-i{\bm r}\wedge {\bm r}^{\prime} 
- i {\bm r}^{\prime} \wedge {\bm r}^{\prime\prime} -i {\bm r}^{\prime\prime} \wedge {\bm r}}, \label{rg-a-1} \\
\frac{dF_{e}({\bm r})}{d\xi} &= F^2_{e}({\bm r}) +  \nonumber \\
& \hspace{-1.4cm} 
+ \int \frac{d{\bm r}^{\prime}d{\bm r}^{\prime\prime}}{(2\pi)^2} \tilde{F}_{g}({\bm r}^{\prime}) 
\tilde{F}_{g}(-{\bm r}^{\prime\prime}) e^{i (r_x r^{\prime}_y+r^{\prime}_x r_y)-i(r_x r^{\prime\prime}_y 
+ r^{\prime\prime}_x r_y)} \nonumber \\ 
& \hspace{-1.6cm} 
- \int\frac{d{\bm r}^{\prime}d{\bm r}^{\prime\prime}}{(2\pi)^2} F_{e}({\bm r}^{\prime}) 
F_{e}({\bm r}^{\prime\prime}) e^{-i{\bm r}\wedge {\bm r}^{\prime} 
- i {\bm r}^{\prime} \wedge {\bm r}^{\prime\prime} -i {\bm r}^{\prime\prime} \wedge {\bm r}}, \label{rg-a-2} \\
\frac{d\tilde{F}_{g}({\bm r})}{d\xi} &= \tilde{F}_{g}({\bm r}) \big(F_{b}({\bm r}) + F_{d}({\bm r})\big) +  \nonumber \\
& \hspace{-1.4cm} 
+ 2 \!\ \int \frac{d{\bm r}^{\prime}d{\bm r}^{\prime\prime}}{(2\pi)^2} F_{e}({\bm r}^{\prime}) 
\tilde{F}_{g}({\bm r}^{\prime\prime}) e^{-i (r^{\prime}_x r_y+r_x r^{\prime}_y)+i(r^{\prime\prime}_x r^{\prime}_y 
+ r^{\prime}_x r^{\prime\prime}_y)}.  \label{rg-a-3} 
\end{align}  
From Eqs.~(\ref{wbd},\ref{we},\ref{wg}), the initial function forms for $F_{\mu}({\bm r})$ ($\mu=b,d,e$) 
and $\tilde{F}_{g}({\bm r})$ are as follows, 
\begin{align}
&F_{b}({\bm r}) = F_{d}({\bm r}) \nonumber \\
& \ \ = \int d{\bm k} e^{-i{\bm k}{\bm r}} \overline{I}_0(q,k) - 
2\pi \overline{I}_{2k_F}(r_x,-r_y), \label{wbd-a} \\
& F_{e}({\bm r}) = \int d{\bm k} e^{-i{\bm k}{\bm r}} \overline{I}_0(q,k), \label{we-a} \\
& \tilde{F}_g({\bm r}) = 2\pi \overline{I}_{2k_F}(r_x,r_y), \label{wg-a} 
\end{align}    
with ${\bm k}\equiv (k,q)$. These initial forms as well as the RG equations in the dual space 
respect the following symmetries,
\begin{align}
&F_{\mu}(r_x,r_y) = F^{*}_{\mu}(r_x,r_y) = F_{\mu}(-r_x,r_y) = F_{\mu}(r_x,-r_y), \label{sym1} \\
&\tilde{F}_{g}(r_x,r_y) =\tilde{F}^{*}_{g}(r_x,r_y) = \tilde{F}_{g}(-r_x,r_y) = \tilde{F}_{g}(r_x,-r_y). \label{sym2}
\end{align}
Accordingly, Eqs.~(\ref{rg-a-1},\ref{rg-a-2},\ref{rg-a-3}) can be rewritten into more symmetric forms,
\begin{align}
\frac{dF_{b/d}({\bm r})}{d\xi} &= F^2_{b/d}({\bm r}) + \tilde{F}_g({\bm r}) \tilde{F}_{g}({\bm r}) \nonumber \\
& \hspace{-1.6cm} 
- \int\frac{d{\bm r}^{\prime}d{\bm r}^{\prime\prime}}{(2\pi)^2} F_{b/d}({\bm r}^{\prime}) 
F_{b/d}({\bm r}^{\prime\prime}) e^{-i{\bm r}\wedge {\bm r}^{\prime} 
- i {\bm r}^{\prime} \wedge {\bm r}^{\prime\prime} -i {\bm r}^{\prime\prime} \wedge {\bm r}}, \label{rg-b-1} \\
\frac{dF_{e}({\bm r})}{d\xi} &= F^2_{e}({\bm r}) +
\int \frac{d{\bm r}^{\prime}d{\bm r}^{\prime\prime}}{(2\pi)^2} \tilde{F}_{g}({\bm r}^{\prime}) 
\tilde{F}_{g}({\bm r}^{\prime\prime}) e^{i ({\bm r}\wedge {\bm r}^{\prime}
-{\bm r}^{\prime\prime}\wedge {\bm r})} \nonumber \\ 
& \hspace{-1.cm} 
- \int\frac{d{\bm r}^{\prime}d{\bm r}^{\prime\prime}}{(2\pi)^2} F_{e}({\bm r}^{\prime}) 
F_{e}({\bm r}^{\prime\prime}) e^{-i{\bm r}\wedge {\bm r}^{\prime} 
- i {\bm r}^{\prime} \wedge {\bm r}^{\prime\prime} -i {\bm r}^{\prime\prime} \wedge {\bm r}}, \label{rg-b-2} \\
\frac{d\tilde{F}_{g}({\bm r})}{d\xi} &= \tilde{F}_{g}({\bm r}) \big(F_{b}({\bm r}) + F_{d}({\bm r})\big) +  \nonumber \\
& \hspace{-0.5cm} 
+ 2 \!\ \int \frac{d{\bm r}^{\prime}d{\bm r}^{\prime\prime}}{(2\pi)^2} F_{e}({\bm r}^{\prime}) 
\tilde{F}_{g}({\bm r}^{\prime\prime}) e^{i ({\bm r}\wedge {\bm r}^{\prime}
+{\bm r}^{\prime}\wedge {\bm r}^{\prime\prime})}.  \label{rg-b-3} 
\end{align}
The RG equations thus obtained as well as the initial forms have the following O(2) symmetry;
\begin{align}
&F_{\mu}(\hat{R}_{\theta} {\bm r}) = F_{\mu}({\bm r}) \equiv \Gamma_{\mu}(r), \label{Gamma1-s} \\ 
&\tilde{F}_{g}(\hat{R}_{\theta} {\bm r}) = \tilde{F}_{g}({\bm r}) \equiv \Gamma_{g}(r),  \label{Gamma2-s} \\
&\hat{R}_{\theta} \equiv \left(\begin{array}{cc} 
\cos \theta & \sin\theta \\
-\sin \theta & \cos \theta \\
\end{array}\right). \label{u1}
\end{align}
with $r\equiv |{\bm r}|$ for $\mu=b,d,e$ and 
arbitrary $\theta \in (0,2\pi]$. Utilizing this symmetry, we can reduce Eqs.~(\ref{rg-b-1},\ref{rg-b-2},\ref{rg-b-3})
into the RG equations for $\Gamma_{\mu}(r)$  ($\mu=b,d,e$) and 
$\Gamma_{g}(r)$, Eqs.~(\ref{gamma-bd},\ref{gamma-e},\ref{gamma-g}).

\end{document}